\documentstyle[12pt]{article}
\newcommand{\ret}{\nonumber \\}

\newcommand{\Section}[1]%
{\section{#1}\setcounter{equation}{0}%
\setcounter{theorem}{0}}
\newtheorem{theorem}{Theorem}
\newtheorem{lemma}[theorem]{Lemma}
\newtheorem{coro}[theorem]{Corollary}
\newtheorem{pro}[theorem]{Proposition}

\newenvironment{proof}[1]%
{\par\noindent{\em #1:\ }}%
{~\rule{2mm}{2mm}\par\bigskip}
\begin{document}
\newpage\thispagestyle{empty}
{\topskip 2cm
\begin{center}
{\Large\bf Insensitivity of Quantized Hall Conductance\\ 
\bigskip
to Disorder and Interactions\\}
\bigskip\bigskip
{\large Tohru Koma$^\ast$\\}
\bigskip
\end{center}
\vfil
\noindent
A two-dimensional quantum Hall system is studied for a wide class of 
potentials including single-body random potentials and repulsive 
electron-electron interactions. We assume that there exists a non-zero 
excitation gap above the ground state(s), and then the conductance is derived 
from the linear perturbation theory with a sufficiently weak electric field. 
Under these two assumptions, we proved that the Hall conductance 
$\sigma_{xy}$ and the diagonal conductance $\sigma_{yy}$ satisfy 
$|\sigma_{xy}+e^2\nu/h|\le{\rm const.}L^{-1/12}$ and 
$|\sigma_{yy}|\le{\rm const.}L^{-1/12}$. 
Here $e^2/h$ is the universal conductance with the charge $-e$ of electron 
and the Planck constant $h$; $\nu$ is the filling factor of the Landau level, 
and $L$ is the linear dimension of the system. In the thermodymanic limit, 
our results show $\sigma_{xy}=-e^2\nu/h$ and $\sigma_{yy}=0$. 
The former implies that integral and fractional filling factors $\nu$ with 
a gap lead to, respectively, integral and fractional quantizations 
of the Hall conductance. 
\par\noindent
\bigskip
\hrule
\bigskip
\noindent
{\bf KEY WORDS:} Integral quantum Hall effect; fractional quantum Hall effect; 
Hall conductance; Landau Hamiltonian; random potential; 
electron-electron interaction. 
\hfill
\bigskip 

\noindent
---$\!$---$\!$---$\!$---
\medskip

\noindent
$^\ast$ {\small Department of Physics, Gakushuin University, 
Mejiro, Toshima-ku, Tokyo 171, JAPAN}
\smallskip

\noindent
{\small\tt e-mail: koma@riron.gakushuin.ac.jp}
\vfil}\newpage
\tableofcontents
\newpage
\Section{Introduction}
\label{Introduction}

The quantum Hall effect\footnote{For the history of the quantum 
Hall effect, see reviews \cite{PGBook,CPBook,SPBook}.} 
is one of the most remarkable phenomena 
discovered in solid state physics. 
The effect is observed in two-dimensional electrons gases 
subjected to a uniform perpendicular magnetic field. 
Experimentally, such systems are realized at interfaces 
in semiconductors. The first experiments of the resistivity 
in a two-dimensional electron system in a magnetic field were performed 
by Kawaji, Igarashi and Wakabayashi \cite{KIW} and Igarashi, Wakabayashi and 
Kawaji \cite{IWK} in 1975. Unfortunately the quality of the samples 
in their early experiments\footnote{See also experiments 
\cite{earlierexperiments}.} had not reached the stage 
where plateaus of finite width for the Hall resistivity could be obtained. 
In the same 1975, Ando, Matsumoto and Uemura \cite{AMU} also were studying 
a two-dimensional electron system with disorder in a magnetic field, 
and theoretically predicted some aspects of the quantum Hall effect. 
However, they did not expect that the Hall resistivity of the plateaus 
is almost precisely quantized.  

Soon after these studies, the integral quantum Hall effect was 
discovered \cite{KDP,KawajiWakabayashi}, and the fractional quantum Hall 
effect was subsequently discovered \cite{TSG,SCTHG}. Most aspects of 
the integral quantum Hall effect may be understood with an essentially 
single electron description, in which electron-electron interactions 
play only a secondary role. Actually early theoretical works 
\cite{earlytheoretical,TKNN,Kohmoto} for explaining the integral quantum Hall 
effect were done along this line. 
In particular, a smashing idea of a topological invariant\footnote{We should 
remark that Dubrovin and Novikov \cite{DN} preceded the article \cite{TKNN} 
and found a topologically nontrivial vector bundle and a topological invariant 
in a two-dimensional periodic Schr\"odinger operator with a magnetic field, 
although they did not treat the conductance.}
for the conductance was introduced by Thouless, Kohmoto, Nightingale and 
den Nijs \cite{TKNN} and Kohmoto \cite{Kohmoto}. 
After their article \cite{TKNN}, there appeared many variants 
\cite{topologicalvariant,NTW,AvronSeiler,AvronYaffe,Ishikawa} 
of their argument. 
In particular, some of the arguments were extended to a quantum Hall 
system with electron-electron interactions 
\cite{NTW,AvronSeiler,AvronYaffe,Ishikawa}. 
However, the results always show an integral quantization 
of the Hall conductance without ad hoc assumptions \cite{NTW}. 
It is questionable that the fractional quantization of the Hall conductance 
can be understood with a topological invariant of the Hall conductance. 

For giving an explanation of the fractional quantum Hall effect, 
the difficulty comes from the fact that the electron-electron 
interaction is essential to this phenomenon. 
In order to overcome the difficulty, it is necessary to clarify 
the nature of the ground state(s) and of the low energy excitations 
for a strongly interacting electrons gas in a uniform magnetic field. 
For such a system, there appeared many approximate theories, 
trial functions for the ground state(s), perturbative approaches, 
mean field approximations, numerical analysis, etc \cite{fractionaltheory}. 
However, the quantum Hall effect, in particular, the fractional quantization 
of the Hall conductance plateaus, is still not explained theoretically 
with a model of an interacting electrons gas in a uniform magnetic field. 

In this paper, we consider a two-dimensional electrons gas in a uniform 
magnetic field with disorder and electron-electron interactions. 
The model is defined on an $L_x\times L_y$ rectangular box with 
periodic boundary conditions. 
The explicit form of the Hamiltonian is given by (\ref{totHam0}) in 
Section~\ref{mainresults}. 
We assume the existence of a non-zero excitation gap 
above the ground state(s). The existence of a gap is believed to be 
essential to the fractional quantization of the Hall conductance. 
Further we assume that an applied electric field to induce a Hall current 
is sufficiently weak so that the conductance is derived as the linear 
response coefficients from the linear perturbation theory. 
Under these two assumptions, we proved that the Hall conductance 
$\sigma_{xy}$ and the diagonal conductance $\sigma_{yy}$ satisfy 
\begin{equation}
\left|\sigma_{xy}+\frac{e^2}{h}\nu\right|\le{\rm const.}L_x^{5/24}L_y^{-7/24}, 
\quad \left|\sigma_{yy}\right|\le{\rm const.}L_x^{5/24}L_y^{-7/24}. 
\label{result10}
\end{equation}
Here $\nu$ is the filling factor of the Landau level, and $e^2/h$ is 
the universal conductance with the charge $-e$ of electron 
and the Planck constant $h$. In particular, 
\begin{equation}
\left|\sigma_{xy}+\frac{e^2}{h}\nu\right|\le{\rm const.}L^{-1/12}, 
\quad \left|\sigma_{yy}\right|\le{\rm const.}L^{-1/12}\quad \mbox{for}\ 
L_x=L_y=L. 
\label{resultssqbox}
\end{equation}
Clearly we have 
\begin{equation}
\sigma_{xy}=-\frac{e^2}{h}\nu, \quad \sigma_{yy}=0 
\end{equation}
in the thermodymanic limit $L\rightarrow +\infty$. 
In the next Section~\ref{mainresults}, the precise statements of these 
results and the precise definitions of the conductance 
$\sigma_{xy},\sigma_{yy}$ and of the filling factor $\nu$ will be given 
in a mathematically rigorous manner, and we will see that our results 
are justified for a wide class of potentials which includes 
single-body potentials with disorder and repulsive electron-electron 
interactions decaying by a power law. But the class does not include 
the standard Coulomb interaction proportional to $1/r$, where $r$ is 
the distance between two electrons. 

A reader may think that the finite-size corrections in the upper bounds 
of (\ref{resultssqbox}) are too large in comparison to 
the precision of the experimentally measured conductance. 
Actually the true finite-size corrections are expected to be 
exponentially small as in ref.~\cite{NiuThouless}. 
But it is very hard to prove the corresponding statement 
in a mathematically rigorous manner. 

Having the result (\ref{result10}) in mind, let us discuss which filling 
factor $\nu$ leads to a spectral gap above the ground state(s). 
For this purpose, we briefly state a result of our separate paper 
\cite{Koma1}. In the paper, we treated a two-dimensional quantum Hall system 
with electron-electron interactions and without disorder. 
The model is defined on an infinitely long strip with a large width, 
and the Hilbert space is restricted to the lowest $(n_{\rm max}+1)$ Landau 
levels with a large integer $n_{\rm max}$. The explicit form 
of the Hamiltonian is given in Section~\ref{mainresults}. 
In the infinite volume, we assumed the absence of non-translationally 
invariant infinite-volume ground state. Then we obtained 
the following result \cite{Koma1}: 
\begin{itemize}
\item If a pure infinite-volume ground state has a non-zero excitation gap, 
then the filling factor $\nu$ must be equal to a rational number.\footnote{
\label{footnoteevendenomi} In ref.~\cite{Koma1}, we also gave 
a phenomenological explanation for the reason 
why odd denominators of filling factions giving the quantized Hall 
conductance, are favor exclusively.}
\end{itemize}
Although we have considered the system without disorder, 
we can expect that, for the presence of weak disorder, the gap persists 
against the disorder. In this situation with the weak disorder, 
we get a rational quantization of the Hall conductance 
\begin{equation}
\sigma_{xy}=-\frac{e^2}{h}\nu \quad \mbox{with a rational filling}\ \nu,
\end{equation}
and the vanishing diagonal conductance 
\begin{equation}
\sigma_{yy}=0
\end{equation}
from the result (\ref{result10}). Further, in order to discuss the 
appearance of the Hall conductance plateaus, 
we change the filling factor $\nu$ slightly from the above rational value. 
Then, if the electrons of low energy 
excitations do not contribute to the current flow owing to the disorder, 
we can expect that the rationally quantized value of the Hall 
conductance $\sigma_{xy}$ remains constant, 
i.e., there appears a plateau of the Hall conductance, with the vanishing 
diagonal conductance. 
The appearance of such a plateau will be discussed in relation to 
localization of wavefunctions in another separate paper \cite{Koma2}. 

This paper is organized as follows: 
In Section~\ref{mainresults}, we give the precise definition of the model 
and describe our main theorems in a mathematically rigorous manner. 
As preliminaries for the proofs of our theorems, we briefly review 
the eigenvalue problem of the single-electron Landau Hamiltonian 
and treat the Landau Hamiltonian with an electric field 
in Section~\ref{preliminaries}. In Section~\ref{relation}, we discuss 
the relation between the electric potential of the present paper 
and the standard time-dependent vector potential. 
In Section~\ref{quantumHallsystem}, 
we calculate the current density by using the Rayleigh-Schr\"odinger 
perturbation theories with a sufficiently weak electric field, 
and give the proofs of our main theorems. 
For the convenience of readers, Appendices~\ref{perturbation}-\ref{estE-E} 
are devoted to technical estimates and calculations of matrix elements 
appeared in our representation of the conductivities (conductance).

\Section{The model and the main results}
\label{mainresults}

We study a two-dimensional interacting $N$ electrons system with 
a disorder potential $V_\omega$ in a unifrom magnetic field $(0,0,B)$ 
perpendicular to the $x$-$y$ plane in which the electrons are confined, 
and in an electric field $(0,F,0)$ oriented along the $y$ axis. 
For simplicity we assume that the electrons do not have spin degrees of 
freedom, although we can treat a quantum Hall system with spin degrees of 
freedom or with multiple layers in the same way. The Hamiltonian we consider 
in this paper is given by 
\begin{eqnarray}
H_\omega^{(N)}
&=&\sum_{j=1}^N\left[\frac{1}{2m_e}\left(p_{x,j}-eBy_j+A_0\right)^2
+\frac{1}{2m_e}p_{y,j}^2+V_\omega({\bf r}_j)\right]\ret
&+&\sum_{1\le i<j\le N}U^{(2)}(x_i-x_j,y_i-y_j)+\sum_{j=1}^NeFy_j
P_{{\rm bulk},j},
\label{totHam0}
\end{eqnarray}
where $-e$ and $m_e$ are, respectively, the charge of electron and the mass 
of electron, and $A_0$ is a real gauge parameter; ${\bf r}_j:=(x_j,y_j)$ 
is the $j$ th Cartesian coordinate of the $N$ electrons. As usual, we define 
\begin{equation}
p_{x,j}=-i\hbar \frac{\partial}{\partial x_j}, \quad \mbox{and} \quad 
p_{y,j}=-i\hbar \frac{\partial}{\partial y_j}
\end{equation}
with the Planck constant $\hbar$. The system is defined on a 
rectangular box 
\begin{equation}
S:=[-L_x/2,L_x/2]\times[-L_y/2,L_y/2] 
\end{equation}
with periodic boundary conditions. 
We have intreoduced a projection operator $P_{{\rm bulk},j}$ so that 
the electrons near boundaries $y=\pm L_y/2$ do not feel the infinitely
strong electric fields at the boundaries. Since the projection 
$(1-P_{{\rm bulk},j})$ acts on a wavefunction at only a neighborhood of the 
boundaries, we can expect that the effect of the projection is 
negligible\footnote{It is very difficult to give a proof for the statement 
that the boundary effect is negligible. In fact, we still cannot prove 
the claim.} in the thermodynamic limit 
$L_y\rightarrow +\infty$. In Section~\ref{Landauhamele}, 
we will give the precise definition of $P_{{\rm bulk},j}$, 
and show that the corresponding electric field is constant except 
for the neighborhood of the boundaries. In Section~\ref{relation}, 
we will discuss the relation between the regularized electric potential and 
the standard time-dependent vector potential. The latter yields the constant 
electric field on the whole torus. 

We assume that the single-body potential $V_\omega$ with disorder satisfies 
the following conditions: periodic boundary conditions 
\begin{equation}
V_\omega(x+L_x,y)=V_\omega(x,y+L_y)=V_\omega(x,y),
\label{PBCVomega}
\end{equation}
and 
\begin{equation}
\Vert V_\omega\Vert<V_0<\infty, 
\label{boundedVomega}
\end{equation}
where $V_0$ is a positive constant 
which is independent of the linear dimensions $L_x,L_y$ of the system. 
The potential $V_\omega$ consists of a random part $V^{\rm ran}_\omega$ 
and a regular part $W$ as 
\begin{equation}
V_\omega(x,y)=V^{\rm ran}_\omega(x,y)+W(x).
\label{decomVomega}
\end{equation}
The regular part $W$ is a function of $x$ only such that $W$ satisfies 
\begin{equation}
W(x+L_x)=W(x)=W(-x).
\end{equation}
A simple example of $W$ is given by\footnote{The question of the applicability 
of our method to a quantum Hall system with a periodic potential was brought 
to the author by Mahito Kohmoto. Thus we have partially answered his 
question, althuough we still cannot treat a periodic potential modulating 
in both $x$ and $y$ directions.} 
\begin{equation}
W(x)=W_0\cos\kappa x \quad\mbox{with}\ \kappa=\frac{2\pi}{L_x}n, \ n\in{\bf Z},
\end{equation}
where $W_0$ is a real constant. 

The electron-electron interaction $U^{(2)}$ satisfies 
\begin{equation}
U^{(2)}(-x,-y)=U^{(2)}(x,y). 
\end{equation}
We impose periodic boundary conditions as 
\begin{equation}
U^{(2)}(x+L_x,y)=U^{(2)}(x,y+L_y)=U^{(2)}(x,y). 
\label{PBCU2}
\end{equation}
We assume that $U^{(2)}$ is two times continuously differentiable 
on ${\bf R}^2$, and satisfies 
\begin{equation}
\left|\frac{\partial^2}{\partial x^2}U^{(2)}(x,y)\right|
+\left|\frac{\partial^2}{\partial y^2}U^{(2)}(x,y)\right|
\le \alpha U^{(2)}(x,y)\quad \mbox{for any}\ (x,y)\in{\bf R}^2, 
\label{U2assumption}
\end{equation}
with a positive constant $\alpha$ which is independent of the linear 
dimensions $L_x,L_y$ of the system. Further we assume that 
\begin{equation}
U^{(2)}(x,y)\le U_0 \left\{1+\left[{\rm dist}(x,y)/r_0\right]^2
\right\}^{-\gamma/2}
\quad \mbox{with} \ U_0>0, \ \gamma>2, \ r_0>0, 
\label{powerdecayassumptionU2}
\end{equation}
where the distance is given by 
\begin{equation}
{\rm dist}(x,y):=\sqrt{\min_{m\in{\bf Z}}\{|x-mL_x|^2\}+
\min_{n\in{\bf Z}}\{|y-nL_y|^2\}}.
\end{equation}
A simple example of $U^{(2)}$ satisfying these conditions is 
\begin{equation}
U^{(2)}(x,y)=\frac{U_0}{\left[1+(r/r_0)^2\right]^{\gamma/2}}\quad 
\mbox{with} \ \gamma>2, 
\end{equation}
where $U_0$ and $r_0$ are posiitve constants, and 
\begin{equation}
r=\sqrt{\left(\frac{L_x}{\pi}\right)^2\sin^2\frac{\pi}{L_x}x
+\left(\frac{L_y}{\pi}\right)^2\sin^2\frac{\pi}{L_y}y}.
\end{equation}
In the limit $L_x,L_y\rightarrow\infty$, we have the usual Euclidean 
distance $r=\sqrt{x^2+y^2}$. 

We take $L_xL_y=2\pi M\ell_B^2$ with a sufficiently large positive 
integer $M$. Here $\ell_B$ is the so-called magnetic length defined as 
$\ell_B:=\sqrt{\hbar/eB}$. The number $M$ is equal to 
the number of the states in a single Landau level of the single-electron 
Hamiltonian in the uniform magnetic field with no single-body potential, 
and with no electric field. For simplicity, we take $M$ even. 
We define the filling factor $\nu$ as $\nu=N/M$. We assume $\nu<\nu_0$, 
where $\nu_0$ is a positive constant which is independent of $L_x,L_y,N$. 
The condition $L_xL_y=2\pi M\ell_B^2$ for $L_x,L_y$ is convenient for imposing 
the following periodic boundary conditions: For an $N$ electrons 
wavefunction $\Phi^{(N)}$, we impose periodic boundary conditions 
\begin{equation}
t_j^{(x)}(L_x)\Phi^{(N)}({\bf r}_1,{\bf r}_2,\ldots,{\bf r}_N)
=\Phi^{(N)}({\bf r}_1,{\bf r}_2,\ldots,{\bf r}_N)\quad \mbox{for}\ 
j=1,2,\ldots,N,
\end{equation}
and
\begin{equation}
t_j^{(y)}(L_y)\Phi^{(N)}({\bf r}_1,{\bf r}_2,\ldots,{\bf r}_N)
=\Phi^{(N)}({\bf r}_1,{\bf r}_2,\ldots,{\bf r}_N)\quad \mbox{for}\ 
j=1,2,\ldots,N.
\end{equation}
Here $t^{(x)}(\cdots)$ and $t^{(y)}(\cdots)$ are magnetic translation 
operators \cite{Zak} defined as 
\begin{equation}
t^{(x)}(x')f(x,y)=f(x-x',y), \quad t^{(y)}(y')f(x,y)=\exp[iy'x/\ell_B^2]
f(x,y-y')
\label{defmgtranslation}
\end{equation}
for a function $f$ on ${\bf R}^2$, and a subscript $j$ of an operator 
indicates that the operator acts on the $j$-th coordinate 
of a function.\footnote{Throughout the present paper, we use this convention.}
The range of $x'$ and $y'$ are given by\footnote{See 
Section~\ref{singleLandauhamSec}.} 
\begin{equation}
x'=m\Delta x\quad \mbox{with}\ m\in{\bf Z}, \quad \mbox{and}\quad
y'=n\Delta y\quad \mbox{with}\ n\in{\bf Z}, 
\end{equation}
where the minimal units of the translations are given by 
\begin{equation}
\Delta x:=\frac{h}{eB}\frac{1}{L_y}, \quad \mbox{and} \quad 
\Delta y:=\frac{h}{eB}\frac{1}{L_x}.
\end{equation}
For a given random potential $V_\omega$, we define a set of random potentials 
as 
\begin{equation}
\Omega(\omega):=\Omega^{\rm T}(\omega)\cup\Omega^{\rm R}(\omega)
\end{equation}
with 
\begin{equation}
\Omega^{\rm T}(\omega):=\left\{\omega'\left|V_{\omega'}(x,y)=V_\omega(x,y-y_0),
y_0=\frac{2\pi\hbar n}{eBL_x}, n\in{\bf Z}\right.\right\}
\label{OmegaT}
\end{equation}
and
\begin{equation}
\Omega^{\rm R}(\omega):=\left\{\omega'\left|V_{\omega'}(x,y)
=V_\omega(-x,y_0-y),y_0=\frac{2\pi\hbar n}{eBL_x}, n\in{\bf Z}\right.\right\}. 
\label{OmegaR}
\end{equation}
Further we define an average with respect to the random potentials 
$\Omega(\omega)$ as
\begin{equation}
{\bf E}_\omega\left[\cdots\right]:=
\frac{1}{\left|\Omega(\omega)\right|}\sum_{\omega'\in\Omega(\omega)}
\left(\cdots\right).
\end{equation}
The regular part $W$ of the single-body potential $V_\omega$ of 
(\ref{decomVomega}) is invariant under the transformations in (\ref{OmegaT}) 
and (\ref{OmegaR}). 

We denote by $H_{\omega,0}^{(N)}$ the Hamiltonian 
$H_\omega^{(N)}$ of (\ref{totHam0}) with $A_0=0$ and $F=0$. We assume that 
the ``ground state" of $H_{\omega,0}^{(N)}$ is finitely $q$-fold degenerate 
in the sense that the lowest-lying $q$ energy eigenvalues 
$E_{\omega,(0,\mu)}^{(N)}$, $\mu=1,2,\ldots,q$ satisfy the condition 
\begin{equation}
\Delta{\cal E}:=
\max_{\mu,\mu'\in\{1,2,\ldots,q\}}
\left\{\left|E_{\omega,(0,\mu)}^{(N)}-E_{\omega,(0,\mu')}^{(N)}\right|\right\}
\rightarrow 0\quad \mbox{as}\ 
\cases{L_x,L_y \rightarrow\infty&;\cr
L_y\rightarrow\infty &for a fixed $L_x$,\cr}
\end{equation}
where the limit is taken for a fixed filling $\nu=N/M$. 
Further we assume that there exists a non-zero excitation gap 
above the ``ground state", i.e., the first excited state has 
an energy eigenvalue $E_{\omega,1}^{(N)}$ such that 
\begin{equation}
\min_{\mu\in\{1,2,\ldots,q\}}
\left\{E_{\omega,1}^{(N)}-E_{\omega,(0,\mu)}^{(N)}\right\}\ge\Delta E,
\label{gapcondition}
\end{equation}
where $\Delta E$ is a positive constant which is independent of $L_x,L_y,N$. 

We denote by ${\tilde \Phi}_{\omega,(0,\mu)}^{(N)}$ with $\mu=1,2,\ldots,q$, 
the ``ground state" eigenvectors of 
the Hamiltonian\footnote{More precisely, we choose $A_0=m_eF/B$ in addition to 
the condition of the small $F$. See Section~\ref{quantumHallsystem}.} 
$H_\omega^{(N)}$ of (\ref{totHam0}) with a sufficienlty 
weak electric field $F$. We take $\left\{{\tilde \Phi}_{\omega,(0,\mu)}^{(N)}
\right\}$ to be an orthonormal system. 
Then the current density ${\bf j}$ at zero temperature is given by 
\begin{equation}
j_s:=-\frac{e^2}{L_xL_y}\frac{1}{q}\sum_{\mu=1}^q{\bf E}_\omega
\left[\left\langle{\tilde \Phi}_{\omega,(0,\mu)}^{(N)},v_{{\rm tot},s}
{\tilde \Phi}_{\omega,(0,\mu)}^{(N)}\right\rangle\right]
\quad \mbox{for}\ s=x,y,
\label{defjs}
\end{equation}
where $\left\langle\cdots,\cdots\right\rangle$ stands for 
the inner product in the $N$ electrons Hilbert space, and the velocity 
operator ${\bf v}_{\rm tot}$ for the $N$ electrons is given by 
\begin{equation}
v_{{\rm tot},s}:=\cases{
\displaystyle{\frac{1}{m_e}\sum_{j=1}^N(p_{x,j}-eBy_j+A_0)}
 & for $s=x$;\cr
\displaystyle{\frac{1}{m_e}\sum_{j=1}^Np_{y,j}} & for $s=y$.\cr}
\end{equation}
The formula (\ref{defjs}) for the current density ${\bf j}$ is justified 
for an inverse temperature $\beta$ satisfying 
$\Delta{\cal E}\ll\beta^{-1}\ll\Delta E$. The conductivities are defined as 
\begin{equation}
\sigma_{sy}:=\lim_{F\downarrow 0}\frac{j_s}{F}\quad \mbox{for}\ s=x,y.
\end{equation}

Now we describe our main theorems for both non-interacting 
and interacting electrons gases. 

\begin{theorem}
\label{theorem1}
Suppose that there is no electron-electron interaction, i.e., $U^{(2)}=0$, 
and that there is a non-zero excitation gap above the ``ground state" 
in the sense of (\ref{gapcondition}). Then 
\begin{equation}
\left|\sigma_{xy}+\frac{e^2}{h}\nu\right|\le{\cal C}_{{\rm con},0}
\left(\frac{\ell_B}{L_y}\right)^{3/5}, \quad
\left|\sigma_{yy}\right|\le{\cal C}_{{\rm con},0}
\left(\frac{\ell_B}{L_y}\right)^{3/5},
\end{equation}
where ${\cal C}_{{\rm con},0}$ is a positive constant which is independent 
of the linear dimensions $L_x,L_y$ of the system and of the number $N$ of the 
electrons. 
\end{theorem}
We remark that the above assumption on the excitation gap is valid 
in the case with $\Vert V_\omega\Vert<V_0<\hbar\omega_c/2$ and 
$\nu\in\{1,2,\ldots\}$. 
Here $\omega_c$ is the cyclotron frequency given by $\omega_c:=eB/m_e$. 
In fact the ground state is unique and has a non-zero excitation gap above it. 
In the thermodynamic limt $L_y\rightarrow\infty$, we have the integral 
quantization of the Hall conductance $\sigma_{xy}=-e^2(n+1)/h$ with the Landau 
level index $n=0,1,2,\ldots$, and the vanishing diagonal conductance 
$\sigma_{yy}=0$. 
\medskip

For the interacting electrons gas, we obtain the following theorem: 

\begin{theorem}
\label{theorem2}
Suppose that the single-body potential $V_\omega$ is two times continuously 
differentiable on ${\bf R}^2$ and satisfies 
\begin{equation}
\left\Vert\frac{\partial^2}{\partial x^2}V_\omega\right\Vert
+\left\Vert\frac{\partial^2}{\partial y^2}V_\omega\right\Vert
<V_0'<\infty
\label{2dbound}
\end{equation}
with a positive constant $V_0'$ which is independent of 
the linear dimensions $L_x,L_y$ of the system, 
and suppose that there is a non-zero excitation gap 
above the ``ground state" in the sense of (\ref{gapcondition}). Then 
there exists a positive number $N_{\rm min}$ such that $N_{\rm min}$ is 
independent of the linear dimensions $L_x,L_y$ of the system and 
of the number $N$ of the electrons, and that the following two bounds 
are valid for $N\ge N_{\rm min}$:
\begin{equation}
\left|\sigma_{xy}+\frac{e^2}{h}\nu\right|\le{\cal C}_{\rm con}
\left(\frac{L_x}{\ell_B}\right)^{5/24}
\left(\frac{\ell_B}{L_y}\right)^{7/24}, \quad
\left|\sigma_{yy}\right|\le{\cal C}_{\rm con}
\left(\frac{L_x}{\ell_B}\right)^{5/24}
\left(\frac{\ell_B}{L_y}\right)^{7/24},
\label{result1}
\end{equation}
where ${\cal C}_{\rm con}$ is a positive constant which is independent 
of the linear dimensions $L_x,L_y$ of the system and of the number $N$ of the 
electrons. In particular, 
\begin{equation}
\left|\sigma_{xy}+\frac{e^2}{h}\nu\right|\le{\cal C}_{\rm con}
\left(\frac{\ell_B}{L}\right)^{1/12}, \quad
\left|\sigma_{yy}\right|\le{\cal C}_{\rm con}
\left(\frac{\ell_B}{L}\right)^{1/12}\quad\mbox{for}\ L_x=L_y=L.
\end{equation}
\end{theorem}
\medskip
The number $N_{\rm min}$ is explicitly given as a function of 
the parameters of the model in (\ref{Nmin}) in Appendix~\ref{decayestInt}. 
\medskip

Having this result in mind, let us discuss which filling factor $\nu$ leads to 
a spectral gap above the ``ground state". For this purpose, 
we briefly state a result of our separate paper \cite{Koma1}. 
Consider first the Hamiltonian 
\begin{equation}
H^{(N)}:=\sum_{j=1}^N\left\{\frac{1}{2m_e}
\left[\left(p_{x,j}-eBy_j\right)^2+p_{y,j}^2\right]+W(x_j)\right\}
+\sum_{1\le i<j\le N}U^{(2)}(x_i-x_j,y_i-y_j)
\label{hamKoma1}
\end{equation}
which is the Hamiltonian (\ref{totHam0}) without the random 
potential $V_\omega^{\rm ran}$ and with $A_0=0$ and $F=0$. Then we restrict 
the Hilbert space to the lowest $(n_{\rm max}+1)$ Landau level 
with a large integer $n_{\rm max}$. Namely the Hamiltonian we treated in 
ref.~\cite{Koma1} is given by 
\begin{equation}
H^{(N)}(n_{\rm max}):=P^{(N)}(n_{\rm max})H^{(N)}P^{(N)}(n_{\rm max})
\end{equation}
with the projection operator $P^{(N)}(n_{\rm max})$. 
We take the thermodynamic limit $L_y\rightarrow +\infty$ 
for a fixed large $L_x$ and a fixed filling factor $\nu$. 
In this infinite-volume limit, we assume the absence of 
non-translationally invariant infinite-volume ground state. Then we obtained 
the following result \cite{Koma1}: 
\begin{itemize}
\item If a pure infinite-volume ground state has a non-zero excitation gap, 
then the filling factor $\nu$ must be equal to a rational number.\footnote{
For the so-called odd denominator rule, see the 
footnote~\ref{footnoteevendenomi} in Section~\ref{Introduction}.}
\end{itemize}
Although the system we treated has no disorder, 
we can expect that, for the presence of weak disorder, the gap persists 
against the disorder. In this situation with the weak disorder, 
we get a rational quantization of the Hall conductance 
\begin{equation}
\sigma_{xy}=-\frac{e^2}{h}\nu \quad \mbox{with a rational filling}\ \nu,
\end{equation}
and the vanishing diagonal conductance 
\begin{equation}
\sigma_{yy}=0
\end{equation}
from the result (\ref{result1}). 

In order to discuss the appearance of plateaus, we change the filling factor 
$\nu$ slightly from a rational value in the non-interacting or 
the interacting cases. Then, if the electrons of low energy 
excitations do not contribute to the current flow owing to the disorder, 
we can expect that the quantized value of the Hall conductance $\sigma_{xy}$ 
remains constant, i.e., there appears a plateau of the Hall conductance, 
with the vanishing diagonal conductance. 
The appearance of such a plateau due to disorder will be discussed 
in relation to localization of wavefunctions in another separate paper 
\cite{Koma2}.

\Section{Single electron Landau systems}
\label{preliminaries}

As preliminaries, we briefly review the properties of the single electron 
systems in a uniform magnetic field with no electric field, 
and then introduce an electric field. 

\subsection{The single electron Landau Hamiltonian}
\label{singleLandauhamSec}

In this subsection, we briefly review the eigenvalue problem of 
the Landau Hamiltonian for a single electron in a uniform magnetic field. 
The Hamiltonian is given by 
\begin{equation}
{\cal H}=\frac{1}{2m_e}\left[(p_x-eBy)^2+p_y^2\right].
\label{singleham}
\end{equation}

Consider first the eigenvalue problem on the infinite plane ${\bf R}^2$. 
In order to obtain an eigenvector of the Hamiltonian ${\cal H}$, put its 
form as 
\begin{equation}
\phi(x,y)=e^{ikx}v(y)
\end{equation}
with a wavenumber $k\in{\bf R}$. Substituting this into the Schr\"odinger 
equation ${\cal H}\phi={\cal E}\phi$, one has 
\begin{equation}
\left[\frac{1}{2m_e}(\hbar k-eBy)^2+\frac{1}{2m_e}p_y^2\right]v(y)
={\cal E}v(y).
\end{equation}
Clearly this is identical to the eigenvalue equation of a quantum harmonic 
oscillator as 
\begin{equation}
\left[-\frac{\hbar^2}{2m_e}\frac{\partial^2}{\partial y^2}+
\frac{e^2B^2}{2m_e}\left(y-\frac{\hbar k}{eB}\right)^2\right]v(y)
={\cal E}v(y). 
\end{equation}
The eigenvectors are 
\begin{equation}
v_{n,k}(y):=v_n(y-y_k):=N_n \exp\left[-(y-y_k)^2/(2\ell_B^2)\right]
H_n\left[(y-y_k)/\ell_B\right],
\end{equation}
where $H_n$ is the Hermite polynomial, $y_k=\hbar k/eB$, and 
$N_n$ is the positive normalization constant so that 
\begin{equation}
\int_{-\infty}^{+\infty}dy|v_{n,k}(y)|^2=1.
\end{equation}
The energy eigenvalues are given by
\begin{equation}
{\cal E}_{n,k}:=\left(n+\frac{1}{2}\right)\hbar \omega_c
\quad \mbox{for} \ \ n=0,1,2,\ldots 
\label{eigenvalueEnk}
\end{equation}
with $\omega_c=eB/m_e$. Thus the eigenvectors of the Hamiltonian 
(\ref{singleham}) are given by 
\begin{equation}
\phi_{n,k}(x,y)=e^{ikx}v_{n,k}(y). 
\end{equation}

Next we consider single electron in $L_x\times L_y$ rectangular box 
$S=[-L_x/2,L_x/2]\times[-L_y/2,L_y/2]$ 
satisfying $L_xL_y=2\pi M\ell_B^2$ with a sufficienlty large positive 
integer $M$. For simplicity we take $M$ even. We impose 
periodic boundary conditions
\begin{equation}
\phi(x,y)=t^{(x)}(L_x)\phi(x,y), \quad \phi(x,y)=t^{(y)}(L_y)\phi(x,y)
\label{PBC}
\end{equation}
for wavefunctions $\phi$ on ${\bf R}^2$. 
Here $t^{(x)}(\cdots)$ and $t^{(y)}(\cdots)$ are the magnetic translation 
operators defined by (\ref{defmgtranslation}). We claim that the functions 
\begin{equation}
f_1(x,y)=t^{(x)}(x')f(x,y)
\end{equation}
and
\begin{equation}
f_2(x,y)=t^{(y)}(y')f(x,y)
\end{equation}
satisfy the periodic boundary conditions (\ref{PBC}) if $f$ satisfies 
(\ref{PBC}). As a result, $x'$ and $y'$ are restricted into the following 
values: 
\begin{equation}
x'=m\Delta x \quad \mbox{with} \ m\in {\bf Z}, \quad \mbox{and}
\quad 
y'=n\Delta y \quad \mbox{with} \ n\in {\bf Z},
\end{equation}
where 
\begin{equation}
\Delta x:=\frac{h}{eB}\frac{1}{L_y}, \quad \mbox{and}\quad 
\Delta y:=\frac{h}{eB}\frac{1}{L_x}.
\end{equation}
In fact one has 
\begin{eqnarray}
f_1(x,y)&=&f(x-x',y)\ret
&=&\exp[iL_y(x-x')/\ell_B^2]f(x-x',y-L_y)\ret
&=&\exp[-iL_yx'/\ell_B^2]\exp[iL_yx/\ell_B^2]f(x-x',y-L_y)\ret
&=&\exp[-iL_yx'/\ell_B^2]\exp[iL_yx/\ell_B^2]f_1(x,y-L_y)\ret
&=&\exp[-iL_yx'/\ell_B^2]t^{(y)}(L_y)f_1(x,y)\ret
&=&\exp[-iL_yx'/\ell_B^2]f_1(x,y).
\end{eqnarray}
by the definitions. This implies $L_yx'/\ell_B^2=2\pi m$ with an integer $m$. 
Similarly 
\begin{eqnarray}
f_2(x,y)&=&\exp[iy'x/\ell_B^2]f(x,y-y')\ret
&=&\exp[iy'x/\ell_B^2]f(x-L_x,y-y')\ret
&=&\exp[iy'L_x/\ell_B^2]\exp[iy'(x-L_x)/\ell_B^2]f(x-L_x,y-y')\ret
&=&\exp[iy'L_x/\ell_B^2]f_2(x-L_x,y)\ret
&=&\exp[iy'L_x/\ell_B^2]t^{(x)}(L_x)f_2(x,y)\ret
&=&\exp[iy'L_x/\ell_B^2]f_2(x,y).
\end{eqnarray}
Thus $y'L_x/\ell_B^2=2\pi n$ with an integer $n$. 
In the following we restrict the ranges of the variables $x',y'$ 
in the magnetic translations to these values. 

Since 
\begin{equation}
t^{(y)}(y')(p_x-eBy)\left[t^{(y)}(y')\right]^{-1}=p_x-eBy 
\end{equation}
for any $y'$, the Hamiltonian (\ref{singleham}) is invariant under 
all the magnetic translations $t^{(x)}(\cdots)$ and $t^{(y)}(\cdots)$. 
Consider wavefunctions 
\begin{equation}
\phi_{n,k}^{\rm P}(x,y)=L_x^{-1/2}\sum_{\ell=-\infty}^{+\infty}
e^{i(k+\ell K)x}v_{n,k}(y-\ell L_y)
\label{phiP}
\end{equation}
for $k=2\pi m/L_x$ with $m=-M/2+1,\ldots,M/2-1,M/2$, and 
with $K=L_y/\ell_B^2$. These wavefunctions are the eigenvectors 
of the Hamiltonian (\ref{singleham}) satisfying the periodic 
boundary conditions (\ref{PBC}), because $L_xL_y=2\pi M\ell_B^2$ 
with the integer $M$. The eigenvalues of $\phi_{n,k}^{\rm P}$ are given by 
(\ref{eigenvalueEnk}). 

We define a reflection operator $R$ as 
\begin{equation}
Rf(x,y)=f(-x,-y)
\label{defR}
\end{equation}
for a function on ${\bf R}^2$. One can easily get the following lemma: 

\begin{lemma} 
\label{Teigenvector}
The vector $\phi_{n,k}^{\rm P}$ of (\ref{phiP}) is an eigenvector of 
the magnetic translation \hfill\break $t^{(x)}(\Delta x)$, i.e., 
\begin{equation}
t^{(x)}(\Delta x)\phi_{n,k}^{\rm P}=e^{-ik\Delta x}\phi_{n,k}^{\rm P}
=e^{-i2\pi m/M}\phi_{n,k}^{\rm P}\quad \mbox{with} \ \ k=\frac{2\pi m}{L_x}, 
\end{equation}
and the magnetic translation $t^{(y)}(\Delta y)$ 
shifts the wavenumber $k$ of the vector $\phi_{n,k}^{\rm P}$ by 
one unit $2\pi/L_x$ as 
\begin{equation}
t^{(y)}(\Delta y)\phi_{n,k}^{\rm P}=\phi_{n,k'}^{\rm P}
\quad \mbox{with} \quad 
k'=k+\frac{\Delta y}{\ell_B^2}=k+\frac{2\pi}{L_x}. 
\end{equation}
Further, 
\begin{equation}
R\phi_{n,k}^{\rm P}=(-1)^n\phi_{n,-k}^{\rm P}.
\end{equation}
\end{lemma}

As usual we denote by $L^2(S)$ the set of functions $f$ on the rectangular 
box $S$ such that 
\begin{equation}
\int_S dxdy \ |f(x,y)|^2=
\int_{-L_x/2}^{L_x/2}dx \int_{-L_y/2}^{L_y/2}dy \ |f(x,y)|^2 <\infty.
\end{equation}
Further we define the associate inner product $(f,g)$ as 
\begin{equation}
(f,g)=\int_S dxdy \ [f(x,y)]^\ast g(x,y)
\end{equation}
for $f,g\in L^2(S)$. 

\begin{lemma} Let $f,g$ be functions on ${\bf R}^2$ such that 
$f,g\in L^2(S)$, and that $f,g$ satisfy the boundary conditions (\ref{PBC}). 
Then 
\begin{equation}
(f,g)=\int_{-L_x/2}^{L_x/2}dx \int_{-L_y/2+y_0}^{L_y/2+y_0}dy \ 
[f(x,y)]^\ast g(x,y) 
\end{equation}
for any $y_0\in {\bf R}$. 
\label{inproind}
\end{lemma}

\begin{proof}{Proof}
By the periodic boundary condition $f(x,y)=t^{(x)}(L_x)f(x,y)$, 
the function $f$ can be expanded in Fourier series as 
\begin{equation}
f(x,y)=L_x^{-1/2}\sum_k e^{ikx}{\hat f}(k,y).
\label{Fourierf}
\end{equation}
Further, since 
\begin{eqnarray}
f(x,y)=t^{(y)}(L_y)f(x,y)&=&L_x^{-1/2}
\sum_k e^{i(k+K)x}{\hat f}(k,y-L_y) \ret
&=&L_x^{-1/2}\sum_k e^{ikx}{\hat f}(k-K,y-L_y),
\end{eqnarray}
one has 
\begin{equation}
{\hat f}(k,y)={\hat f}(k-K,y-L_y).
\label{PBCk}
\end{equation}
Using this relation repeatedly, the function $f$ of (\ref{Fourierf}) 
can be rewritten as 
\begin{equation}
f(x,y)=\sum_{\{k=2\pi n/L_x\left|-M/2+1\le n\le M/2\right.\}}
L_x^{-1/2}\sum_{\ell=-\infty}^{+\infty}
e^{i(k+\ell K)x}{\hat f}(k,y-\ell L_y).
\end{equation}
By the help of this expression, one has 
\begin{eqnarray}
(f,g)&=&\int_{-L_x/2}^{L_x/2}dx \int_{-L_y/2}^{L_y/2}dy \
\left[f(x,y)\right]^\ast g(x,y)\ret
&=&\sum_{\{k=2\pi n/L_x\left|-M/2+1\le n\le M/2\right.\}}\ 
\sum_{\ell=-\infty}^{+\infty} \int_{-L_y/2}^{L_y/2} dy \ 
\left[{\hat f}(k,y-\ell L_y)\right]^\ast
{\hat g}(k,y-\ell L_y)\ret
&=&\sum_{\{k=2\pi n/L_x\left|-M/2+1\le n\le M/2\right.\}}\ 
\int_{-\infty}^{+\infty} dy \ \left[{\hat f}(k,y)\right]^\ast
{\hat g}(k,y)\ret
&=&\sum_{\{k=2\pi n/L_x\left|-M/2+1\le n\le M/2\right.\}}\ 
\sum_{\ell=-\infty}^{+\infty} \int_{-L_y/2+y_0}^{L_y/2+y_0} dy \ 
\left[{\hat f}(k,y-\ell L_y)\right]^\ast
{\hat g}(k,y-\ell L_y)\ret
&=&\int_{-L_x/2}^{L_x/2}dx \int_{-L_y/2+y_0}^{L_y/2+y_0}dy \
\left[f(x,y)\right]^\ast g(x,y).
\label{exprin}
\end{eqnarray}
\end{proof}

Let us see that the set of the eigenvectors $\{\phi_{n,k}^{\rm P}\}$ 
of (\ref{phiP}) forms an orthonormal complete system. 
{From} (\ref{exprin}) in Lemma~\ref{inproind}, one has 
\begin{eqnarray}
\left(\phi_{n',k'}^{\rm P},\phi_{n,k}^{\rm P}\right)
=\int_{-\infty}^{+\infty} dy \ 
v_{n',k}^\ast(y)v_{n,k}(y)\delta_{k,k'}=\delta_{n,n'}\delta_{k,k'}.
\label{innerprophiP}
\end{eqnarray}
Here $\delta_{k,k'}$ is the Kronecker delta. 
To show the completeness, consider a function $f$ satisfying 
the boundary conditions (\ref{PBC}). 
In the same way, 
\begin{equation}
\left(\phi_{n,k}^{\rm P},f\right)
=\int_{-\infty}^{+\infty} dy \ v_{n,k}^\ast(y){\hat f}(k,y).
\end{equation}
This implies that the function $f$ must be zero if the inner 
product $\left(\phi_{n,k}^{\rm P},f\right)$ is vanishing for all 
the vectors $\phi_{n,k}^{\rm P}$. 

\subsection{The Landau Hamiltonian in an electric field}
\label{Landauhamele}

Next we consider a single electron in magnetic and electric fields in 
the rectangular box $S=[-L_x/2,L_x/2]\times[-L_y/2,L_y/2]$. 
The Hamiltonian is given by 
\begin{equation}
{\cal H}=\frac{1}{2m_e}\left[{\bf p}+e{\bf A}({\bf r})\right]^2
+eFyP_{\rm bulk}-{m_e \over 2}\left({F \over B}\right)^2.
\label{hamj}
\end{equation}
We take the vector potential as 
\begin{equation}
e{\bf A}({\bf r})=(-eBy+A_0,0,0)
\end{equation}
which gives the constant magnetic field ${\bf B}=(0,0,B)$
perpendicular to the $x$-$y$ plane. We also applied the constant electric 
field ${\bf F}=(0,F,0)$ oriented along the $y$ axis. We have introduced 
the projection operator $P_{\rm bulk}$ so that the electrons 
near the boundaries $y=\pm L_y/2$ do not feel the infinitely strong 
electric field at the boundaries. 
The precise definition of $P_{\rm bulk}$ is given as follows: 
We define a projection operator $P(k)$ onto the Fourier component 
with a wavenumber $k$ for a function $f\in L^2(S)$ as 
\begin{equation}
P(k)f(x,y)=L_x^{-1/2}e^{ikx}{\hat f}(k,y)
\label{defP(k)}
\end{equation}
with the Fourier coefficient 
\begin{equation}
{\hat f}(k,y)=L_x^{-1/2}\int_{-L_x/2}^{L_x/2}dx \ e^{-ikx}f(x,y). 
\end{equation}
For an interval $I$ we define a projection operator as 
\begin{equation}
P(I):=\sum_{k\in {\cal F}(I)}P(k)
\end{equation}
with
\begin{equation}
{\cal F}(I):=\left\{k=\frac{2\pi n}{L_x}\left|\ n\in{\bf Z} \quad \mbox{and}
\quad \frac{\hbar k}{eB}\in I\right.\right\}. 
\end{equation}
We define $P_{\rm bulk}$ as 
\begin{equation}
P_{\rm bulk}:=P(I_{\rm bulk})
\label{defPbulk}
\end{equation}
with the interval 
\begin{equation}
I_{\rm bulk}:=\bigcup_{n=-\infty}^{+\infty}
[-L_y/2+\delta+nL_y,L_y/2-\delta+nL_y]
\label{Ibulk}
\end{equation}
with a positive number $\delta$. We choose $\delta$ satisfying 
$\delta/L_y\rightarrow 0$ as $L_y\rightarrow +\infty$ so that 
the effect of the projection $(1-P_{\rm bulk})$ at the boundaries 
is negligible\footnote{Unfortunately we cannot prove this claim.} 
in the thermodynamic limit $L_y\rightarrow +\infty$. 
Here we stress that the operator $eFyP_{\rm bulk}$ in 
the Hamiltonian ${\cal H}$ of (\ref{hamj}) is self-adjoint 
because $y$ and $P_{\rm bulk}$ commute with each other 
from their definitions.  

Next let us show the locality of $(1-P_{\rm bulk})$. 
Namely the operator is vanishing on the bulk region 
which is at a distance from the boundaries. 
Therefore the electric field is constant on the bulk region. 

Let $\Phi^{(N)}$ be an $N$ electrons wavefunction satisfying 
\begin{equation}
\frac{1}{N}\left\langle\Phi^{(N)},H_{\omega,0}^{(N)}\Phi^{(N)}\right\rangle
<{\tilde {\cal E}}<\infty,
\label{enenrgyassume}
\end{equation}
where the positive constant ${\tilde {\cal E}}$ is independent 
of the linear dimensions $L_x,L_y$ and of the number $N$ of electrons, 
and $H_{\omega,0}^{(N)}$ is the Hamiltonian $H_\omega^{(N)}$ of 
(\ref{totHam0}) with $A_0=0$ and $F=0$. Then we have the following bound:
\begin{equation}
\left|\left\langle\Phi^{(N)},\chi_{\rm bulk}(y_j)(1-P_{{\rm bulk},j})
\Phi^{(N)}\right\rangle\right|\le
\frac{2({\tilde {\cal E}}+\Vert V_\omega\Vert)}{\hbar \omega_c}
\left(\frac{\ell_B}{\delta}\right)^2, 
\label{locality}
\end{equation}
where $\chi_{\rm bulk}$ is a characteristic function given by 
\begin{equation}
\chi_{\rm bulk}(y):=\cases{1, & for $y\in [-L_y/2+2\delta,L_y/2-2\delta]$;\cr
0, & otherwise.\cr}
\end{equation}
Since we can choose $\delta\rightarrow\infty$ as $L_y\rightarrow\infty$, 
this bound clearly implies the locality of $(1-P_{{\rm bulk},j})$. 
Let us prove the bound. Note that 
\begin{equation}
(eB\delta)^2\chi_{\rm bulk}(y_j)(1-P_{{\rm bulk},j})\le 
\chi_{\rm bulk}(y_j)(1-P_{{\rm bulk},j})(p_{x,j}-eBy_j)^2 
\end{equation}
from the definitions. Using this inequality and the assumption 
(\ref{enenrgyassume}), we have 
\begin{eqnarray}
& &(eB\delta)^2\left\langle\Phi^{(N)},\chi_{\rm bulk}(y_j)(1-P_{{\rm bulk},j})
\Phi^{(N)}\right\rangle\ret&\le& 
\left\langle\Phi^{(N)},\chi_{\rm bulk}(y_j)(1-P_{{\rm bulk},j})
(p_{x,j}-eBy_j)^2\Phi^{(N)}\right\rangle\ret
&\le& \left\langle\Phi^{(N)},(p_{x,j}-eBy_j)^2\Phi^{(N)}\right\rangle\ret
&\le&2m_e\left(\frac{1}{N}\left\langle\Phi^{(N)},H_{\omega,0}^{(N)}\Phi^{(N)}
\right\rangle+\Vert V_\omega\Vert\right)
\le2m_e\left({\tilde {\cal E}}+\Vert V_\omega\Vert\right), 
\end{eqnarray}
where we have used the positivity of the electron-electron interaction 
$U^{(2)}$ for getting the third inequality. 
This is nothing but the desired bound (\ref{locality}). 

For the convenience of the following calculations, we choose $A_0=m_eF/B$. 
Then the Hamiltonian ${\cal H}$ of (\ref{hamj}) becomes 
\begin{eqnarray}
{\cal H}&=&\frac{1}{2m_e}\left[(p_x-eBy+A_0)^2+p_y^2\right]+eFyP_{\rm bulk}
-{m_e \over 2}\left({F \over B}\right)^2 \ret
&=&{1 \over 2m_e}\left[(p_x-eBy)^2+p_y^2\right]+\frac{1}{2m_e}[2A_0(p_x-eBy)
+A_0^2]
+eFyP_{\rm bulk}-{m_e \over 2}\left({F \over B}\right)^2 \ret
&=&{1 \over 2m_e}\left[(p_x-eBy)^2+p_y^2\right]
+{F \over B}p_xP_{\rm bulk}+\frac{F}{B}(p_x-eBy)
\left(1-P_{\rm bulk}\right),
\label{PhiM}
\end{eqnarray}
and the velocity operator $v_x$ is given by 
\begin{equation}
m_ev_x=p_x-eBy+{m_eF \over B}.
\end{equation}
In the following we will treat the second and the third terms in 
the last line of (\ref{PhiM}) as a perturbation. 

\Section{The relation between the present ``electric potential" 
and the standard time-dependent vector potential}
\label{relation}

In the same setting as in Section~\ref{mainresults}, 
we introduce the standard time-dependent vector potential ${\bf A}(t)$ 
instead of the regularized electric potential so that the vector potential 
yields the constant electric field on the whole 
torus. Since the electric potential gives the constant electric field 
except for the neighborhood of the boundaries as we have seen 
in Section~\ref{Landauhamele}, we can expect that 
these two different potentials yield the same transport properties 
in the large volume limit. In this section, we shall discuss this issue. 

The time-dependent Schr\"odinger equation with the vector potential 
is given by 
\begin{equation}
i\hbar\frac{\partial}{\partial t}\Phi^{(N)}(t)
=H_\omega^{(N)}(t)\Phi^{(N)}(t)
\end{equation}
with the time-dependent Hamiltonian 
\begin{equation}
H_\omega^{(N)}(t)=\sum_{j=1}^N\frac{1}{2m_e}
\left\{(p_{x,j}-eBy_j)^2+[p_{y,j}+eA(t)]^2\right\}
+{\tilde U}^{(N)}_\omega({\bf r}_1,\ldots,{\bf r}_N)
\end{equation}
with the vector potential ${\bf A}(t)=(0,A(t),0)$ with 
\begin{equation}
A(t)=-Fte^{\eta t}\quad \mbox{for}\ \ -\infty<t\le 0, 
\end{equation}
and with the potentials 
\begin{equation}
{\tilde U}^{(N)}_\omega({\bf r}_1,\ldots,{\bf r}_N)
=\sum_{j=1}^N V_\omega({\bf r}_j)+
\sum_{i<j}U^{(2)}({\bf r}_i-{\bf r}_j). 
\label{deftildeU}
\end{equation}
Here $\eta$ is a small positive parameter switching the corresponding 
electric field adiabatically. 

In order to show the equivalence between the two systems, 
we introduce a unitary transformation $\Phi^{(N)}(t)=G^{(N)}(t)\Psi^{(N)}(t)$ 
with 
\begin{equation}
G^{(N)}(t)=\exp\left[-\frac{i}{\hbar}\sum_{j=1}^N
ey_jP_{{\rm bulk},j}{\tilde \chi}_{\rm bulk}(y_j)A(t)\right],
\end{equation}
where we have introduced the function ${\tilde \chi}_{\rm bulk}$ 
so that the wavefunctions satisfy the periodic boundary conditions. 
We take the function ${\tilde \chi}_{\rm bulk}$ to be 
an infinitely differentiable function satisfying 
\begin{equation}
{\tilde \chi}_{\rm bulk}(y)=\cases{1, & for $y\in 
[-L_y/2+2\delta/3,L_y/2-2\delta/3]$;\cr
0, & for $y\in [-L_y/2,-L_y/2+\delta/3]\cup[L_y/2-\delta/3,L_y/2]$.\cr}
\end{equation}
Namely it is equal to the identity on the bulk region and 
vanishing near the boundaries. The effect of ${\tilde \chi}_{\rm bulk}$ 
to the conductance is negligible 
for the large volume\footnote{One can prove the statement by using 
the method in the present paper, although we do not give the proof here.}
by the locality of $P_{\rm bulk}$. 

Note that 
\begin{equation}
i\hbar\frac{\partial}{\partial t}\Phi^{(N)}(t)
=G^{(N)}(t)i\hbar\frac{\partial}{\partial t}\Psi^{(N)}(t)
+\left[i\hbar\frac{\partial}{\partial t}G^{(N)}(t)\right]
\Psi^{(N)}(t),
\end{equation}
and
\begin{equation}
\left[G^{(N)}(t)\right]^\ast i\hbar\frac{\partial}{\partial t}G^{(N)}(t)
=-\sum_{j=1}^Ney_jF(1+\eta t)e^{\eta t}P_{{\rm bulk},j}
{\tilde \chi}_{\rm bulk}(y_j). 
\end{equation}
Since $\eta$ is an infinitesimally small parameter, 
the right-hand side leads to the regularized electric potential 
of the present paper. We also have 
\begin{eqnarray}
& &\left[G^{(N)}(t)\right]^\ast[p_{y,j}+eA(t)]G^{(N)}(t)\ret
&=&p_{y,j}+eA(t)\left[1-P_{{\rm bulk},j}{\tilde \chi}_{\rm bulk}(y_j)\right]
-eA(t)y_jP_{{\rm bulk},j}\frac{\partial}{\partial y_j}
{\tilde \chi}_{\rm bulk}(y_j). 
\end{eqnarray}
The second and third terms in the right-hand side are 
vanishing in the large volume limit for getting the conductance. 
This statement can be proved in the same way as in the present paper. 
{From} these observations, we obtain 
\begin{equation}
i\hbar\frac{\partial}{\partial t}\Psi^{(N)}(t)
=\left[H_\omega^{(N)}+\Delta {\tilde U}_\omega^{(N)}({\bf r}_1,\ldots,
{\bf r}_N;t)\right]\Psi^{(N)}(t)
\end{equation}
with the Hamiltonian $H_\omega^{(N)}$ of (\ref{totHam0}) and 
\begin{equation}
\Delta {\tilde U}_\omega^{(N)}({\bf r}_1,\ldots,{\bf r}_N;t)
=\left[G^{(N)}(t)\right]^\ast
{\tilde U}^{(N)}_\omega({\bf r}_1,\ldots,{\bf r}_N)G^{(N)}(t)
-{\tilde U}^{(N)}_\omega({\bf r}_1,\ldots,{\bf r}_N).
\end{equation}
Here we have dropped some terms which do not contribute to the conductance 
in the large volume limit. If we can drop the  potential 
$\Delta {\tilde U}_\omega^{(N)}({\bf r}_1,\ldots,{\bf r}_N;t)$, 
then we get the desired result, i.e., the unitary equivalence between 
the two systems with the different potentials in the large volume limit. 
Unfortunately we can not dropp the potential. 
But we can expect that the contribution to conductance 
is of order of $\delta/L_y$ which is vanishing in the limit 
$L_y\rightarrow\infty$. 

Let us estimate the correction from 
$\Delta {\tilde U}_\omega^{(N)}({\bf r}_1,\ldots,{\bf r}_N;t)$ 
to the conductance. 
Note that the unitary operator $G^{(N)}(t)$ can be rewritten as 
\begin{equation}
G^{(N)}(t)=\exp\left[-\frac{i}{\hbar}\sum_{j=1}^N ey_jA(t)
{\tilde \chi}_{\rm bulk}(y_j)\right]G_{\rm edge}^{(N)}(t)
\end{equation}
with 
\begin{equation}
G_{\rm edge}^{(N)}(t)
=\exp\left[\frac{i}{\hbar}\sum_{j=1}^N ey_jA(t)(1-P_{{\rm bulk},j})
{\tilde \chi}_{\rm bulk}(y_j)\right]. 
\end{equation}
Using this, we have 
\begin{equation}
\left[G^{(N)}(t)\right]^\ast
{\tilde U}^{(N)}_\omega({\bf r}_1,\ldots,{\bf r}_N)G^{(N)}(t)
=\left[G_{\rm edge}^{(N)}(t)\right]^\ast
{\tilde U}^{(N)}_\omega({\bf r}_1,\ldots,{\bf r}_N)G_{\rm edge}^{(N)}(t). 
\end{equation}
Immediately, we get 
\begin{equation}
\Delta {\tilde U}_\omega^{(N)}({\bf r}_1,\ldots,{\bf r}_N;t)
=\left[G_{\rm edge}^{(N)}(t)\right]^\ast
{\tilde U}^{(N)}_\omega({\bf r}_1,\ldots,{\bf r}_N)G_{\rm edge}^{(N)}(t)
-{\tilde U}^{(N)}_\omega({\bf r}_1,\ldots,{\bf r}_N). 
\end{equation}
This implies, owing to the definition of $G_{\rm edge}^{(N)}(t)$ and 
the locality of $P_{\rm bulk}$, that, if the potential 
${\tilde U}^{(N)}_\omega({\bf r}_1,\ldots,{\bf r}_N)$ is 
vanishing near the boundaries, then 
$\Delta {\tilde U}_\omega^{(N)}({\bf r}_1,\ldots,{\bf r}_N;t)$ 
is almost vanishing on the whole torus. 
In fact, if we take the potential 
\begin{equation}
{\tilde U}^{(N)}_\omega({\bf r}_1,\ldots,{\bf r}_N)
=\sum_{j=1}^N P_{{\rm bulk},j}V_\omega({\bf r}_j)P_{{\rm bulk},j}
+\sum_{i<j}P_{{\rm bulk},i}P_{{\rm bulk},j}U^{(2)}({\bf r}_i-{\bf r}_j)
P_{{\rm bulk},i}P_{{\rm bulk},j}
\end{equation}
instead of (\ref{deftildeU}), then 
the potential difference 
$\Delta {\tilde U}_\omega^{(N)}({\bf r}_1,\ldots,{\bf r}_N;t)$ 
is exactly equal to zero. 
Thus $\Delta {\tilde U}_\omega^{(N)}({\bf r}_1,\ldots,{\bf r}_N;t)$ is 
vanishing on the bulk region, and the correction to the conductance is 
expected to be of order of $\delta/L_y$. Unfortunately 
we could not estimate the correction in a mathematically rigorous sense. 

\Section{Proofs of the main theorems}
\label{quantumHallsystem}

In this section, we calculate the conductivities which are derived as 
the linear response coefficients for the weak electric field. 
For this purpose, we use the Rayleigh-Schr\"odinger perturbation theories. 
Our goal is to give proofs of our main Theorems~\ref{theorem1} and 
\ref{theorem2}. Namely we show that the conductivities (conductance) 
so obtained satisfy the bounds in the theorems. 
For the convenience of readers, in Appendices~\ref{perturbation}-\ref{estE-E} 
we give technical estimates and 
calculations of matrix elements appeared in our representation of the 
conductivities. 

By choosing $A_0=m_eF/B$ as in (\ref{PhiM}) in Section~\ref{Landauhamele}, 
the $N$ electrons Hamiltonian $H_\omega^{(N)}$ 
of (\ref{totHam0}) which we mainly treat in this paper can be rewritten as 
\begin{equation}
H_\omega^{(N)}=H_{\omega,0}^{(N)}+\lambda{\tilde H}^{(N)}
\label{totHam}
\end{equation}
with
\begin{equation}
H_{\omega,0}^{(N)}=\sum_{j=1}^N\left[\frac{1}{2m_e}\left(p_{x,j}-eBy_j\right)^2
+\frac{1}{2m_e}p_{y,j}^2+V_\omega({\bf r}_j)\right]
+U^{(N)}({\bf r}_1,\ldots,{\bf r}_N) 
\label{unperturbedHam}
\end{equation}
and
\begin{equation}
{\tilde H}^{(N)}=\sum_{j=1}^N\left[p_{x,j}P_{{\rm bulk},j}
+(p_{x,j}-eBy_j)\left(1-P_{{\rm bulk},j}\right)\right]. 
\end{equation}
Here $\lambda=F/B$ is a real parameter, $U^{(N)}$ is written in a sum of 
two-body interactions as 
\begin{equation}
U^{(N)}({\bf r}_1,\ldots,{\bf r}_N)
=\sum_{1\le i<j\le N}U^{(2)}(x_i-x_j,y_i-y_j), 
\end{equation}
and we have dropped the constant term. 

We treat ${\tilde H}^{(N)}$ as a perturbation, and apply 
the Rayleigh-Schr\"odinger perturbation theory 
to the eigenvalue problem of the Hamiltonian $H_\omega^{(N)}$ of 
(\ref{totHam}) for getting a ``ground state" eigenvector of $H_\omega^{(N)}$. 

\subsection{The Hall and diagonal conductivities}

\subsubsection{Non-degenerate ground state}

Consider first the case when the ground state $\Phi_{\omega,0}^{(N)}$ 
of the unperturbed Hamiltonian $H_{\omega,0}^{(N)}$ is non-degenerate. 
Let ${\tilde \Phi}_{\omega,0}^{(N)}$ be the corresponding normalized 
ground state eigenvector of the full Hamiltonian $H_\omega^{(N)}$. 
Since the electric field $F$ is assumed to be sufficiently weak, 
the ground state eigenvector ${\tilde \Phi}_{\omega,0}^{(N)}$ is unique. 
Then the current density averaged 
over the random potentials $\Omega(\omega)$ at zero temperature is given by 
\begin{equation}
j_s=-\frac{e}{L_xL_y}{\bf E}_\omega\left[
\left\langle{\tilde \Phi}_{\omega,0}^{(N)},v_{{\rm tot},s}
{\tilde \Phi}_{\omega,0}^{(N)}
\right\rangle\right] \quad \mbox{for}\ s=x,y,
\end{equation}
where the velocity operator $v_{{\rm tot},s}$ for the $N$ electrons is given 
by 
\begin{eqnarray}
v_{{\rm tot},x}:=\sum_{j=1}^N v_{x,j}=N{F \over B}+{1 \over m_e}\sum_{j=1}^N 
\left(p_{x,j}-eBy_j\right),
\end{eqnarray}
and
\begin{equation}
v_{{\rm tot},y}:=\frac{1}{m_e}\sum_{j=1}^N p_{y,j}.
\end{equation}
We rewrite the current density ${\bf j}=(j_x,j_y)$ as 
\begin{equation}
j_s=\cases{\displaystyle{-\frac{e^2}{h}\nu F+\Delta j_x} & for $s=x$; \cr
\Delta j_y & for $s=y$\cr}
\label{defcurrentdensitynon}
\end{equation}
with
\begin{equation}
\Delta j_s:=-\frac{e}{m_eL_xL_y}\sum_{j=1}^N{\bf E}_\omega\left[
\left\langle{\tilde \Phi}_{\omega,0}^{(N)},\pi_{s,j}
{\tilde \Phi}_{\omega,0}^{(N)}\right\rangle\right],
\label{Deltaj}
\end{equation}
where 
\begin{equation}
\pi_s:=\cases{p_x-eBy & for $s=x$ ; \cr
p_y & for $s=y$,\cr}
\end{equation}
and $\nu$ is the filling factor for the Landau level. Namely 
$N=\nu M$ with the number $M={eB}L_xL_y/h$ which is equal to the number of 
states in a single Landau level of the non-interacting Landau Hamiltonian with 
no disorder. 

{From} the standard formula of the perturbation theory, 
the ground state eigenvector ${\tilde \Phi}_{\omega,0}^{(N)}$ of 
$H_\omega^{(N)}$ is expanded as 
\begin{equation}
{\tilde \Phi}_{\omega,0}^{(N)}=\Phi_{\omega,0}^{(N)}+\lambda\sum_{\ell \ne 0}
\Phi_{\omega,\ell}^{(N)}
{\left\langle\Phi_{\omega,\ell}^{(N)},{\tilde H}^{(N)}
\Phi_{\omega,0}^{(N)}\right\rangle \over 
E_{\omega,0}^{(N)}-E_{\omega,\ell}^{(N)}}+{\cal O}(\lambda^2) 
\label{expandPhi}
\end{equation}
in powers of $\lambda$. Here $\Phi_{\omega,\ell}^{(N)}$ are 
the orthonormal eigenvectors of the unperturbed Hamiltonian 
$H_{\omega,0}^{(N)}$ with the energy eigenvalues $E_{\omega,\ell}^{(N)}$. 
For the detail, see Appendix~\ref{nondegenerateperturbation}. 
Using this expansion (\ref{expandPhi}), we have 
\begin{eqnarray}
\sum_{j=1}^N
\left\langle{\tilde \Phi}_{\omega,0}^{(N)}, \pi_{s,j}
{\tilde \Phi}_{\omega,0}^{(N)}\right\rangle
&=&\sum_{j=1}^N
\left\langle\Phi_{\omega,0}^{(N)},\pi_{s,j}
\Phi_{\omega,0}^{(N)}\right\rangle \nonumber \\
&+&2\lambda\sum_{\ell \ne 0}\sum_{j=1}^N{\rm Re}\left[{
\left\langle\Phi_{\omega,\ell}^{(N)},{\tilde H}^{(N)}\Phi_{\omega,0}^{(N)}
\right\rangle
\over E_{\omega,0}^{(N)}-E_{\omega,\ell}^{(N)}}
\left\langle\Phi_{\omega,0}^{(N)},\pi_{s,j}
\Phi_{\omega,\ell}^{(N)}\right\rangle\right]+{\cal O}(\lambda^2),\ret
\label{peByexpand}
\end{eqnarray}
where ${\rm Re}$ stands for a real part. 
By Lemma~\ref{Phipi-} in Appendix~\ref{MatElements}, the average of the first 
sum is vanishing as 
\begin{equation}
\sum_{j=1}^N{\bf E}_\omega\left[
\left\langle\Phi_{\omega,0}^{(N)},\pi_{s,j}
\Phi_{\omega,0}^{(N)}\right\rangle\right]=0. 
\label{avfirstzero}
\end{equation}
Substituting (\ref{peByexpand}) and (\ref{avfirstzero}) into the right-hand 
side of (\ref{Deltaj}), we have 
\begin{equation}
\Delta j_s=\Delta j_s^{(1)}+{\cal O}\left((F/B)^2\right)
\label{expandDeltaj}
\end{equation}
with
\begin{equation}
\Delta j_s^{(1)}:=-\frac{2e^2}{h}\nu F \ 
{\rm Re}{\bf E}_\omega\left[{\cal M}_s\right], 
\label{Deltaj1}
\end{equation}
where 
\begin{equation}
{\cal M}_s:=\frac{1}{m_eN}\sum_{j=1}^N
\left\langle\Phi_{\omega,0}^{(N)},\pi_{s,j}
{[1-G_\omega^{(N)}] \over E_{\omega,0}^{(N)}-H_{\omega,0}^{(N)}}
{\tilde H}^{(N)}\Phi_{\omega,0}^{(N)}\right\rangle . 
\label{M}
\end{equation}
Here $G_\omega^{(N)}$ is the orthogonal projection onto 
the ground state $\Phi_{\omega,0}^{(N)}$. 
{From} (\ref{defcurrentdensitynon}), (\ref{Deltaj}), (\ref{expandDeltaj}) 
and (\ref{Deltaj1}), the Hall and diagonal conductivities can be written as 
\begin{equation}
\sigma_{xy}:=\lim_{F\rightarrow 0}\frac{j_x}{F}=-\frac{e^2}{h}\nu
-\frac{2e^2}{h}\nu 
{\rm Re}{\bf E}_\omega\left[{\cal M}_x\right]
\label{expsigmaxy}
\end{equation}
and
\begin{equation}
\sigma_{yy}:=\lim_{F\rightarrow 0}\frac{j_y}{F}=-\frac{2e^2}{h}\nu 
{\rm Re}{\bf E}_\omega\left[{\cal M}_y\right],
\label{expsigmayy}
\end{equation}
respectively. 

\subsubsection{Degenerate ``ground state"}

Next consider the case when the ``ground state" of the Hamiltonian 
$H_{\omega,0}^{(N)}$ is $q$-fold degenerate. 
Let $\Phi_{\omega,(0,\mu)}^{(N)}$ be the ``ground state" 
eigenvectors with the energy eigenvalue $E_{\omega,(0,\mu)}^{(N)}$ 
for $\mu=1,2,\ldots,q$. We take $\left\{\Phi_{\omega,(0,\mu)}^{(N)}\right\}$ 
to be an orthonormal system. In this case, the current desity is given by 
\begin{equation}
j_s:=-\frac{e}{L_xL_y}{\bf E}_\omega
\left[\frac{1}{q}\sum_{\mu=1}^q\left\langle
{\tilde \Phi}_{\omega,(0,\mu)}^{(N)},
v_{{\rm tot},s}{\tilde \Phi}_{\omega,(0,\mu)}^{(N)}\right\rangle
\right],
\end{equation}
where ${\tilde \Phi}_{\omega,(0,\mu)}^{(N)}$ are the corresponding normalized 
ground state eigenvectors of the Hamiltonian $H_\omega^{(N)}$, 
with the corresponding energy eigenvalues ${\tilde E}_{\omega,(0,\mu)}^{(N)}$. 
Similarly to the non-degenerate case, the corrections for the current density 
${\bf j}$ are given by 
\begin{equation}
\Delta j_s:=-\frac{e}{m_eL_xL_y}
{\bf E}_\omega
\left[\frac{1}{q}\sum_{\mu=1}^q\left\langle
{\tilde \Phi}_{\omega,(0,\mu)}^{(N)},
\sum_{j=1}^N\pi_{s,j}{\tilde \Phi}_{\omega,(0,\mu)}^{(N)}\right\rangle
\right]. 
\label{Djs}
\end{equation}
The ``ground state" eigenvectors ${\tilde \Phi}_{\omega,(0,\mu)}^{(N)}$ are 
expanded as 
\begin{equation}
{\tilde \Phi}_{\omega,(0,\mu)}^{(N)}=\Phi_{\omega,(0,\mu)}^{(N,0)}
+\lambda\sum_{\ell\ne 0}\Phi_{\omega,\ell}^{(N)}
\frac{1}{E_{\omega,(0,\mu)}^{(N)}-E_{\omega,\ell}^{(N)}}
\left\langle\Phi_{\omega,\ell}^{(N)},{\tilde H}^{(N)}
\Phi_{\omega,(0,\mu)}^{(N,0)}\right\rangle+\cdots
\end{equation}
by using the degenerate perturbation theory. Here 
$\Phi_{\omega,(0,\mu)}^{(N,0)}$ are orthonormal vectors which span 
the sector spanned by the ``ground states" eigenvectors 
$\Phi_{\omega,(0,\mu)}^{(N)}$ of the unperturbed Hamiltonian 
$H_{\omega,0}^{(N)}$. For the detail of the degenerate perturbation theory, 
see Appendix~\ref{degenerateperturbation}. Using this expansion, we have 
\begin{eqnarray}
& &\left\langle{\tilde \Phi}_{\omega,(0,\mu)}^{(N)},\sum_{j=1}^N
\pi_{s,j}{\tilde \Phi}_{\omega,(0,\mu)}^{(N)}\right\rangle\ret
&=&\left\langle\Phi_{\omega,(0,\mu)}^{(N,0)},\sum_{j=1}^N
\pi_{s,j}\Phi_{\omega,(0,\mu)}^{(N,0)}\right\rangle
+2\lambda{\rm Re}
\left[\left\langle\Phi_{\omega,(0,\mu)}^{(N,0)},
\sum_{j=1}^N\pi_{s,j}\frac{1-G_\omega^{(N)}}{E_{\omega,(0,\mu)}^{(N)}-
H_{\omega,0}^{(N)}}{\tilde H}^{(N)}\Phi_{\omega,(0,\mu)}^{(N)}\right\rangle
\right]\ret
& &\qquad +{\cal O}(\lambda^2). 
\label{piavexpansiondegenerate}
\end{eqnarray}
Here $G_\omega^{(N)}$ is the orthogonal projection onto the sector of 
the degenerate ``ground state" whose space is 
spanned by the $q$ energy eigenvectors $\Phi_{\omega,(0,\mu)}^{(N)}$, 
$\mu=1,2,\ldots,q$. 
Substituting (\ref{piavexpansiondegenerate}) into (\ref{Djs}), we obtain 
\begin{eqnarray}
\Delta j_s&=&-\frac{e}{m_eL_xL_y}
{\bf E}_\omega
\left[\frac{1}{q}\sum_{\mu=1}^q\left\langle\Phi_{\omega,(0,\mu)}^{(N,0)},
\sum_{j=1}^N\pi_{s,j}\Phi_{\omega,(0,\mu)}^{(N,0)}\right\rangle
\right]\ret 
&-&\frac{2e\lambda}{m_eL_xL_y}{\rm Re}{\bf E}_\omega
\left[\frac{1}{q}\sum_{\mu=1}^q\left\langle\Phi_{\omega,(0,\mu)}^{(N,0)},
\sum_{j=1}^N\pi_{s,j}\frac{1-G_\omega^{(N)}}{E_{\omega,(0,\mu)}^{(N)}-
H_{\omega,0}^{(N)}}{\tilde H}^{(N)}\Phi_{\omega,(0,\mu)}^{(N,0)}\right\rangle
\right]+{\cal O}(\lambda^2) \ret
&=&-\frac{e}{m_eL_xL_y}
{\bf E}_\omega
\left[\frac{1}{q}\sum_{\mu=1}^q\left\langle\Phi_{\omega,(0,\mu)}^{(N)},
\sum_{j=1}^N\pi_{s,j}\Phi_{\omega,(0,\mu)}^{(N)}\right\rangle
\right]\ret 
&-&\frac{2e\lambda}{m_eL_xL_y}{\rm Re}{\bf E}_\omega
\left[\frac{1}{q}\sum_{\mu=1}^q\left\langle\Phi_{\omega,(0,\mu)}^{(N)},
\sum_{j=1}^N\pi_{s,j}\frac{1-G_\omega^{(N)}}{E_{\omega,(0,\mu)}^{(N)}-
H_{\omega,0}^{(N)}}{\tilde H}^{(N)}\Phi_{\omega,(0,\mu)}^{(N)}\right\rangle
\right]+{\cal O}(\lambda^2).\ret
\label{Djbeta2}
\end{eqnarray}
Since the first term in the right-hand side of the second equality 
is vanishing owing to Lemma~\ref{Phipi-}, we obatin 
\begin{equation}
\Delta j_s=\Delta j_s^{(1)}+{\cal O}(F^2)
\end{equation}
with 
\begin{equation}
\Delta j_s^{(1)}:=-\frac{2e^2}{h}\nu F{\rm Re}
{\bf E}_\omega\left[{\cal M}_s\right]
\end{equation}
and
\begin{equation}
{\cal M}_s:=\frac{1}{m_e N}
\frac{1}{q}\sum_{\mu=1}^q\left\langle\Phi_{\omega,(0,\mu)}^{(N)},
\sum_{j=1}^N\pi_{s,j}\frac{1-G_\omega^{(N)}}{E_{\omega,(0,\mu)}^{(N)}-
H_{\omega,0}^{(N)}}{\tilde H}^{(N)}\Phi_{\omega,(0,\mu)}^{(N)}\right\rangle. 
\end{equation}
In consequence, we have the expressions of the conductivities as 
\begin{equation}
\sigma_{xy}:=\lim_{F\rightarrow 0}\frac{j_x}{F}=-\frac{e^2}{h}\nu
-\frac{2e^2}{h}\nu 
{\rm Re}{\bf E}_\omega\left[{\cal M}_x\right]
\label{expsigmaxydegenerate}
\end{equation}
and
\begin{equation}
\sigma_{yy}:=\lim_{F\rightarrow 0}\frac{j_y}{F}=-\frac{2e^2}{h}\nu 
{\rm Re}{\bf E}_\omega\left[{\cal M}_y\right]
\label{expsigmayydegenerate}
\end{equation}
with the above ${\cal M}_s$. These have the same forms as 
(\ref{expsigmaxy}) and (\ref{expsigmayy}) in the non-degenerate case. 

\subsection{Estimate of ${\bf E}_\omega[{\cal M}_s]$}

{From} the expressions of the conductivities (\ref{expsigmaxy}), 
(\ref{expsigmayy}), (\ref{expsigmaxydegenerate}) and 
(\ref{expsigmayydegenerate}), we want to estimate 
${\bf E}_\omega[{\cal M}_s]$, in order to prove Theorems~\ref{theorem1} and 
\ref{theorem2}. In the following, we treat only the non-degenerate case 
because one can treat the degenerate case in the same way. 

We define two projection operators $P_{\rm in}$ and $P_{\rm out}$ as 
\begin{equation}
P_{\rm in}:=P(I_{\rm in}), \quad \mbox{and} \quad 
P_{\rm out}:=P(I_{\rm out})
\end{equation}
with the intervals 
\begin{equation}
I_{\rm in}=[-L_y/2+\delta,L_y/2-\delta], \quad \mbox{and} \quad 
I_{\rm out}=I_{\rm bulk}\backslash I_{\rm in}.
\end{equation}
Clearly we have $P_{\rm bulk}=P_{\rm in}+P_{\rm out}$ from the definition
(\ref{defPbulk}) of $P_{\rm bulk}$ with (\ref{Ibulk}). 
We write ${\cal M}_s$ of (\ref{M}) as 
\begin{equation}
{\cal M}_s={\cal M}_{s,{\rm in}}+{\cal M}_{s,{\rm out}}
+{\cal M}_{s,{\rm edge}}, 
\end{equation}
where
\begin{equation}
{\cal M}_{s,{\rm in}}:=\frac{1}{m_eN}\sum_{i=1}^N\sum_{j=1}^N
\left\langle\Phi_{\omega,0}^{(N)},\pi_{s,j}
{[1-G_\omega^{(N)}] \over E_{\omega,0}^{(N)}-H_{\omega,0}^{(N)}}
P_{{\rm in},i}p_{x,i}\Phi_{\omega,0}^{(N)}\right\rangle , 
\label{Min}
\end{equation}
\begin{equation}
{\cal M}_{s,{\rm out}}:=\frac{1}{m_eN}\sum_{i=1}^N\sum_{j=1}^N
\left\langle\Phi_{\omega,0}^{(N)},\pi_{s,j}
{[1-G_\omega^{(N)}] \over E_{\omega,0}^{(N)}-H_{\omega,0}^{(N)}}
P_{{\rm out},i}p_{x,i}\Phi_{\omega,0}^{(N)}\right\rangle , 
\label{Mout}
\end{equation}
and
\begin{equation}
{\cal M}_{s,{\rm edge}}:=\frac{1}{m_eN}\sum_{i=1}^N\sum_{j=1}^N
\left\langle\Phi_{\omega,0}^{(N)},\pi_{s,j}
{[1-G_\omega^{(N)}] \over E_{\omega,0}^{(N)}-H_{\omega,0}^{(N)}}
(1-P_{{\rm bulk},i})(p_{x,i}-eBy_i)\Phi_{\omega,0}^{(N)}\right\rangle. 
\label{Medge1}
\end{equation}

Let us sketch the idea of the proofs of Theorems~\ref{theorem1} and 
\ref{theorem2}. Since one can expect that 
the contributions of ${\cal M}_{s,{\rm out}}$ and ${\cal M}_{s,{\rm edge}}$ 
become small for a large volume, we explain the idea only for 
${\cal M}_{s,{\rm in}}$. Consider the random average of the matrix element 
in (\ref{Min}). It is written as 
\begin{eqnarray}
& &{\bf E}_\omega\left[\left\langle\Phi_{\omega,0}^{(N)},\pi_{s,j}
{[1-G_\omega^{(N)}] \over E_{\omega,0}^{(N)}-H_{\omega,0}^{(N)}}
P_{{\rm in},i}p_{x,i}\Phi_{\omega,0}^{(N)}\right\rangle\right]\ret
&=&\sum_k\hbar k{\bf E}_\omega\left[\left\langle\Phi_{\omega,0}^{(N)},\pi_{s,j}
{[1-G_\omega^{(N)}] \over E_{\omega,0}^{(N)}-H_{\omega,0}^{(N)}}
P_i(k)\Phi_{\omega,0}^{(N)}\right\rangle\right]
\label{identityEmat}
\end{eqnarray}
by using the projection operator $P(k)$ onto the Fourier 
component with the wavenumber $k$. We introduce a transformation 
consisting of a reflection and a magnetic translation as 
\begin{equation}
x\rightarrow -x\quad , \quad y\rightarrow 2y_k-y
\end{equation}
with $y_k=\hbar k/(eB)$. In particular, $y=y_k$ is the fixed point for 
the second part of the transformation. This yields that 
the wavenumber $k$ also is the fixed point in the space of the wavenumbers. 
Using the transformation, we have 
\begin{eqnarray}
& &\hbar k{\bf E}_\omega\left[\left\langle\Phi_{\omega,0}^{(N)},\pi_{s,j}
{[1-G_\omega^{(N)}] \over E_{\omega,0}^{(N)}-H_{\omega,0}^{(N)}}
P_i(k)\Phi_{\omega,0}^{(N)}\right\rangle\right]\ret
&=&-\hbar k{\bf E}_\omega\left[\left\langle\Phi_{\omega,0}^{(N)},\pi_{s,j}
{[1-G_\omega^{(N)}] \over E_{\omega,0}^{(N)}-H_{\omega,0}^{(N)}}
P_i(k)\Phi_{\omega,0}^{(N)}\right\rangle\right]\ret
&+&(\mbox{corrections from the boundaries}\ y=\pm L_y/2)
\label{approEMinmat}
\end{eqnarray}
for the summand with $k$ in the right-hand side of (\ref{identityEmat}). 
{From} these observations, we conclude that the contributions of 
${\bf E}_\omega\left[{\cal M}_{s,{\rm in}}\right]$ for the conductivities 
are small if the corrections from the boundaries give small contributions 
for a large volume. In fact, the corrections are small as we will show 
in Appendix~\ref{EstEMsin}. 
\medskip 

In order to give the proofs of Theorems~\ref{theorem1} and 
\ref{theorem2}, let us summarize the results of the estimates 
for ${\bf E}_\omega\left[{\cal M}_{s,{\rm in}}\right],
{\bf E}_\omega\left[{\cal M}_{s,{\rm out}}\right]$ and 
${\bf E}_\omega\left[{\cal M}_{s,{\rm edge}}\right]$. 
For the details of the calculations, see the corresponding 
Appendices.

\subsubsection{Non-interacting case}

Consider first the non-interacting case $U^{(2)}=0$. 
We obtain the following estimates: 
\begin{equation}
\left|{\bf E}_\omega\left[{\cal M}_{s,{\rm in}}\right]\right|
\le{\cal C}_{{\rm in},0}\frac{L_y}{\ell_B}
\left(\frac{\ell_B}{\delta}\right)^4
\end{equation}
from (\ref{estEMinNon}) in Appendix~\ref{App:EstEMinNon}, 
\begin{equation}
\left|{\bf E}_\omega\left[{\cal M}_{s,{\rm out}}\right]\right|
\le{\cal C}_{{\rm out},0}^{(1)}\frac{L_y}{\ell_B}
\left(\frac{\ell_B}{\delta}\right)^4+{\cal C}_{{\rm out},0}^{(2)}
\left(\frac{\ell_B}{\delta}\right)^2
\end{equation}
from (\ref{estMoutNon}) in Appendix~\ref{App:EstMoutNon}, and 
\begin{equation}
\left|{\bf E}_\omega\left[{\cal M}_{s,{\rm edge}}\right]\right|
\le{\cal C}_{{\rm edge},0}\frac{\delta}{L_y}
\end{equation}
from (\ref{EMsedge}) in Appendix~\ref{EstMedge}. 
Here ${\cal C}_{{\rm in},0},{\cal C}_{{\rm out},0}^{(1)},
{\cal C}_{{\rm out},0}^{(2)}$ and ${\cal C}_{{\rm edge},0}$ are positive 
constants which are independent of $L_x,L_y$. By choosing 
\begin{equation}
\delta=\ell_B\left(\frac{L_y}{\ell_B}\right)^{2/5}, 
\end{equation}
we get 
\begin{equation}
\left|{\bf E}_\omega\left[{\cal M}_s\right]\right|
\le \left|{\bf E}_\omega\left[{\cal M}_{s,{\rm in}}\right]\right|
+\left|{\bf E}_\omega\left[{\cal M}_{s,{\rm out}}\right]\right|
+\left|{\bf E}_\omega\left[{\cal M}_{s,{\rm edge}}\right]\right|
\le {\cal C}_0\left(\frac{\ell_B}{L_y}\right)^{3/5} 
\end{equation}
with a positive constant ${\cal C}_0$. 
Combining this bound, (\ref{expsigmaxy}) and 
(\ref{expsigmayy}), we obtain 
\begin{equation}
\left|\sigma_{xy}+\frac{e^2}{h}\nu\right|
\le{\cal C}_{{\rm con},0}\left(\frac{\ell_B}{L_y}\right)^{3/5}, \quad 
\left|\sigma_{yy}\right|
\le{\cal C}_{{\rm con},0}\left(\frac{\ell_B}{L_y}\right)^{3/5},
\end{equation}
where ${\cal C}_{{\rm con},0}$ is a positive constant. 

\subsubsection{Interacting case}

Next consider the interacting case $U^{(2)}\ne 0$. 
We take large $L_x,L_y$ so that $N\ge N_{\rm min}$, 
and assume that the single-body potential $V_\omega$ 
is two times continuously differentiable on ${\bf R}^2$, i.e., 
$V_\omega\in C^2({\bf R}^2)$, and satisfies the bound (\ref{2dbound}). 
Here $N_{\rm min}$ is 
a positive number given by (\ref{Nmin}) in Appendix~\ref{decayestInt}. 
Then we obtain the following estimates: 
\begin{equation}
\left|{\bf E}_\omega\left[{\cal M}_{s,{\rm in}}\right]\right|
\le{\cal C}_{\rm in}\left(\frac{L_x}{\ell_B}\right)^{5/6}
\left(\frac{L_y}{\ell_B}\right)^{11/6}
\left(\frac{\ell_B}{\delta}\right)^3
\end{equation}
from Proposition~\ref{pro:estEMin} in Appendix~\ref{IntMin}, 
\begin{equation}
\left|{\bf E}_\omega\left[{\cal M}_{s,{\rm out}}\right]\right|
\le{\cal C}_{\rm out}\left(\frac{L_x}{\ell_B}\right)^{5/6}
\left(\frac{L_y}{\ell_B}\right)^{11/6}
\left(\frac{\ell_B}{\delta}\right)^3
\end{equation}
from Proposition~\ref{Pro:Moutnon} in Appendix~\ref{IntMout}, and 
\begin{equation}
\left|{\bf E}_\omega\left[{\cal M}_{s,{\rm edge}}\right]\right|
\le{\cal C}_{\rm edge}\frac{\delta}{L_y}
\end{equation}
from (\ref{EMsedge}) in Appendix~\ref{EstMedge}. 
Here ${\cal C}_{\rm in},{\cal C}_{\rm out}$ and ${\cal C}_{\rm edge}$ 
are positive constants which are independent of $L_x,L_y$. We choose 
\begin{equation}
\delta=\ell_B\left(\frac{L_x}{\ell_B}\right)^{5/24}
\left(\frac{L_y}{\ell_B}\right)^{17/24}. 
\end{equation}
Then we get 
\begin{equation}
\left|{\bf E}_\omega\left[{\cal M}_s\right]\right|
\le \left|{\bf E}_\omega\left[{\cal M}_{s,{\rm in}}\right]\right|
+\left|{\bf E}_\omega\left[{\cal M}_{s,{\rm out}}\right]\right|
+\left|{\bf E}_\omega\left[{\cal M}_{s,{\rm edge}}\right]\right|
\le {\cal C}\left(\frac{L_x}{\ell_B}\right)^{5/24}
\left(\frac{\ell_B}{L_y}\right)^{7/24} 
\end{equation}
with a positive constant ${\cal C}$. In particular, we have 
\begin{equation}
\left|{\bf E}_\omega\left[{\cal M}_s\right]\right|
\le{\cal C}\left(\frac{\ell_B}{L}\right)^{1/12} 
\quad \mbox{and}\quad \delta=\ell_B\left(\frac{L}{\ell_B}\right)^{11/12}
\quad\mbox{for}\ L_x=L_y=L. 
\end{equation}
Combining these bounds, (\ref{expsigmaxy}) and (\ref{expsigmayy}), we obtain 
\begin{equation}
\left|\sigma_{xy}+\frac{e^2}{h}\nu\right|
\le{\cal C}_{\rm con}\left(\frac{L_x}{\ell_B}\right)^{5/24}
\left(\frac{\ell_B}{L_y}\right)^{7/24}, \quad 
\left|\sigma_{yy}\right|
\le{\cal C}_{\rm con}\left(\frac{L_x}{\ell_B}\right)^{5/24}
\left(\frac{\ell_B}{L_y}\right)^{7/24},
\end{equation}
and
\begin{equation}
\left|\sigma_{xy}+\frac{e^2}{h}\nu\right|
\le{\cal C}_{\rm con}\left(\frac{\ell_B}{L}\right)^{1/12}, \quad 
\left|\sigma_{yy}\right|
\le{\cal C}_{\rm con}\left(\frac{\ell_B}{L}\right)^{1/12}
\quad \mbox{for}\ L_x=L_y=L.
\end{equation}
Here ${\cal C}_{\rm con}$ is a positive constant.

\appendix

\Section{The Rayleigh-Schr\"odinger perturbation theories}
\label{perturbation}

In this appendix, we apply the Rayleigh-Schr\"odinger perturbation 
theories to the non-degenerate and degenerate ``ground states" 
of the present quantum Hall Hamiltonian $H_\omega^{(N)}$ of (\ref{totHam}). 
Since there is an excitation gap above the ``ground state(s)", this 
perturbative treatment is justified mathematically in the sense 
of an asymptotic expansion with respect to a sufficiently weak electric 
field.\footnote{See ref.~\cite{Kato} 
for the mathematically rigorous perturbation theories.}

Recall the Hamiltonian 
\begin{equation}
H_\omega^{(N)}=H_{\omega,0}^{(N)}+\lambda{\tilde H}^{(N)},
\label{hamnontot}
\end{equation}
where $\lambda$ is a sufficiently small real parameter. 
The Schr\"odinger equation is 
\begin{equation}
H_\omega^{(N)}{\tilde \Phi}_\omega^{(N)}
={\tilde E}_\omega^{(N)}{\tilde \Phi}_\omega^{(N)} 
\label{Schronontot}
\end{equation}
with an energy eigenvalue ${\tilde E}_\omega^{(N)}$. 
In order to obtain a ground state eigenvector ${\tilde \Phi}_\omega^{(N)}$ 
and the eigenvalue ${\tilde E}_\omega^{(N)}$ in powers of $\lambda$, 
we treat the Hamiltonian 
${\tilde H}^{(N)}$ in (\ref{hamnontot}) as a perturbation. 

\subsection{Non-degenerate case}
\label{nondegenerateperturbation}

Consider first the case when the ground state $\Phi_{\omega,0}^{(N)}$ 
of the unperturbed Hamiltonian $H_{\omega,0}^{(N)}$ is non-degenerate. 
As usual we expand 
the eigenvector ${\tilde \Phi}_{\omega,0}^{(N)}$ of 
the ground state of $H_\omega^{(N)}$ in powers of $\lambda$ as 
\begin{equation}
{\tilde \Phi}_{\omega,0}^{(N)}=\Phi_{\omega,0}^{(N)}
+\lambda \sum_{\ell \ne 0} a_\ell \Phi_{\omega,\ell}^{(N)}+\cdots 
\label{expansionPhi}
\end{equation}
in terms of the eigenvectors $\Phi_{\omega,\ell}^{(N)}$ of the 
unperturbed Hamiltonian $H_{\omega,0}^{(N)}$, 
and expand the corresponding eigenvalue ${\tilde E}_{\omega,0}^{(N)}$ 
in powers of $\lambda$ as 
\begin{equation}
{\tilde E}_{\omega,0}^{(N)}=E_{\omega,0}^{(N)}+\lambda E_{\omega,0}^{(N,1)}+
\cdots. 
\end{equation}
Here $E_{\omega,0}^{(N)}$ is the energy eigenvalue for the ground state 
eigenvector $\Phi_{\omega,0}^{(N)}$ of $H_{\omega,0}^{(N)}$. 
Substituting these expansions and (\ref{hamnontot}) into the Schr\"odinger 
equation (\ref{Schronontot}), one has 
\begin{eqnarray}
& & \left[H_{\omega,0}^{(N)}+\lambda 
{\tilde H}^{(N)}\right]\left[\Phi_{\omega,0}^{(N)}+\lambda
\sum_{\ell \ne 0}a_\ell\Phi_{\omega,\ell}^{(N)}+\cdots\right]
\nonumber\\
&=& \left[E_{\omega,0}^{(N)} +\lambda E_{\omega,0}^{(N,1)}
+\cdots\right]\left[\Phi_{\omega,0}^{(N)}+\lambda
\sum_{\ell \ne 0}a_\ell\Phi_{\omega,\ell}^{(N)}+\cdots\right]. 
\end{eqnarray}
Immediately, 
\begin{equation}
H_{\omega,0}^{(N)}\Phi_{\omega,0}^{(N)}=
E_{\omega,0}^{(N)}\Phi_{\omega,0}^{(N)},
\end{equation}
in the zero-th order of $\lambda$, and 
\begin{equation}
{\tilde H}^{(N)}\Phi_{\omega,0}^{(N)}+\sum_{\ell \ne 0}
a_\ell H_{\omega,0}^{(N)}\Phi_{\omega,\ell}^{(N)}
=E_{\omega,0}^{(N,1)}\Phi_{\omega,0}^{(N)}
+E_{\omega,0}^{(N)}\sum_{\ell \ne 0} 
a_\ell \Phi_{\omega,\ell}^{(N)} 
\label{lambdafirst}
\end{equation}
in the first order of $\lambda$. Taking the inner product 
with $\Phi_{\omega,\ell}^{(N)} \ (\ell\ne 0)$ in both sides of 
(\ref{lambdafirst}), one has 
\begin{equation}
\left\langle\Phi_{\omega,\ell}^{(N)},{\tilde H}^{(N)}
\Phi_{\omega,0}^{(N)}
\right\rangle+a_\ell E_{\omega,\ell}^{(N)}=
E_{\omega,0}^{(N)}a_\ell. 
\end{equation}
Here we have taken $\{\Phi_{\omega,\ell}^{(N)}\}$ to be the orthonormal 
complete system. As a result, the coefficient $a_\ell$ is 
\begin{equation}
a_\ell={1 \over E_{\omega,0}^{(N)}
-E_{\omega,\ell}^{(N)}}
\left\langle\Phi_{\omega,\ell}^{(N)},{\tilde H}^{(N)}
\Phi_{\omega,0}^{(N)}\right\rangle. 
\label{aell}
\end{equation}
Substituting this into (\ref{expansionPhi}), one has 
\begin{equation}
{\tilde \Phi}_{\omega,0}^{(N)}=\Phi_{\omega,0}^{(N)}
+\lambda \sum_{\ell \ne 0} \Phi_{\omega,\ell}^{(N)}{1 \over E_{\omega,0}^{(N)}
-E_{\omega,\ell}^{(N)}}
\left\langle\Phi_{\omega,\ell}^{(N)},{\tilde H}^{(N)}
\Phi_{\omega,0}^{(N)}\right\rangle+{\cal O}(\lambda^2). 
\end{equation}

\subsection{Degenerate case}
\label{degenerateperturbation}

In order to treat the ``degenerate ground state", we first rewrite 
the Hamiltonian $H_\omega^{(N)}$ of (\ref{hamnontot}) as 
\begin{equation}
H_\omega^{(N)}=H_{\omega,0}^{(N)}+\lambda G_\omega^{(N)}{\tilde H}^{(N)}
G_\omega^{(N)}+\lambda{\tilde H}_{\rm G}^{(N)}
\label{decomperturbHam}
\end{equation}
with
\begin{equation}
{\tilde H}_{\rm G}^{(N)}:=
G_\omega^{(N)}{\tilde H}^{(N)}
(1-G_\omega^{(N)})+(1-G_\omega^{(N)}){\tilde H}^{(N)}G_\omega^{(N)}+
(1-G_\omega^{(N)}){\tilde H}^{(N)}(1-G_\omega^{(N)}),
\end{equation}
where $G_\omega^{(N)}$ is the orthogonal projection onto the sector 
spanned by the ``ground state" eigenvectors 
$\Phi_{\omega,(0,\mu)}^{(N)}$ of $H_{\omega,0}^{(N)}$. 
In the present case, we formally treat the Hamiltonian 
${\tilde H}_{\rm G}^{(N)}$ as a perturbation, although the second term 
in the right-hand side of (\ref{decomperturbHam}) is still a small 
perturbation. Let $\Phi_{\omega,(0,\mu)}^{(N,0)}$ be the $q$ 
eigenvectors of the ``unperturbed" Hamiltonian $H_{\omega,0}^{(N)}+\lambda 
G_\omega^{(N)}{\tilde H}^{(N)}G_\omega^{(N)}$, 
and let $E_{\omega,(0,\mu)}^{(N,0)}$ be the corresponding energy eigenvalues. 
We take $\left\{\Phi_{\omega,(0,\mu)}^{(N,0)}\right\}$ to be an orthonormal 
system. Clearly 
\begin{equation}
E_{\omega,(0,\mu)}^{(N,0)}=E_{\omega,(0,\mu)}^{(N)}+{\cal O}(\lambda),
\label{Eomega0muexpand}
\end{equation}
where $E_{\omega,(0,\mu)}^{(N)}$ are the ``ground state" energy eigenvalues 
of the Hamiltonian $H_{\omega,0}^{(N)}$. 
In the same way as in the preceding Section~\ref{nondegenerateperturbation}, 
the ``ground state" eigenvector ${\tilde \Phi}_{\omega,(0,\mu)}^{(N)}$ of 
$H_\omega^{(N)}$ is expanded as 
\begin{equation}
{\tilde \Phi}_{\omega,(0,\mu)}^{(N)}=\Phi_{\omega,(0,\mu)}^{(N,0)}
+\lambda\sum_{\ell\ne 0}a_\ell\Phi_{\omega,\ell}^{(N)}+\cdots
\label{degeneratePhiexpand}
\end{equation}
and expand the corresponding energy eigenvalue 
${\tilde E}_{\omega,(0,\mu)}^{(N)}$ as 
\begin{equation}
{\tilde E}_{\omega,(0,\mu)}^{(N)}=E_{\omega,(0,\mu)}^{(N,0)}
+\lambda E_{\omega,(0,\mu)}^{(N,0)}+\cdots.
\end{equation}
Substituting these into the Schr\"odinger equation, one has 
\begin{equation}
{\tilde H}_{\rm G}^{(N)}\Phi_{\omega,(0,\mu)}^{(N,0)}
+H_{\omega,0}^{(N)}\sum_{\ell\ne 0}a_\ell\Phi_{\omega,\ell}^{(N)}
=E_{\omega,(0,\mu)}^{(N)}\sum_{\ell\ne 0}a_\ell\Phi_{\omega,\ell}^{(N)}
+E_{\omega,0}^{(N,1)}\Phi_{\omega,(0,\mu)}^{(N)},
\end{equation}
where we have used (\ref{Eomega0muexpand}). 
Taking the inner product with $\Phi_{\omega,\ell}^{(N)}$ with $\ell\ne 0$ 
in both sides, one gets 
\begin{equation}
a_\ell=\frac{1}{E_{\omega,(0,\mu)}^{(N)}-E_{\omega,\ell}^{(N)}}
\left\langle\Phi_{\omega,\ell}^{(N)},{\tilde H}^{(N)}
\Phi_{\omega,(0,\mu)}^{(N,0)}\right\rangle. 
\end{equation}
Substituting this into (\ref{degeneratePhiexpand}), one has 
\begin{equation}
{\tilde \Phi}_{\omega,(0,\mu)}^{(N)}=\Phi_{\omega,(0,\mu)}^{(N,0)}
+\lambda\sum_{\ell\ne 0}\Phi_{\omega,\ell}^{(N)}
\frac{1}{E_{\omega,(0,\mu)}^{(N)}-E_{\omega,\ell}^{(N)}}
\left\langle\Phi_{\omega,\ell}^{(N)},{\tilde H}^{(N)}
\Phi_{\omega,(0,\mu)}^{(N,0)}\right\rangle+\cdots. 
\label{Phiexpansiondegenerate}
\end{equation}

\Section{Matrix elements of the quantum Hall systems with disorder}
\label{MatElements}

In this appendix, we study the properties of some matrix elements (\ref{M}) 
appeared in the expressions (\ref{expsigmaxy}), (\ref{expsigmayy}) 
of the conductivities. 

\subsection{The single electron Landau Hamiltonian with disorder} 
\label{Landauhamdisorder}

Consider first the single electron Landau Hamiltonian 
\begin{equation}
{\cal H}_\omega=
\frac{1}{2m_e}\left[(p_x-eBy)^2+p_y^2\right]+V_\omega(x,y)
\label{singlehamomega}
\end{equation}
with the periodic boundary conditions (\ref{PBC}). 
The single-electron potential $V_\omega$ with disorder satisfies the periodic 
boundary conditions (\ref{PBCVomega}) and the condition (\ref{boundedVomega}) 
of boundedness. 

\begin{lemma} 
\label{lemmaTRV}
Let $\varphi_\omega$ be an eigenvector of the Hamiltonian ${\cal H}_\omega$ of 
(\ref{singlehamomega}). Then the translate $t^{(y)}(y_k)\varphi_\omega$ 
is an eigenvector of the Hamiltonian ${\cal H}_{\omega'}$ 
with the potential $V_{\omega'}$ given by 
\begin{equation}
V_{\omega'}(x,y)=V_\omega(x,y-y_k).
\end{equation}
Here $y_k=\hbar k/(eB)$ with $k=2\pi n/L_x$, $(n\in{\bf Z})$.
\end{lemma}

\begin{proof}{Proof} {From} the assumption 
${\cal H}_\omega\varphi_\omega={\cal E}_\omega\varphi_\omega$, we have 
\begin{eqnarray}
{\cal E}_\omega t^{(y)}(y_k)\varphi_\omega
&=&t^{(y)}(y_k){\cal H}_\omega\varphi_\omega
=t^{(y)}(y_k){\cal H}_\omega \left[t^{(y)}(y_k)\right]^{-1}
t^{(y)}(y_k)\varphi_\omega
={\cal H}_{\omega'}t^{(y)}(y_k)\varphi_\omega.\ret
\end{eqnarray}
\end{proof}
In the same way, we have 

\begin{lemma}
Let $\varphi_{\omega,\ell}$ be an eigenvector of the Hamiltonian 
${\cal H}_\omega$ of (\ref{singlehamomega}) 
with the eigenvalue ${\cal E}_{\omega,\ell}$. Let $\varphi_{\omega',\ell}=
t^{(y)}(2y_0)R\varphi_{\omega,\ell}$, where $R$ is the reflection operator 
defined in (\ref{defR}). Then $\varphi_{\omega',\ell}$ 
is an eigenvector of the Hamiltonian ${\cal H}_{\omega'}$ with the random 
potential $V_{\omega'}$ given by 
\begin{equation}
V_{\omega'}(x,y)=V_\omega(-x,2y_0-y). 
\label{defV'}
\end{equation}
Here $y_0=\hbar k_0/(eB)$ with $k_0=2\pi n_0/L_x$, $(n_0\in {\bf Z})$. 
The corresponding eigenvalue ${\cal E}_{\omega',\ell}$ is equal to 
${\cal E}_{\omega,\ell}$. Further 
the system $\left\{\varphi_{\omega',\ell}\right\}$ is an orthogonal complete 
system if the original system $\left\{\varphi_{\omega,\ell}\right\}$ of the 
eigenvectors is an orthogonal complete system. 
\label{lemmaV}
\end{lemma}

Since $\pi_s$ is invariant under the magnetic translations 
$t^{(x)}(\cdots)$ and $t^{(y)}(\cdots)$, one can easily obtain the following 
lemma:

\begin{lemma} 
\label{lemram2}
Let $V_\omega$ be a random potential, 
and let $V_{\omega'}$ be the random potential given by (\ref{defV'}). 
Let $\varphi_{\omega,\ell}$ be the eigenvectors of 
the Hamiltonian ${\cal H}_\omega$, and 
let $\varphi_{\omega',\ell}=t^{(y)}(2y_0)R\varphi_{\omega,\ell}$. 
Then the following relation is valid: 
\begin{equation}
\left(\varphi_{\omega,\ell},\pi_s\varphi_{\omega,\ell'}\right)
=-\left(\varphi_{\omega',\ell},\pi_s\varphi_{\omega',\ell'}\right).
\end{equation}
\end{lemma}

Let $\varphi_{\omega,\ell}$ be an eigenvector of the Hamiltonian 
${\cal H}_\omega$. We expand $\varphi_{\omega,\ell}$ in Fourier series as 
\begin{equation}
\varphi_{\omega,\ell}(x,y)=L_x^{-1/2}\sum_k e^{ikx}
{\hat \varphi}_{\omega,\ell}(k,y). 
\label{phiomega}
\end{equation}
Since the vector $\varphi_{\omega,\ell}$ satisfies the periodic boundary 
condition $\varphi_{\omega,\ell}(x,y)=
t_y(L_y)\varphi_{\omega,\ell}(x,y)$, we have 
\begin{equation}
{\hat \varphi}_{\omega,\ell}(k,y)=
{\hat \varphi}_{\omega,\ell}(k-K,y-L_y) 
\label{phiPBC}
\end{equation}
as in (\ref{PBCk}) in the proof of Lemma~\ref{inproind}. 
We define a projection operator as 
\begin{equation}
{\tilde P}(k):=\sum_{\ell\in{\bf Z}}P(k+\ell K), 
\end{equation}
where $P(k)$ is given in (\ref{defP(k)}). 

\begin{lemma} 
\label{translateVomega}
Let $V_\omega$ be a random potential, and let 
$V_{\omega'}$ be the translate given by 
\begin{equation}
V_{\omega'}(x,y)=V_\omega(x,y-y_0),
\end{equation}
where $eBy_0=\hbar k_0=2\pi\hbar n_0/L_x$ with an integer $n_0$. 
Then
\begin{equation}
\left(\varphi_{\omega',m},{\tilde P}(k)\pi_s\varphi_{\omega',n}\right)
=\left(\varphi_{\omega,m},{\tilde P}(k-k_0)\pi_s\varphi_{\omega,n}\right). 
\label{translatek0}
\end{equation}
Here $\varphi_{\omega,n}$ is an eigenvector of the Hamiltonian 
${\cal H}_\omega$, and $\varphi_{\omega',n}=t^{(y)}(y_0)\varphi_{\omega,n}$ 
which is the corresponding eigenvector of ${\cal H}_{\omega'}$ as 
we showed in Lemma~\ref{lemmaTRV}. 
\end{lemma}

\begin{proof}{Proof} 
Since the vector ${\tilde P}(k)\pi_s\varphi_{\omega',n}$ satisfies 
the periodic boundary conditions (\ref{PBC}), one has 
\begin{eqnarray}
\left(\varphi_{\omega',m},{\tilde P}(k)\pi_s\varphi_{\omega',n}\right)
&=&\left(\varphi_{\omega,m},\left[t^{(y)}(y_0)\right]^{-1}{\tilde P}(k)\pi_s
t^{(y)}(y_0)\varphi_{\omega,n}\right)\ret
&=&\left(\varphi_{\omega,m},\left[t^{(y)}(y_0)\right]^{-1}{\tilde P}(k)
t^{(y)}(y_0)\pi_s\varphi_{\omega,n}\right). 
\end{eqnarray}
Therefore it is sufficinet to show 
\begin{equation}
\left[t^{(y)}(y_0)\right]^{-1}{\tilde P}(k)
t^{(y)}(y_0)={\tilde P}(k-k_0).
\label{tildePktrans}
\end{equation}
Let $f$ be a function on ${\bf R}^2$ such that it has a Fourier expansion 
\begin{equation}
f(x,y)=\sum_{k'} e^{ik'x}{\hat f}(k',y). 
\end{equation}
Then 
\begin{eqnarray}
\left[t^{(y)}(y_0)\right]^{-1}{\tilde P}(k)t^{(y)}(y_0)
f(x,y)&=&\left[t^{(y)}(y_0)\right]^{-1}{\tilde P}(k)
\sum_{k'}e^{i(k'+k_0)x}{\hat f}(k',y-y_0)\ret
&=&\left[t^{(y)}(y_0)\right]^{-1}\sum_{\ell}e^{i(k+\ell K)}
{\hat f}(k-k_0+\ell K,y-y_0)\ret
&=&\sum_{\ell}e^{i(k-k_0+\ell K)}
{\hat f}(k-k_0+\ell K,y)\ret
&=&{\tilde P}(k-k_0)f(x,y).
\end{eqnarray}
\end{proof}

\subsection{The $N$ electrons Landau Hamiltonian with disorder}

We define the magnetic translation operators for $N$ electrons as 
\begin{equation}
T^{(N,x)}(x'):=\bigotimes_{j=1}^N t_j^{(x)}(x')
\end{equation}
and 
\begin{equation}
T^{(N,y)}(y'):=\bigotimes_{j=1}^N t_j^{(y)}(y'). 
\end{equation}
Further we define the reflection operator for $N$ electrons as 
\begin{equation}
R^{(N)}:=\bigotimes_{j=1}^N R_j. 
\end{equation}
In the same way as in Section~\ref{Landauhamdisorder}, 
we have the following two lemmas: 

\begin{lemma}
\label{eigenPhip}
Let $\Phi_\omega^{(N)}$ be an eigenvector of the Hamiltoian 
$H_{\omega,0}^{(N)}$ of (\ref{unperturbedHam}) with a ramdom potential 
$V_\omega$, and let $E_\omega^{(N)}$ be the corresponding energy eigenvalue. 
Let $V_{\omega'}$ be the reflection of the random potentail $V_\omega$ 
with respect to the axes $x=0$ and $y=y_0$, i.e., 
\begin{equation}
V_{\omega'}(x,y)=V_\omega(-x,2y_0-y),
\end{equation}
where $y_0=\hbar k_0/(eB)$ with $k_0=2\pi n_0/L_x$, $(n_0\in {\bf Z})$. 
Set $\Phi_{\omega'}^{(N)}=T^{(N,y)}(2y_0)R^{(N)}\Phi_\omega^{(N)}$. 
Then $\Phi_{\omega'}^{(N)}$ is an eigenvector of $H_{\omega',0}^{(N)}$ with 
the random potential $V_{\omega'}$, and the energy 
eigenvalue $E_{\omega'}^{(N)}$ is equal to $E_\omega^{(N)}$.
\end{lemma}

\begin{lemma} 
\label{Phipi-}
Let $V_\omega$ be a ramdom potential, and let 
$V_{\omega'}$ be the reflection given by 
\begin{equation}
V_{\omega'}(x,y)=V_\omega(-x,2y_0-y).
\end{equation}
Here $y_0$ is the same as in Lemma~\ref{eigenPhip}. 
Let $\Phi_{\omega,\ell}^{(N)}$ be eigenvectors of the Hamiltonian 
$H_{\omega,0}^{(N)}$. Then 
\begin{equation}
\left\langle\Phi_{\omega',\ell}^{(N)},\pi_{x,j}
\Phi_{\omega',\ell'}^{(N)}\right\rangle=
-\left\langle\Phi_{\omega,\ell}^{(N)},\pi_{x,j}
\Phi_{\omega,\ell'}^{(N)}\right\rangle,
\label{reflecpi}
\end{equation}
where the vector $\Phi_{\omega',\ell}^{(N)}=
T^{(N,y)}(2y_0)R^{(N)}\Phi_\omega^{(N)}$ which are the eigenvectors 
of the Hamiltonian $H_{\omega',0}^{(N)}$ 
with the random potential $V_{\omega'}$ as we showed in the preceding 
Lemma~\ref{eigenPhip}. 
\end{lemma}

\begin{lemma}
\label{translateVomegaPhi}
Let $V_\omega$ be a random potential, and let 
$V_{\omega'}$ be the translate given by 
\begin{equation}
V_{\omega'}(x,y)=V_\omega(x,y-y_0),
\end{equation}
where $eBy_0=\hbar k_0=2\pi\hbar n_0/L_x$ with an integer $n_0$. 
Then the following equalities are valid: 
\begin{equation}
\left\langle\Phi_{\omega',\ell}^{(N)},\pi_{s,j}\Phi_{\omega',\ell'}^{(N)}
\right\rangle=
\left\langle\Phi_{\omega,\ell}^{(N)},\pi_{s,j}\Phi_{\omega,\ell'}^{(N)}
\right\rangle
\label{Phiomega'piPhiomega}
\end{equation}
and
\begin{equation}
\left\langle\Phi_{\omega',\ell}^{(N)},{\tilde P}_j(k)\pi_{s,j}
\Phi_{\omega',\ell'}^{(N)}\right\rangle=
\left\langle\Phi_{\omega,\ell}^{(N)},{\tilde P}_j(k-k_0)
\pi_{s,j}\Phi_{\omega,\ell'}^{(N)}\right\rangle.
\label{tildePktransN}
\end{equation}
Here $\Phi_{\omega,n}^{(N)}$ are the eigenvectors of the Hamiltonian 
$H_{\omega,0}^{(N)}$, and $\Phi_{\omega',n}^{(N)}=
T^{(N,y)}(y_0)\Phi_{\omega,n}^{(N)}$ which are the eigenvectors of 
$H_{\omega',0}^{(N)}$ with the random potential $V_{\omega'}$. 
\end{lemma}

\begin{proof}{Proof}
Since $\pi_{s,j}$ is invariant the magnetic translations, 
one can easily obtain (\ref{Phiomega'piPhiomega}). 
The relation (\ref{tildePktransN}) follows from 
the identity (\ref{tildePktrans}) in Lemma~\ref{translateVomega}
\end{proof}

\Section{Estimate of ${\bf E}_\omega\left[{\cal M}_{s,{\rm in}}\right]$}
\label{EstEMsin}

In this appendix, we estimate the random average of ${\cal M}_{s,{\rm in}}$ 
of (\ref{Min}), which is given by 
\begin{equation}
{\bf E}_\omega\left[{\cal M}_{s,{\rm in}}\right]
=\frac{1}{m_e N}\sum_{i=1}^N\sum_{j=1}^N
{\bf E}_\omega\left[\left\langle\Phi_{\omega,0}^{(N)},\pi_{s,j}
\frac{\left[1-G_\omega^{(N)}\right]}{E_{\omega,0}^{(N)}-H_{\omega,0}^{(N)}}
P_{{\rm in},i}p_{x,i}\Phi_{\omega,0}^{(N)}\right\rangle\right]. 
\label{expMsin}
\end{equation}
For this purpose, we first want to get the explicit forms 
of the ``corrections from the boundaries" in (\ref{approEMinmat}). 

To begin with, we note the following: 
Let $\Phi_\omega^{(N)}$ be an $N$ electrons eigenvector of the unperturbed 
Hamiltonian $H_{\omega,0}^{(N)}$ of (\ref{unperturbedHam}). 
Clearly this vector can be expanded as 
\begin{equation}
\Phi_\omega^{(N)}=\sum_{\{\ell_j\}}a_{\omega,\{\ell_j\}}
{\rm Asym}\left[\varphi_{\omega,\ell_1}\otimes\varphi_{\omega,\ell_2}\otimes
\cdots\otimes\varphi_{\omega,\ell_N}\right]
\label{PhiexpandSlater}
\end{equation}
in terms of the normalized eigenvectors $\{\varphi_{\omega,\ell}\}$ of 
the single electron Hamiltonian ${\cal H}_\omega$ of (\ref{singlehamomega}), 
where ${\rm Asym}[\cdots]$ stands for the antisymmetrization of 
a wavefunction, i.e., 
\begin{equation}
{\rm Asym}\left[\Phi\right]({\bf r}_1,{\bf r}_2,\ldots,{\bf r}_N)
:=\frac{1}{\sqrt{N!}}\sum_\sigma (-1)^{\ell(\sigma)}
\Phi({\bf r}_{\sigma(1)},{\bf r}_{\sigma(2)},\ldots,{\bf r}_{\sigma(N)})
\label{defAsym}
\end{equation}
for a function $\Phi$ of $({\bf r}_1,{\bf r}_2,\ldots,{\bf r}_N)$. 
Here the sum runs over all the permutations $\sigma$ of $(1,2,\ldots,N)$, 
and $\ell(\sigma)$ is the number of binary permutations in the permutation 
$\sigma$. 

\begin{lemma}
\label{E+}
The following relation is valid:
\begin{eqnarray}
& &{\bf E}_\omega\left[\left\langle\Phi_{\omega,0}^{(N)},\pi_{s,j}
\frac{1-G_\omega^{(N)}}{E_{\omega,0}^{(N)}-H_{\omega,0}^{(N)}}
P_{{\rm in},i}p_{x,i}\Phi_{\omega,0}^{(N)}\right\rangle
\right]\ret
&=&2{\bf E}_\omega\left[\left\langle\Phi_{\omega,0}^{(N)},\pi_{s,j}
\frac{1-G_\omega^{(N)}}{E_{\omega,0}^{(N)}-H_{\omega,0}^{(N)}}
P_{{\rm in},i}^{(+)}p_{x,i}\Phi_{\omega,0}^{(N)}\right\rangle\right],
\label{PinPin+}
\end{eqnarray}
where
\begin{equation}
P_{\rm in}^{(+)}=\sum_{k \in {\cal F}(I_{\rm in}^{(+)})}P(k)
\end{equation}
with the interval $I_{\rm in}^{(+)}=(0,L_y/2-\delta]$. 
\end{lemma}

\begin{proof}{Proof}
Let $V_\omega$ be a random potential, and let $V_{\omega'}$ be the reflection 
of $V_\omega$ with respect to the $x$ and $y$ axes, i.e., 
\begin{equation}
V_{\omega'}(x,y)=V_\omega(-x,-y). 
\label{Vomega-}
\end{equation}
Let $\varphi_{\omega,n}$ and $\varphi_{\omega',n}$ are the normalized 
eigenvectors of the single electron Hamiltonian ${\cal H}_\omega$ of 
(\ref{singlehamomega})
with the random potentials $V_\omega$ and $V_{\omega'}$, respectively. 
{From} Lemma~\ref{lemmaV}, we can take 
$\varphi_{\omega',n}=R\varphi_{\omega,n}$. By using the Fourier expansion 
(\ref{phiomega}) for $\varphi_{\omega,n}$, we have
\begin{eqnarray}
\varphi_{\omega',n}(x,y)
&=&RL_x^{-1/2}\sum_k e^{ikx}{\hat \varphi}_{\omega,n}(k,y)\ret
&=&L_x^{-1/2}\sum_k e^{-ikx}{\hat \varphi}_{\omega,n}(k,-y)\ret
&=&L_x^{-1/2}\sum_k e^{ikx}{\hat \varphi}_{\omega,n}(-k,-y).
\end{eqnarray}
This implies 
\begin{equation}
{\hat \varphi}_{\omega',n}(k,y)={\hat \varphi}_{\omega,n}(-k,-y).
\label{kmkrelation}
\end{equation}
Thereby we have 
\begin{equation}
\left(\varphi_{\omega',m},P(k)\varphi_{\omega',n}\right)
=\left(\varphi_{\omega,m},P(-k)\varphi_{\omega,n}\right).
\label{eq-omega}
\end{equation}

Let $\Phi_{\omega,n}^{(N)}$ be an eigenvector of 
the Hamiltonian $H_{\omega,0}^{(N)}$ 
with the random potential $V_\omega$, 
and let $\Phi_{\omega',n}^{(N)}$ be an eigenvector of 
the Hamiltonian $H_{\omega',0}^{(N)}$ 
with the random potential $V_{\omega'}$ of (\ref{Vomega-}). 
{From} Lemma~\ref{eigenPhip}, we can take 
$\Phi_{\omega',n}^{(N)}=R^{(N)}\Phi_{\omega,n}^{(N)}$. 
Combining this with the expansion (\ref{PhiexpandSlater}) 
for the vector $\Phi_{\omega,n}^{(N)}$, we have 
\begin{equation}
\Phi_{\omega',n}^{(N)}=\sum_{\{\ell_j\}}a_{\omega,\{\ell_j\}}^{(n)}
{\rm Asym}\left[\varphi_{\omega',\ell_1}\otimes\varphi_{\omega',\ell_2}\otimes
\cdots\otimes\varphi_{\omega',\ell_N}\right]
\end{equation}
with $\varphi_{\omega',\ell}=R\varphi_{\omega,\ell}$. 
Using this expression, one can easily obtain 
$$
\left\langle\Phi_{\omega',m}^{(N)},P_i(k)\Phi_{\omega',n}^{(N)}\right\rangle
\hspace{11cm}
$$
\begin{equation}
=\sum_{\{\ell_j\},\{\ell_j'\}}
{a_{\omega,\{\ell_j\}}^{(m)}}^\ast
a_{\omega,\{\ell_j'\}}^{(n)}
\left\langle{\rm Asym}\left[\varphi_{\omega',\ell_1}\otimes\cdots\otimes
\varphi_{\omega',\ell_N}\right],P_i(k)
{\rm Asym}\left[\varphi_{\omega',\ell_1'}\otimes\cdots\otimes
\varphi_{\omega',\ell_N'}\right]\right\rangle.
\label{expPkmatPhi}
\end{equation}
The matrix elements in the right-hand side are written as 
\begin{eqnarray}
& &\left\langle{\rm Asym}
\left[\varphi_{\omega',\ell_1}\otimes\varphi_{\omega',\ell_2}\otimes\cdots,
\varphi_{\omega',\ell_N}\right],P_i(k)
{\rm Asym}\left[\varphi_{\omega',\ell_1'}\otimes\varphi_{\omega',\ell_2'}
\otimes\cdots\otimes\varphi_{\omega',\ell_N'}\right]\right\rangle\ret
&=&\frac{1}{N}\cases{\displaystyle{\sum_{\ell\in\{\ell_1,\ldots,\ell_N\}}
\left(\varphi_{\omega',\ell},P(k)\varphi_{\omega',\ell}\right)}& 
if $\ell_j=\ell_j'$ for all $j=1,2,\ldots,N$;\cr
& \cr
\qquad\quad \pm
\left(\varphi_{\omega',\ell_j},P(k)\varphi_{\omega',\ell_m'}\right)& 
if $\{\ell_k\}_{k=1}^N\backslash\{\ell_j\}=\{\ell_k'\}_{k=1}^N\backslash
\{\ell_m'\}$ and 
$\ell_j\ne \ell_m'$;\cr
& \cr
\qquad\qquad\qquad\quad 0, & otherwise.   \cr}\ret
\label{detmatPk}
\end{eqnarray}
Combining (\ref{eq-omega}), (\ref{expPkmatPhi}) and (\ref{detmatPk}), 
we obtain 
\begin{equation}
\left\langle\Phi_{\omega',m}^{(N)},P_i(-k)\Phi_{\omega',n}^{(N)}\right\rangle
=\left\langle\Phi_{\omega,m}^{(N)},P_i(k)\Phi_{\omega,n}^{(N)}\right\rangle.
\end{equation}
{From} this and (\ref{reflecpi}) in Lemma~\ref{Phipi-}, we obtain 
the desired result (\ref{PinPin+}). 
\end{proof}

\begin{lemma} 
\label{innomegap}
Let $V_\omega$ be a random potential, and let $V_{\omega'}$ be 
the reflection given by 
\begin{equation}
V_{\omega'}(x,y)=V_\omega(-x,2y_k-y) 
\label{VVrelation}
\end{equation}
with $y_k>0$. Let $\varphi_{\omega,n}$ be the normalized 
eigenvectors of the single electron Hamiltonian ${\cal H}_\omega$ of 
(\ref{singlehamomega})with the potential $V_\omega$, and let 
$\varphi_{\omega',n}=t^{(y)}(2y_k)R\varphi_{\omega,n}$ which are 
the eigenvectors of ${\cal H}_{\omega'}$ with $V_{\omega'}$ 
from Lemma~\ref{lemmaV}. Then the following relation is valid: 
\begin{eqnarray}
\left(\varphi_{\omega',\ell},P(k)\varphi_{\omega',\ell'}\right)
&=&\left(\varphi_{\omega,\ell},P(k)\varphi_{\omega,\ell'}\right)
+\left(\varphi_{\omega,\ell},{\tilde \chi}_k P(k-K)
\varphi_{\omega,\ell'}\right)-
\left(\varphi_{\omega,\ell},{\tilde \chi}_k P(k)\varphi_{\omega,\ell'}\right),
\ret
\label{phiomegap}
\end{eqnarray}
where $K=L_y/\ell_B^2$, and 
\begin{equation}
{\tilde \chi}_k(y):=\cases{1 & if \ $y \in [-L_y/2,-L_y/2+2y_k]$; \cr 
0, & otherwise.\cr}
\end{equation}
\end{lemma}

\begin{proof}{Proof} 
By using the Fourier expansion (\ref{phiomega}) for $\varphi_{\omega,n}$, 
we have 
\begin{eqnarray}
\varphi_{\omega',n}(x,y)&=&t^{(y)}(2y_k)L_x^{-1/2}\sum_{k'}
e^{-ik'x}{\hat \varphi}_{\omega,n}(k',-y)\ret
&=&L_x^{-1/2}\sum_{k'}
e^{i(2k-k')x}{\hat \varphi}_{\omega,n}(k',2y_k-y)\ret
&=&L_x^{-1/2}\sum_{k''}
e^{ik''x}{\hat \varphi}_{\omega,n}(2k-k'',2y_k-y).
\end{eqnarray}
Thereby we get 
\begin{equation}
\left(\varphi_{\omega',\ell},P(k)\varphi_{\omega',\ell'}\right)
=\int_{-L_y/2}^{L_y/2}dy \ 
\left[{\hat \varphi}_{\omega,\ell}(k,2y_k-y)\right]^\ast
{\hat \varphi}_{\omega,\ell'}(k,2y_k-y). 
\end{equation}
Further we can rewrite the right-hand side as 
\begin{eqnarray}
& &\left(\varphi_{\omega',\ell},P(k)\varphi_{\omega',\ell'}\right)\ret
&=&\int_{-L_y/2+2y_k}^{L_y/2+2y_k}d{\tilde y} \ 
\left[{\hat \varphi}_{\omega,\ell}(k,{\tilde y})\right]^\ast
{\hat \varphi}_{\omega,\ell'}(k,{\tilde y})\ret
&=&\left(\varphi_{\omega,\ell},P(k)\varphi_{\omega,\ell'}\right)\ret
&+&\int_{L_y/2}^{L_y/2+2y_k}d{\tilde y} \ 
\left[{\hat \varphi}_{\omega,\ell}(k,{\tilde y})\right]^\ast
{\hat \varphi}_{\omega,\ell'}(k,{\tilde y})
-\int_{-L_y/2}^{-L_y/2+2y_k}d{\tilde y} \ 
\left[{\hat \varphi}_{\omega,\ell}(k,{\tilde y})\right]^\ast
{\hat \varphi}_{\omega,\ell'}(k,{\tilde y})\ret 
&=&\left(\varphi_{\omega,\ell},P(k)\varphi_{\omega,\ell'}\right)
+\int_{-L_y/2}^{-L_y/2+2y_k}d{\tilde y} \ 
\left[{\hat \varphi}_{\omega,\ell}(k-K,{\tilde y})\right]^\ast
{\hat \varphi}_{\omega,\ell'}(k-K,{\tilde y})\ret
& &-\int_{-L_y/2}^{-L_y/2+2y_k}d{\tilde y} \ 
\left[{\hat \varphi}_{\omega,\ell}(k,{\tilde y})\right]^\ast
{\hat \varphi}_{\omega,\ell'}(k,{\tilde y})\ret
&=&\left(\varphi_{\omega,\ell},P(k)\varphi_{\omega,\ell'}\right)
+\left(\varphi_{\omega,\ell},{\tilde \chi}_k P(k-K)
\varphi_{\omega,\ell'}\right)-
\left(\varphi_{\omega,\ell},{\tilde \chi}_k P(k)\varphi_{\omega,\ell'}\right).
\end{eqnarray}
Here we have used (\ref{phiPBC}) for getting the third equality. 
\end{proof}

\begin{lemma} Let $V_\omega$ be a ramdom potential, and let 
$V_{\omega'}$ be the reflection given by 
\begin{equation}
V_{\omega'}(x,y)=V_\omega(-x,2y_k-y).
\label{randomomegap}
\end{equation}
Here $y_k$ is the same as in the preceding Lemma~\ref{innomegap}. 
Let $\Phi_{\omega,n}^{(N)}$ be the eigenvectors of the Hamiltonian 
$H_{\omega,0}^{(N)}$ with the random potential $V_\omega$, and 
let $\Phi_{\omega',n}^{(N)}=T^{(N,y)}(2y_k)R^{(N)}\Phi_{\omega,n}^{(N)}$ 
which are the eigenvectors of the Hamiltonian $H_{\omega',0}^{(N)}$ 
with the random potential $V_{\omega'}$ of (\ref{randomomegap}), 
as we showed in Lemma~\ref{eigenPhip}. Then the following relation is valid:
\begin{eqnarray}
& &\left\langle\Phi_{\omega',n}^{(N)},P_i(k)\Phi_{\omega',0}^{(N)}\right\rangle
\ret
&=&\left\langle\Phi_{\omega,n}^{(N)},P_i(k)\Phi_{\omega,0}^{(N)}\right\rangle
+\left\langle\Phi_{\omega,n}^{(N)},{\tilde \chi}_{k,i}
P_i(k-K)\Phi_{\omega,0}^{(N)}\right\rangle
-\left\langle\Phi_{\omega,n}^{(N)},{\tilde \chi}_{k,i}
P_i(k)\Phi_{\omega,0}^{(N)}\right\rangle.\ret
\label{matoop}
\end{eqnarray}
\end{lemma}

\begin{proof}{Proof}
In the same way as in the proof of Lemma~\ref{E+}, we have the expressions 
(\ref{expPkmatPhi}) and (\ref{detmatPk}) also for the random potential 
$V_{\omega'}$ of (\ref{randomomegap}). Combining these with (\ref{phiomegap}), 
we obtain the desired result (\ref{matoop}). 
\end{proof}

Using the above result (\ref{matoop}) and Lemma~\ref{Phipi-}, we have 
\begin{eqnarray}
& &\left\langle\Phi_{\omega',0}^{(N)},\pi_{s,j}\Phi_{\omega',\ell}^{(N)}
\right\rangle\frac{1}{E_{\omega',0}^{(N)}-E_{\omega',\ell}^{(N)}}
\left\langle\Phi_{\omega',\ell}^{(N)},P_{{\rm in},i}^{(+)}p_{x,i}
\Phi_{\omega',0}^{(N)}\right\rangle\ret
&=&\sum_{k\in {\cal F}(I_{\rm in}^{(+)})}\hbar k
\left\langle\Phi_{\omega',0}^{(N)},\pi_{s,j}\Phi_{\omega',\ell}^{(N)}
\right\rangle\frac{1}{E_{\omega',0}^{(N)}-E_{\omega',\ell}^{(N)}}
\left\langle\Phi_{\omega',\ell}^{(N)},P_i(k)\Phi_{\omega',0}^{(N)}\right\rangle
\ret
&=&-\sum_{k\in {\cal F}(I_{\rm in}^{(+)})}\hbar k
\left\langle\Phi_{\omega,0}^{(N)},\pi_{s,j}\Phi_{\omega,\ell}^{(N)}
\right\rangle\frac{1}{E_{\omega,0}^{(N)}-E_{\omega,\ell}^{(N)}}\ret
&\times&\left[\left\langle\Phi_{\omega,\ell}^{(N)},P_i(k)\Phi_{\omega,0}^{(N)}
\right\rangle+\left\langle\Phi_{\omega,\ell}^{(N)},{\tilde \chi}_{k,i}P_i(k-K)
\Phi_{\omega,0}^{(N)}
\right\rangle-\left\langle\Phi_{\omega,\ell}^{(N)},{\tilde \chi}_{k,i}P_i(k)
\Phi_{\omega,0}^{(N)}\right\rangle\right]\ret
\end{eqnarray}
for $\ell\ne 0$. Taking the random average in both sides, we get 
\begin{eqnarray}
& &2{\bf E}_\omega\left[
\sum_{k\in {\cal F}(I_{\rm in}^{(+)})}\hbar k
\left\langle\Phi_{\omega,0}^{(N)},\pi_{s,j}\Phi_{\omega,\ell}^{(N)}
\right\rangle\frac{1}{E_{\omega,0}^{(N)}-E_{\omega,\ell}^{(N)}}
\left\langle\Phi_{\omega,\ell}^{(N)},P_i(k)\Phi_{\omega,0}^{(N)}\right\rangle
\right]\ret
&=&{\bf E}_\omega\left[
\sum_{k\in {\cal F}(I_{\rm in}^{(+)})}\hbar k
\left\langle\Phi_{\omega,0}^{(N)},\pi_{s,j}\Phi_{\omega,\ell}^{(N)}
\right\rangle\frac{1}{E_{\omega,0}^{(N)}-E_{\omega,\ell}^{(N)}}
\left\langle\Phi_{\omega,\ell}^{(N)},{\tilde \chi}_{k,i}P_i(k-K)
\Phi_{\omega,0}^{(N)}
\right\rangle\right]\ret
&-&{\bf E}_\omega\left[
\sum_{k\in {\cal F}(I_{\rm in}^{(+)})}\hbar k
\left\langle\Phi_{\omega,0}^{(N)},\pi_{s,j}\Phi_{\omega,\ell}^{(N)}
\right\rangle\frac{1}{E_{\omega,0}^{(N)}-E_{\omega,\ell}^{(N)}}
\left\langle\Phi_{\omega,\ell}^{(N)},{\tilde \chi}_{k,i}P_i(k)
\Phi_{\omega,0}^{(N)}\right\rangle\right]\ret
\label{cancelE}
\end{eqnarray}
for $\ell\ne 0$. {From} (\ref{expMsin}), (\ref{PinPin+}) and 
(\ref{cancelE}), we have 
\begin{eqnarray}
{\bf E}_\omega\left[{\cal M}_{s,{\rm in}}\right]&=&
\frac{1}{m_eN}\sum_{i=1}^N\sum_{j=1}^N
{\bf E}_\omega\left[\left\langle\Phi_{\omega,0}^{(N)},\pi_{s,j}
\frac{[1-G_\omega^{(N)}]}{E_{\omega,0}^{(N)}-H_{\omega,0}^{(N)}}
P_{{\rm in},i}p_{x,i}\Phi_{\omega,0}^{(N)}\right\rangle\right]\ret
&=&\frac{1}{m_eN}\sum_{i=1}^N\sum_{j=1}^N
{\bf E}_\omega\left[\left\langle\Phi_{\omega,0}^{(N)},\pi_{s,j}
\frac{[1-G_\omega^{(N)}]}{E_{\omega,0}^{(N)}-H_{\omega,0}^{(N)}}
{\tilde Q}_{{\rm in},i}^{(+)}(p_{x,i}+\hbar K)
\Phi_{\omega,0}^{(N)}\right\rangle\right]\ret
&-&\frac{1}{m_eN}\sum_{i=1}^N\sum_{j=1}^N
{\bf E}_\omega\left[\left\langle\Phi_{\omega,0}^{(N)},\pi_{s,j}
\frac{[1-G_\omega^{(N)}]}{E_{\omega,0}^{(N)}-H_{\omega,0}^{(N)}}
Q_{{\rm in},i}^{(+)}p_{x,i}\Phi_{\omega,0}^{(N)}\right\rangle\right],
\label{EMin}
\end{eqnarray}
where 
\begin{equation}
Q_{\rm in}^{(+)}:=\sum_{k\in {\cal F}(I_{\rm in}^{(+)})}P(k){\tilde \chi}_k, 
\end{equation}
and
\begin{equation}
{\tilde Q}_{\rm in}^{(+)}:=\sum_{k\in {\cal F}(I_{\rm in}^{(+)})}P(k-K)
{\tilde \chi}_k. 
\end{equation}
\medskip

Now we estimate ${\bf E}_\omega\left[{\cal M}_{s,{\rm in}}\right]$ 
by using the expression (\ref{EMin}). 

\subsection{Non-interacting case}
\label{App:EstEMinNon}

Consider first the non-interacting case, i.e., $U^{(2)}=0$. 
Then ${\bf E}_\omega\left[{\cal M}_{s,{\rm in}}\right]$ of (\ref{EMin}) 
can be written as 
\begin{eqnarray}
{\bf E}_\omega\left[{\cal M}_{s,{\rm in}}\right]
&=&\frac{1}{m_eN}\sum_{n\le N}{\bf E}_\omega
\left[\left(\varphi_{\omega,n},\pi_s
\frac{{\cal P}_>}{{\cal E}_{\omega,n}-{\cal H}_\omega}
{\tilde Q}_{\rm in}^{(+)}(p_x+\hbar K)\varphi_{\omega,n}\right)\right]\ret
&-&\frac{1}{m_eN}\sum_{n\le N}{\bf E}_\omega
\left[\left(\varphi_{\omega,n},\pi_s
\frac{{\cal P}_>}{{\cal E}_{\omega,n}-{\cal H}_\omega}
Q_{\rm in}^{(+)}p_x\varphi_{\omega,n}\right)\right]
\label{expMin}
\end{eqnarray}
in terms of the eigenvectors $\varphi_{\omega,n}$ of the single electron 
Hamiltonian ${\cal H}_\omega$ of (\ref{singlehamomega}), with the energy 
eigenvalues ${\cal E}_{\omega,n}$, $n=1,2,\ldots$. Here 
we have taken order ${\cal E}_{\omega,m}\le{\cal E}_{\omega,n}$ for 
$m<n$, and ${\cal P}_>$ is the projection onto 
the subspace spanned by all states above the Fermi level, 
i.e., all the vectors $\varphi_{\omega,n}$ with $n>N$. 
Without loss of generality, we can assume $V_\omega\ge 0$. Then 
we have ${\cal E}_{\omega,n}\ge 0$ for all indices $n$. 

Let us estimate the matrix elements in the second sum in the right-hand side 
of (\ref{expMin}). Using the Schwarz inequality we have 
\begin{eqnarray}
& &\left|\left(\varphi_{\omega,n},\pi_s
\frac{{\cal P}_>}{{\cal E}_{\omega,n}-{\cal H}_\omega}
Q_{\rm in}^{(+)}p_x\varphi_{\omega,n}\right)\right|\ret
&\le&\frac{eBL_y}{2}\sqrt{\left(\psi_{\omega,n}^{(s)},
Q_{\rm in}^{(+)}\psi_{\omega,n}^{(s)}\right)
\left(\varphi_{\omega,n},Q_{\rm in}^{(+)}
\varphi_{\omega,n}\right)}, 
\label{pQpbound}
\end{eqnarray}
where 
\begin{equation}
\psi_{\omega,n}^{(s)}:=\frac{{\cal P}_>}{{\cal E}_{\omega,n}-{\cal H}_\omega}
\pi_s\varphi_{\omega,n}.
\label{defpsi}
\end{equation}

\begin{lemma}
\label{phiQphibound}
\begin{equation}
\left(\varphi_{\omega,n},Q_{\rm in}^{(+)}
\varphi_{\omega,n}\right)\le 
{\cal C}_1\left(\frac{\ell_B}{\delta}\right)^4\quad\mbox{for}\ n\le N
\label{Qestimate}
\end{equation}
with the positive constant 
\begin{equation}
{\cal C}_1
:=\left(\frac{2}{\hbar\omega_c}\right)^2
\left({\cal E}_{0,>}+\Vert V_\omega\Vert\right)
\left({\cal E}_{0,>}+\Vert V_\omega\Vert+4\hbar\omega_c\right).
\end{equation}
Here ${\cal E}_{0,>}:=\min_{m>N}\{{\cal E}_{\omega,m}\}$. 
\end{lemma}

\begin{proof}{Proof} Note that 
\begin{eqnarray}
& &\sum_{k\in{\cal F}(I_{\rm in}^{(+)})}\int_{-L_y/2}^{-L_y/2+2y_k}dy
\ (y-y_k)^4|{\hat \varphi}_{\omega,n}(k,y)|^2 \ret
&\ge&
\sum_{k\in{\cal F}(I_{\rm in}^{(+)})}\int_{-L_y/2}^{-L_y/2+2y_k}dy
\ \left(\frac{L_y}{2}-y_k\right)^4
|{\hat \varphi}_{\omega,n}(k,y)|^2 \ret
&\ge&\delta^4
\sum_{k\in{\cal F}(I_{\rm in}^{(+)})}\int_{-L_y/2}^{-L_y/2+2y_k}dy
\ |{\hat \varphi}_{\omega,n}(k,y)|^2 \ret
&=&\delta^4\left(\varphi_{\omega,n},Q_{\rm in}^{(+)}
\varphi_{\omega,n}\right).
\end{eqnarray}
Combining this with the bound (\ref{4momentBound}) 
in Appendix~\ref{DecaywaveNon}, we get (\ref{Qestimate}). 
\end{proof}

A similar bound for 
$\left(\psi_{\omega,n}^{(s)},Q_{\rm in}^{(+)}\psi_{\omega,n}^{(s)}\right)$ 
in (\ref{pQpbound}) can be obtained as follows: 
In the same way as in the proof of Lemma~\ref{phiQphibound}, we have 
\begin{eqnarray}
& &\delta^4\left(\psi_{\omega,n}^{(s)},Q_{\rm in}^{(+)}\psi_{\omega,n}^{(s)}
\right)\ret
&\le&\sum_{k\in {\cal F}(I_{\rm in}^{(+)})}\int_{-L_y/2}^{-L_y/2+2y_k}
dy\ (y-y_k)^4\left|{\hat \psi}_{\omega,n}^{(s)}(k,y)\right|^2\ret
&\le&\ell_B^4\left(\frac{2}{\hbar\omega_c}\right)^2
\left\{\left(\Vert{\cal H}_\omega\psi_{\omega,n}^{(s)}\Vert
+\Vert V_\omega\Vert\Vert\psi_{\omega,n}^{(s)}\Vert\right)^2
+4\hbar\omega_c\left[\left(\psi_{\omega,n}^{(s)},{\cal H}_\omega
\psi_{\omega,n}^{(s)}\right)
+\Vert V_\omega\Vert\Vert\psi_{\omega,n}^{(s)}\Vert^2\right]\right\}.\ret
\label{psiQpsibound2}
\end{eqnarray}
Note that 
\begin{eqnarray}
\frac{{\cal H}_\omega}{{\cal H}_\omega-{\cal E}_{\omega,n}}{\cal P}_>
&=&
\frac{{\cal H}_\omega-{\cal E}_{\omega,n}+{\cal E}_{\omega,n}}
{{\cal H}_\omega-{\cal E}_{\omega,n}}{\cal P}_>\ret
&=&\left(1+\frac{{\cal E}_{\omega,n}}{{\cal H}_\omega-{\cal E}_{\omega,n}}
\right){\cal P}_>\ret
&\le& \min_{m>N}
\left(1+\frac{{\cal E}_{\omega,n}}{{\cal E}_{\omega,m}-{\cal E}_{\omega,n}}
\right){\cal P}_>\ret
&=&\frac{{\cal E}_{0,>}}{{\cal E}_{0,>}-{\cal E}_{\omega,n}}{\cal P}_>
\le \frac{{\cal E}_{0,>}}{\Delta E}{\cal P}_>
\label{H/H-E}
\end{eqnarray}
for the indices $n\le N$. Here $\Delta E$ is the lower bound for 
the energy gap given in (\ref{gapcondition}). 
Clearly $\Delta E\le\min_{n\le N}\{{\cal E}_{0,>}-{\cal E}_{\omega,n}\}$ 
which we have used for getting the last inequality in (\ref{H/H-E}). 
Using the bound (\ref{H/H-E}), we have 
\begin{equation}
\left\Vert{\cal H}_\omega\psi_{\omega,n}^{(s)}\right\Vert^2
\le \left(\frac{{\cal E}_{0,>}}{\Delta E}\right)^2
\left(\varphi_{\omega,n},\pi_s^2\varphi_{\omega,n}\right)
\le 2m_e{\cal E}_{\omega,n}
\left(\frac{{\cal E}_{0,>}}{\Delta E}\right)^2.
\end{equation}
Similarly we obtain 
\begin{equation}
\left(\psi_{\omega,n}^{(s)},{\cal H}_\omega\psi_{\omega,n}^{(s)}\right)
\le \frac{2m_e{\cal E}_{\omega,n}{\cal E}_{0,>}}{(\Delta E)^2},
\end{equation}
and 
\begin{equation}
\left\Vert\psi_{\omega,n}^{(s)}\right\Vert^2
\le \frac{2m_e{\cal E}_{\omega,n}}{(\Delta E)^2}.
\end{equation}
Substituting these bounds into (\ref{psiQpsibound2}), we have 

\begin{lemma}
\label{psiQpsibound}
\begin{equation}
\left(\psi_{\omega,n}^{(s)},Q_{\rm in}^{(+)}\psi_{\omega,n}^{(s)}\right)
\le {\cal C}_1\frac{2m_e{\cal E}_{\omega,n}}{(\Delta E)^2}
\left(\frac{\ell_B}{\delta}\right)^4\quad\mbox{for} \ n\le N.
\end{equation}
\end{lemma}
{From} the bound (\ref{pQpbound}) and Lemmas~\ref{phiQphibound} and 
\ref{psiQpsibound}, we have 
\begin{lemma}
\label{pQpbound2}
\begin{eqnarray}
& &\left|\left(\varphi_{\omega,n},\pi_s
\frac{{\cal P}_>}{{\cal E}_{\omega,n}-{\cal H}_\omega}
Q_{\rm in}^{(+)}p_x\varphi_{\omega,n}\right)\right|
\le \frac{m_e\sqrt{\hbar\omega_c{\cal E}_{0,>}}}{\sqrt{2}\Delta E}
{\cal C}_1\frac{L_y}{\ell_B}\left(\frac{\ell_B}{\delta}\right)^4 
\quad \mbox{for}\ n\le N.\ret
\end{eqnarray}
\end{lemma}

In the same way, we have the following lemma: 
\begin{lemma}
\label{pQTpbound2}
\begin{equation}
\left|\left(\varphi_{\omega,n},\pi_s
\frac{{\cal P}_>}{{\cal E}_{\omega,n}-{\cal H}_\omega}
{\tilde Q}_{\rm in}^{(+)}(p_x+\hbar K)
\varphi_{\omega,n}\right)\right|
\le\frac{m_e\sqrt{\hbar\omega_c{\cal E}_{0,>}}}{\sqrt{2}\Delta E}
{\cal C}_1\frac{L_y}{\ell_B}\left(\frac{\ell_B}{\delta}\right)^4\quad 
\mbox{for}\ n\le N.
\end{equation}
\end{lemma}

Combining these Lemmas with (\ref{expMin}),  we obtain 
\begin{equation}
\left|{\bf E}_\omega\left[{\cal M}_{s,{\rm in}}\right]\right|
\le  \frac{\sqrt{2\hbar\omega_c{\cal E}_{0,>}}}{\Delta E}
{\cal C}_1\frac{L_y}{\ell_B}\left(\frac{\ell_B}{\delta}\right)^4.
\label{estEMinNon}
\end{equation}

\subsection{Interacting case}
\label{IntMin}

Next we estimate ${\bf E}_\omega\left[{\cal M}_{s,{\rm in}}\right]$ 
of (\ref{EMin}) in the interacting case $U^{(2)}\ne 0$. 
As a result we obtain the following proposition:

\begin{pro}
\label{pro:estEMin}
Suppose that $V_\omega\in C^2({\bf R}^2)$ and 
$V_\omega$ satisfies the bound (\ref{2dbound}) in Theorem~\ref{theorem2}. Then 
\begin{equation}
\left|{\bf E}_\omega\left[{\cal M}_{s,{\rm in}}\right]\right|
\le {\cal C}_{\rm in}
\left(\frac{L_x}{\ell_B}\right)^{5/6}
\left(\frac{L_y}{\ell_B}\right)^{11/6}
\left(\frac{\ell_B}{\delta}\right)^3\quad\mbox{for}\ N\ge N_{\rm min}, 
\label{estEMin}
\end{equation}
where $N_{\rm min}$ is a positive number which is independent of 
the linear dimensions $L_x,L_y$ of the system, and 
${\cal C}_{\rm in}$ is a positive constant which is independent of 
the linear dimensions $L_x,L_y$ of the system. 
\end{pro}
The number $N_{\rm min}$ is given explicitly in (\ref{Nmin}) 
in Appendix~\ref{decayestInt}. 
In the rest of this appendix, we assume $V_\omega\in C^2({\bf R}^2)$. 

Let $A$ be a symmetric operator. Then one formally has 
\begin{eqnarray}
\left\langle\Phi_{\omega,0}^{(N)},A[1-G_\omega^{(N)}]A\Phi_{\omega,0}^{(N)}
\right\rangle
&=&\left\langle A\Phi_{\omega,0}^{(N)},[1-G_\omega^{(N)}]A\Phi_{\omega,0}^{(N)}
\right\rangle\ret
&\le&\left\langle A\Phi_{\omega,0}^{(N)},\frac{H_{\omega,0}^{(N)}-
E_{\omega,0}^{(N)}}{\Delta E}A\Phi_{\omega,0}^{(N)}
\right\rangle\ret
&=&\frac{1}{2\Delta E}\left\langle\Phi_{\omega,0}^{(N)},
\left[A,[H_{\omega,0}^{(N)},A]\right]\Phi_{\omega,0}^{(N)}\right\rangle 
\label{transitionestimate}
\end{eqnarray}
for the ground state $\Phi_{\omega,0}^{(N)}$ of the Hamiltonian 
$H_{\omega,0}^{(N)}$ of (\ref{unperturbedHam}). Using the techniques developed 
in \cite{HvL,KomaTasaki} 
with this bound, we obtain the following lemma: 

\begin{lemma}
\label{Psixbound}
The following bound is valid:
\begin{equation}
\left\langle\Phi_{\omega,0}^{(N)},\sum_{i=1}^N\pi_{s,i}
[1-G_\omega^{(N)}]\sum_{j=1}^N\pi_{s,j}\Phi_{\omega,0}^{(N)}\right\rangle
\le m_e\hbar\omega_c {\cal C}_2N
\label{normboundPsi}
\end{equation}
with the positive constant 
\begin{equation}
{\cal C}_2:=\frac{1}{2\Delta E}
\left[\hbar\omega_c+\ell_B^2
\left\Vert\frac{\partial^2}{\partial x^2}V_\omega\right\Vert\right]. 
\label{C2}
\end{equation}
\end{lemma}
\begin{proof}{Proof}
We treat only the case with $s=x$ because the other can be treated in the same 
way. Note that 
\begin{eqnarray}
& &\left[\sum_{i=1}^N\pi_{x,i},\left[H_{\omega,0}^{(N)},
\sum_{j=1}^N\pi_{x,j}\right]\right]\ret
&=&\sum_{i=1}^N\left\{\left[\pi_{x,i},\left[\frac{p_{y,i}^2}{2m_e},
\pi_{x,i}\right]\right]+\left[\pi_{x,i},\left[V_\omega({\bf r}_i),
\pi_{x,i}\right]\right]\right\}
+\sum_{i,j}\left[\pi_{x,i},\left[U^{(N)}({\bf r}_1,{\bf r}_2,\ldots,{\bf r}_N),
\pi_{x,j}\right]\right]\ret
&=&\frac{N(\hbar eB)^2}{m_e}+\sum_{i=1}^N
\hbar^2\frac{\partial^2}{\partial x_i^2}V_\omega({\bf r}_i),
\end{eqnarray}
where we have used the identity 
$\left[U^{(N)}({\bf r}_1,{\bf r}_2,\ldots,{\bf r}_N),\sum_j\pi_{x,j}\right]=0$ 
which is due to the assumption that the potential $U^{(N)}$ is 
a function of only the relative coordinates ${\bf r}_{ij}=(x_i-x_j,y_i-y_j)$. 
Combining this with (\ref{transitionestimate}), we have the desired bound 
\begin{eqnarray}
& &\left\langle\Phi_{\omega,0}^{(N)},\sum_{i=1}^N\pi_{x,i}
[1-G_\omega^{(N)}]
\sum_{j=1}^N\pi_{x,j}\Phi_{\omega,0}^{(N)}\right\rangle\ret
&\le&\frac{N}{2\Delta E}\left[
\frac{(\hbar eB)^2}{m_e}+\hbar^2\left\langle\Phi_{\omega,0}^{(N)},
\frac{\partial^2}{\partial x_i^2}V_\omega({\bf r}_i)\Phi_{\omega,0}^{(N)}
\right\rangle\right]\le\frac{N}{2\Delta E}\left[
\frac{(\hbar eB)^2}{m_e}+\hbar^2
\left\Vert\frac{\partial^2}{\partial x^2}V_\omega\right\Vert\right].\ret
\end{eqnarray}
\end{proof}

We write 
\begin{equation}
\Psi_\omega^{(N,s)}=\frac{[1-G_\omega^{(N)}]}{E_{\omega,0}^{(N)}-
H_{\omega,0}^{(N)}}\sum_{j=1}^N\pi_{s,j}\Phi_{\omega,0}^{(N)}.
\label{PsiomegaNs}
\end{equation}

\begin{lemma}
The following bound is valid: 
\begin{equation}
\left\langle\Psi_\omega^{(N,s)},\pi_{x,j}^2\Psi_\omega^{(N,s)}
\right\rangle\le\frac{2m_e^2\hbar\omega_c}{\Delta E}{\cal C}_2
\left(1+\frac{N{\tilde {\cal E}}}{\Delta E}\right),
\label{Psipi2Psi}
\end{equation}
where ${\tilde {\cal E}}$ is a positive constant which is independent of 
the linear dimensions $L_x,L_y$ of the system, and the constant ${\cal C}_2$ 
is given by (\ref{C2}). 
\end{lemma}

\begin{proof}{Proof}
Using an identity 
\begin{equation}
\frac{H_{\omega,0}^{(N)}}{E_{\omega,0}^{(N)}-H_{\omega,0}^{(N)}}
[1-G_\omega^{(N)}]=\left(-1+\frac{E_{\omega,0}^{(N)}}
{E_{\omega,0}^{(N)}-H_{\omega,0}^{(N)}}\right)[1-G_\omega^{(N)}],
\end{equation}
we have 
\begin{eqnarray}
\frac{1}{2m_e}
\left\langle\Psi_\omega^{(N,s)},\pi_{x,j}^2\Psi_\omega^{(N,s)}\right\rangle
&\le&\frac{1}{N}\left\langle\Psi_\omega^{(N,s)},H_{\omega,0}^{(N)}
\Psi_\omega^{(N,s)}\right\rangle\ret
&\le&\frac{1}{N}\left\langle\Phi_{\omega,0}^{(N)},\sum_{i=1}^N
\pi_{s,i}\frac{[1-G_\omega^{(N)}]}{H_{\omega,0}^{(N)}-E_{\omega,0}^{(N)}}
\sum_{j=1}^N\pi_{s,j}\Phi_{\omega,0}^{(N)}\right\rangle\ret
&+&\frac{E_{\omega,0}^{(N)}}{N}
\left\langle\Phi_{\omega,0}^{(N)},\sum_{i=1}^N\pi_{s,i}
\left[\frac{1-G_\omega^{(N)}}{E_{\omega,0}^{(N)}-H_{\omega,0}^{(N)}}\right]^2
\sum_{j=1}^N\pi_{s,j}\Phi_{\omega,0}^{(N)}\right\rangle\ret
&\le&\frac{1}{N\Delta E}\left(1+\frac{E_{\omega,0}^{(N)}}{\Delta E}\right)
\left\langle\Phi_{\omega,0}^{(N)},\sum_{i=1}^N\pi_{s,i}
\left[1-G_\omega^{(N)}\right]\sum_{j=1}^N\pi_{s,j}
\Phi_{\omega,0}^{(N)}\right\rangle.\ret 
\end{eqnarray}
Combining this, the bound (\ref{normboundPsi}) of Lemma~\ref{Psixbound} 
and Lemma~\ref{EUbound} in Appendix~\ref{EstEU}, 
we obtain the desired bound (\ref{Psipi2Psi}). 
\end{proof}

\begin{proof}{Proof of Proposition~\ref{pro:estEMin}}
In terms of the vector $\Psi_\omega^{(N,s)}$ of 
(\ref{PsiomegaNs}), ${\bf E}_\omega\left[{\cal M}_{s,{\rm in}}\right]$ 
can be written as 
\begin{eqnarray}
{\bf E}_\omega\left[{\cal M}_{s,{\rm in}}\right]&=&
\frac{1}{m_eN}\sum_{i=1}^N{\bf E}_\omega\left[
\left\langle\Psi_\omega^{(N,s)},{\tilde Q}_{{\rm in},i}^{(+)}
(p_{x,i}+\hbar K)\Phi_{\omega,0}^{(N)}\right\rangle\right]\ret
&-&\frac{1}{m_eN}\sum_{i=1}^N{\bf E}_\omega\left[
\left\langle\Psi_\omega^{(N,s)},Q_{{\rm in},i}^{(+)}
p_{x,i}\Phi_{\omega,0}^{(N)}\right\rangle\right]. 
\label{expEMin}
\end{eqnarray}
Using the Schwarz inequality, we have 
\begin{equation}
\left|\left\langle\Psi_\omega^{(N,s)},Q_{{\rm in},i}^{(+)}p_{x,i}
\Phi_{\omega,0}^{(N)}\right\rangle\right|
\le\frac{eBL_y}{2}\sqrt{\left\langle\Psi_\omega^{(N,s)},
Q_{{\rm in},i}^{(+)}\Psi_\omega^{(N,s)}\right\rangle
\left\langle\Phi_{\omega,0}^{(N)},Q_{{\rm in},i}^{(+)}
\Phi_{\omega,0}^{(N)}\right\rangle}. 
\label{PsiQPhi}
\end{equation}
In the same way as in the proof of Lemma~\ref{phiQphibound}, we obtain 
\begin{equation}
\left\langle\Psi_\omega^{(N,s)},
Q_{{\rm in},i}^{(+)}\Psi_\omega^{(N,s)}\right\rangle
\le \frac{2m_e}{\Delta E}{\cal C}_2
\left(1+\frac{N{\tilde {\cal E}}}{\Delta E}\right)
\left(\frac{\ell_B}{\delta}\right)^2,
\end{equation}
and 
\begin{equation}
\left\langle\Phi_{\omega,0}^{(N)},Q_{{\rm in},i}^{(+)}
\Phi_{\omega,0}^{(N)}\right\rangle
\le\left(C_3N^{2/3}+{\cal C}_4\right)
\left(\frac{\ell_B}{\delta}\right)^4, 
\end{equation}
where we have used the bound (\ref{Psipi2Psi}) and 
Proposition~\ref{proPhipi4Phibound}. 
Substituting these bounds into (\ref{PsiQPhi}), we get 
\begin{equation}
\left|\left\langle\Psi_\omega^{(N,s)},Q_{{\rm in},i}^{(+)}p_{x,i}
\Phi_{\omega,0}^{(N)}\right\rangle\right|
\le m_e\sqrt{\frac{\hbar\omega_c}{2\Delta E}
\left(1+\frac{N{\tilde {\cal E}}}{{\Delta E}}\right){\cal C}_2(C_3+{\cal C}_4
N^{-2/3})}\times N^{1/3}\left(\frac{L_y}{\ell_B}\right)
\left(\frac{\ell_B}{\delta}\right)^3. 
\end{equation}
Similarly we have 
\begin{eqnarray}
& &\left|\left\langle\Psi_\omega^{(N,s)},{\tilde Q}_{{\rm in},i}^{(+)}
(p_{x,i}+\hbar K)\Phi_{\omega,0}^{(N)}\right\rangle\right|\ret
&\le&m_e\sqrt{\frac{\hbar\omega_c}{2\Delta E}
\left(1+\frac{N{\tilde {\cal E}}}{{\Delta E}}\right){\cal C}_2(C_3+{\cal C}_4
N^{-2/3})}\times N^{1/3}\left(\frac{L_y}{\ell_B}\right)
\left(\frac{\ell_B}{\delta}\right)^3. 
\end{eqnarray}
Using these bounds for (\ref{expEMin}), we obtain 
\begin{equation}
\left|{\bf E}_\omega\left[{\cal M}_{s,{\rm in}}\right]\right|\le
C_{\rm in}'N^{5/6}\frac{L_y}{\ell_B}\left(\frac{\ell_B}{\delta}\right)^3 
\end{equation}
with the constant 
\begin{equation}
C_{\rm in}'=\sqrt{\frac{2\hbar\omega_c}{\Delta E}
\left(\frac{1}{N}+\frac{{\tilde {\cal E}}}{\Delta E}\right)
{\cal C}_2(C_3+{\cal C}_4N^{-2/3})}.
\end{equation}
Consequently we obtain the desired bound (\ref{estEMin}) from $N=\nu M$ with 
$M=L_xL_y eB/h$. 
\end{proof}

\Section{Estimate of ${\bf E}_\omega\left[{\cal M}_{s,{\rm out}}\right]$}

In this appendix we estimate ${\cal M}_{s,{\rm out}}$ of (\ref{Mout}). 
It can be divided into two parts as 
\begin{equation}
{\cal M}_{s,{\rm out}}={\cal M}_{s,{\rm out}}^{(1)}+
{\cal M}_{s,{\rm out}}^{(2)}
\end{equation}
with 
\begin{equation}
{\cal M}_{s,{\rm out}}^{(1)}=
\frac{1}{m_eN}\sum_{i=1}^N\sum_{j=1}^N\left\langle
\Phi_{\omega,0}^{(N)},\pi_{s,j}\frac{[1-G_\omega^{(N)}]}
{E_{\omega,0}^{(N)}-H_{\omega,0}^{(N)}}P_{{\rm out},i}\pi_{x,i}
\Phi_{\omega,0}^{(N)}\right\rangle
\label{Msout1}
\end{equation}
and
\begin{equation}
{\cal M}_{s,{\rm out}}^{(2)}=
\frac{eB}{m_eN}\sum_{i=1}^N\sum_{j=1}^N\left\langle
\Phi_{\omega,0}^{(N)},\pi_{s,j}\frac{[1-G_\omega^{(N)}]}
{E_{\omega,0}^{(N)}-H_{\omega,0}^{(N)}}P_{{\rm out},i}y_i
\Phi_{\omega,0}^{(N)}\right\rangle.
\label{Msout2}
\end{equation}

\subsection{Non-interacting case}
\label{App:EstMoutNon}

Consider first the non-interacting case, $U^{(2)}=0$. 
Then ${\cal M}_{s,{\rm out}}^{(1)}$ and ${\cal M}_{s,{\rm out}}^{(2)}$ can be 
written as 
\begin{eqnarray}
{\cal M}_{s,{\rm out}}^{(1)}&=&\frac{1}{m_eN}\sum_{n\le N}
\left(\varphi_{\omega,n},\pi_s
\frac{{\cal P}_>}{{\cal E}_{\omega,n}-{\cal H}_\omega}P_{\rm out}
\pi_x\varphi_{\omega,n}\right)=\frac{1}{m_eN}\sum_{n\le N}
\left(\psi_{\omega,n}^{(s)},P_{\rm out}\pi_x\varphi_{\omega,n}\right),\ret
\label{Moutrep}
\end{eqnarray}
and
\begin{equation}
{\cal M}_{s,{\rm out}}^{(2)}
=\frac{eB}{m_eN}\sum_{n\le N}
\left(\psi_{\omega,n}^{(s)},P_{\rm out}y\varphi_{\omega,n}\right)
\label{Moutrep2}
\end{equation}
in terms of the eigenvectors $\varphi_{\omega,n}$ of the single-electron 
Hamiltonian ${\cal H}_\omega$. Here the vector $\psi_{\omega,n}^{(s)}$ is 
given by (\ref{defpsi}). 

Since we have 
\begin{eqnarray}
& & \sum_{k\in K(I_{\rm out})}\int_{-L_y/2}^{L_y/2}dy \ 
(y-y_k)^4\left|{\hat \varphi}_{\omega,n}(k,y)\right|^2\ret
&\ge&\delta^4\sum_{k\in K(I_{\rm out})}\int_{-L_y/2}^{L_y/2}dy \ 
\left|{\hat \varphi}_{\omega,n}(k,y)\right|^2
=\delta^4\left(\varphi_{\omega,n},P_{\rm out}\varphi_{\omega,n}\right),
\end{eqnarray}
we obtain 
\begin{equation}
\left(\varphi_{\omega,n},P_{\rm out}\varphi_{\omega,n}\right)
\le {\cal C}_1\left(\frac{\ell_B}{\delta}\right)^4\quad\mbox{for}\ n\le N
\label{phiPoutphi}
\end{equation}
in the same way as in the proof of Lemma~\ref{phiQphibound}. Further we get 
\begin{equation}
\left(\psi_{\omega,n}^{(s)},P_{\rm out}\psi_{\omega,n}^{(s)}\right)
\le \frac{2m_e{\cal E}_{\omega,n}}{(\Delta E)^2}{\cal C}_1
\left(\frac{\ell_B}{\delta}\right)^4\quad\mbox{for}\ n\le N. 
\label{psiPoutpsi}
\end{equation}
Using the Schwarz inequality and the bound (\ref{psiPoutpsi}), we have 
\begin{eqnarray}
\frac{1}{m_e}
|(\psi_{\omega,n}^{(s)},P_{\rm out}\pi_x\varphi_{\omega,n})|
\le\frac{1}{m_e}\sqrt{(\psi_{\omega,n}^{(s)},P_{\rm out}
\psi_{\omega,n}^{(s)})}\left\Vert\pi_x\varphi\right\Vert
\le\frac{2{\cal E}_{\omega,n}}{\Delta E}
\sqrt{{\cal C}_1}\left(\frac{\ell_B}{\delta}\right)^2.
\end{eqnarray}
Therefore we obtain 
\begin{equation}
\left|{\cal M}_{s,{\rm out}}^{(1)}\right|
\le \frac{2{\cal E}_{0,>}}{\Delta E}
\sqrt{{\cal C}_1}\left(\frac{\ell_B}{\delta}\right)^2
\end{equation}
for ${\cal M}_{s,{\rm out}}^{(1)}$ of (\ref{Moutrep}).

On the other hand we have 
\begin{eqnarray}
\frac{eB}{m_e}
\left|\left(\psi_{\omega,n}^{(s)},P_{\rm out}y\varphi_{\omega,n}\right)\right|
&\le&\frac{eBL_y}{2m_e}\sqrt{(\psi_{\omega,n}^{(s)},P_{\rm out}
\psi_{\omega,n}^{(s)})(\varphi_{\omega,n}P_{\rm out}\varphi_{\omega,n})}\ret
&\le&\frac{\sqrt{\hbar\omega_c{\cal E}_{0,>}}}{\sqrt{2}\Delta E}
{\cal C}_1\frac{L_y}{\ell_B}\left(\frac{\ell_B}{\delta}\right)^4
\label{psiPphibound}
\end{eqnarray}
by using the Schwarz inequality, (\ref{phiPoutphi}) and (\ref{psiPoutpsi}). 
Therefore we obtain 
\begin{equation}
\left|{\cal M}_{s,{\rm out}}^{(2)}\right|\le 
\frac{\sqrt{\hbar\omega_c{\cal E}_{0,>}}}{\sqrt{2}\Delta E}
{\cal C}_1\frac{L_y}{\ell_B}\left(\frac{\ell_B}{\delta}\right)^4
\end{equation}
for ${\cal M}_{s,{\rm out}}^{(2)}$ of (\ref{Moutrep2}).

Consequently we get 
\begin{equation}
\left|{\cal M}_{s,{\rm out}}\right|\le 
\left|{\cal M}_{s,{\rm out}}^{(1)}\right|
+\left|{\cal M}_{s,{\rm out}}^{(2)}\right|\le
\frac{2{\cal E}_{0,>}}{\Delta E}
\sqrt{{\cal C}_1}\left(\frac{\ell_B}{\delta}\right)^2
+\frac{\sqrt{\hbar\omega_c{\cal E}_{0,>}}}{\sqrt{2}\Delta E}
{\cal C}_1\frac{L_y}{\ell_B}\left(\frac{\ell_B}{\delta}\right)^4.
\label{estMoutNon}
\end{equation}

\subsection{Interacting case}
\label{IntMout}

In this case we have the following estimate 
for ${\bf E}_\omega\left[{\cal M}_{s,{\rm out}}\right]$: 

\begin{pro}
\label{Pro:Moutnon}
Suppose that $V_\omega\in C^2({\bf R}^2)$ and 
$V_\omega$ satisfies the bound (\ref{2dbound}) in Theorem~\ref{theorem2}. Then 
\begin{equation}
\left|{\bf E}_\omega\left[{\cal M}_{s,{\rm out}}\right]\right|
\le {\cal C}_{\rm out}\left(\frac{L_x}{\ell_B}\right)^{5/6}
\left(\frac{L_y}{\ell_B}\right)^{11/6}
\left(\frac{\ell_B}{\delta}\right)^3\quad \mbox{for}\ N\ge N_{\rm min},
\label{EstEMoutInt}
\end{equation}
where ${\cal C}_{\rm out}$ and $N_{\rm min}$ are positive constants 
which are independent of the linear dimensions $L_x,L_y$ of the system. 
The number $N_{\rm min}$ is given explicitly by (\ref{Nmin}) in 
Appendix~\ref{decayestInt}. 
\end{pro}

\begin{proof}{Proof}
In terms of the vector $\Psi_\omega^{(N,s)}$ of (\ref{PsiomegaNs}), 
we write ${\cal M}_{s,{\rm out}}^{(1)}$ of (\ref{Msout1}) as 
\begin{equation}
{\cal M}_{s,{\rm out}}^{(1)}=\frac{1}{m_eN}
\sum_{i=1}^N\left\langle\Psi_\omega^{(N,s)},P_{{\rm out},i}
\pi_{x,i}\Phi_{\omega,0}^{(N)}\right\rangle.
\label{Mout1int}
\end{equation}
Using the Schwarz inequality, we evaluate the matrix element in the right-hand 
side as 
\begin{equation}
\left|\left\langle\Psi_\omega^{(N,s)},P_{{\rm out},i}
\pi_{x,i}\Phi_{\omega,0}^{(N)}\right\rangle\right|
\le\sqrt{\left\langle\Psi_\omega^{(N,s)},P_{{\rm out},i}\Psi_\omega^{(N,s)}
\right\rangle\left\langle\Phi_{\omega,0}^{(N)}
P_{{\rm out},i}\pi_{x,i}^2\Phi_{\omega,0}^{(N)}\right\rangle}. 
\end{equation}
In the same way as in Section~\ref{IntMin}, we have 
\begin{equation}
\left\langle\Psi_\omega^{(N,s)},P_{{\rm out},i}\Psi_\omega^{(N,s)}
\right\rangle\le\frac{2m_e}{\Delta E}{\cal C}_2
\left(1+\frac{N{\tilde {\cal E}}}{\Delta E}\right)
\left(\frac{\ell_B}{\delta}\right)^2,
\label{PsiNsPoutPsiNs}
\end{equation}
and
\begin{equation}
\left\langle\Phi_{\omega,0}^{(N)}
P_{{\rm out},i}\pi_{x,i}^2\Phi_{\omega,0}^{(N)}\right\rangle
\le\hbar eB\left({\cal C}_3N^{2/3}+{\cal C}_4
\right)\left(\frac{\ell_B}{\delta}\right)^2. 
\end{equation}
{From} these three bounds, we estimate ${\cal M}_{s,{\rm out}}^{(1)}$ of 
(\ref{Mout1int}) as 
\begin{equation}
\left|{\cal M}_{s,{\rm out}}^{(1)}\right|\le
\sqrt{\frac{2\hbar\omega_c}{\Delta E}
\left(\frac{1}{N}+\frac{{\tilde {\cal E}}}{\Delta E}\right)
{\cal C}_2\left({\cal C}_3+{\cal C}_4 N^{-2/3}\right)}\times N^{5/6}
\left(\frac{\ell_B}{\delta}\right)^2. 
\label{Msout1bound}
\end{equation}

Similarly we can write ${\cal M}_{s,{\rm out}}^{(2)}$ of (\ref{Msout2}) as 
\begin{equation}
{\cal M}_{s,{\rm out}}^{(2)}=\frac{eB}{m_eN}
\sum_{i=1}^N\left\langle\Psi_\omega^{(N,s)},P_{{\rm out},i}y_i
\Phi_{\omega,0}^{(N)}\right\rangle
\label{Mout2int}
\end{equation}
in terms of the vector $\Psi_\omega^{(N,s)}$ of (\ref{PsiomegaNs}). 
Using the Schwarz inequality, we evaluate the matrix element in the right-hand 
side as 
\begin{eqnarray}
\left|\left\langle\Psi_\omega^{(N,s)},P_{{\rm out},i}y_i
\Phi_{\omega,0}^{(N)}\right\rangle\right|&\le&
\frac{L_y}{2}\sqrt{\left\langle\Psi_\omega^{(N,s)},P_{{\rm out},i}
\Psi_\omega^{(N,s)}\right\rangle\left\langle\Phi_{\omega,0}^{(N)},
P_{{\rm out},i}\Phi_{\omega,0}^{(N)}\right\rangle}\ret
&\le&\frac{m_e}{eB}
\sqrt{\frac{\hbar\omega_c}{2\Delta E}\left(\frac{1}{N}+\frac{{\tilde {\cal E}}}
{\Delta E}\right){\cal C}_2({\cal C}_3+{\cal C}_4N^{-2/3})}\times N^{5/6}
\frac{L_y}{\ell_B}\left(\frac{\ell_B}{\delta}\right)^3,\ret 
\label{PsiPoutPhi}
\end{eqnarray}
where we have used the bound (\ref{PsiNsPoutPsiNs}) and the bound 
\begin{equation}
\left\langle\Phi_{\omega,0}^{(N)},P_{{\rm out},i}\Phi_{\omega,0}^{(N)}
\right\rangle\le
\left({\cal C}_3N^{2/3}+{\cal C}_4\right)\left(\frac{\ell_B}{\delta}\right)^4.
\end{equation}
The second bound can be derived in the same way as in Section~\ref{IntMin}. 
Substituting (\ref{PsiPoutPhi}) into (\ref{Mout2int}), we get 
\begin{equation}
\left|{\cal M}_{s,{\rm out}}^{(2)}\right|
\le\sqrt{\frac{\hbar\omega_c}{2\Delta E}
\left(\frac{1}{N}+\frac{{\tilde {\cal E}}}{\Delta E}\right)
{\cal C}_2({\cal C}_3+{\cal C}_4N^{-2/3})}\times N^{5/6}\frac{L_y}{\ell_B}
\left(\frac{\ell_B}{\delta}\right)^3. 
\end{equation}
Combining this with (\ref{Msout1bound}), we obtain 
\begin{eqnarray}
\left|{\cal M}_{s,{\rm out}}\right|
&\le&\left|{\cal M}_{s,{\rm out}}^{(1)}\right|+
\left|{\cal M}_{s,{\rm out}}^{(2)}\right|\ret
&\le&\sqrt{\frac{2\hbar\omega_c}{\Delta E}\left(\frac{1}{N}
+\frac{{\tilde {\cal E}}}{\Delta E}\right)
{\cal C}_2({\cal C}_3+{\cal C}_4N^{-2/3})}\times
\left(1+\frac{L_y}{2\delta}\right)N^{5/6}
\left(\frac{\ell_B}{\delta}\right)^2.\ret
\end{eqnarray}
Consequently we have the desired bound (\ref{EstEMoutInt}) 
with $N=\nu M$ and $M=L_xL_yeB/h$. 
\end{proof}

\Section{Estimate of ${\bf E}_\omega\left[{\cal M}_{s,{\rm edge}}\right]$}
\label{EstMedge}

In this appendix, we estimate the random average 
${\bf E}_\omega\left[{\cal M}_{s,{\rm edge}}\right]$ of (\ref{Medge1}) 
which is the contribution near the edges of the system. It can be written as 
\begin{equation}
{\cal M}_{s,{\rm edge}}=
\frac{1}{m_eN}\sum_{i=1}^N\sum_{j=1}^N\sum_{\ell\ne 0}
\left\langle\Phi_{\omega,0}^{(N)},\pi_{s,j}\Phi_{\omega,\ell}^{(N)}
\right\rangle\frac{1}{E_{\omega,0}^{(N)}-E_{\omega,\ell}^{(N)}}
\left\langle\Phi_{\omega,\ell}^{(N)},P_i(I_{\rm edge})
\pi_{x,i}\Phi_{\omega,0}^{(N)}\right\rangle,
\label{Msedge1}
\end{equation}
where
\begin{equation}
I_{\rm edge}=\bigcup_{\ell\in{\bf Z}}
(L_y/2-\delta+\ell L_y,L_y/2+\delta+\ell L_y). 
\end{equation}
As a result, we will obtain 
\begin{equation}
\left|{\bf E}_\omega\left[{\cal M}_{s,{\rm edge}}\right]\right|
\le{\cal C}_{\rm edge}\frac{\delta}{L_y}, 
\label{EMsedge}
\end{equation}
where ${\cal C}_{\rm edge}$ is a positive constant which is independent 
of the linear dimensions $L_x,L_y$ of the system 
in both non-interacting and interacting cases. 

{From} Lemma~\ref{translateVomegaPhi}, we have 
\begin{eqnarray}
& &{\bf E}_\omega
\left[\left\langle\Phi_{\omega,0}^{(N)},\pi_{s,j}\Phi_{\omega,\ell}^{(N)}
\right\rangle
\frac{1}{E_{\omega,0}^{(N)}-E_{\omega,\ell}^{(N)}}
\left\langle\Phi_{\omega,\ell}^{(N)},{\tilde P}_i(k)\pi_{x,i}
\Phi_{\omega,0}^{(N)}\right\rangle\right]\ret
&=&\frac{1}{M}{\bf E}_\omega
\left[\left\langle\Phi_{\omega,0}^{(N)},\pi_{s,j}\Phi_{\omega,\ell}^{(N)}
\right\rangle
\frac{1}{E_{\omega,0}^{(N)}-E_{\omega,\ell}^{(N)}}
\left\langle\Phi_{\omega,\ell}^{(N)},\pi_{x,i}
\Phi_{\omega,0}^{(N)}\right\rangle\right]
\end{eqnarray}
for $\ell\ne 0$, where $M=eBL_xL_y/h$. Combining this with 
(\ref{Msedge1}), we obtain 
\begin{equation}
\left|{\bf E}_\omega\left[{\cal M}_{s,{\rm edge}}\right]\right|
\le\frac{2\delta}{m_eL_yN}\left|\sum_{i=1}^N\sum_{j=1}^N
{\bf E}_\omega\left[\left\langle\Phi_{\omega,0}^{(N)},\pi_{s,j}
\frac{[1-G_\omega^{(N)}]}{E_{\omega,0}^{(N)}-H_{\omega,0}^{(N)}}
\pi_{x,i}\Phi_{\omega,0}^{(N)}\right\rangle\right]\right|. 
\label{Msedge1bound}
\end{equation}

\subsection{Non-interacting case}

Consider first the non-interacting case, $U^{(2)}=0$. 
Then (\ref{Msedge1bound}) can be evaluated as 
\begin{eqnarray}
\left|{\bf E}_\omega\left[{\cal M}_{s,{\rm edge}}\right]\right|
&\le&\frac{2\delta}{m_eL_yN}\left|\sum_{n\le N}
{\bf E}_\omega\left[\left(\varphi_{\omega,n},\pi_s
\frac{{\cal P}_>}{{\cal E}_{\omega,n}-{\cal H}_\omega}
\pi_x\varphi_{\omega,n}\right)\right]\right|\ret
&\le&\frac{2\delta}{m_eL_yN}\sum_{n\le N}{\bf E}_\omega\left[
\frac{1}{\Delta E}
\sqrt{\left(\varphi_{\omega,n},\pi_s^2\varphi_{\omega,n}\right)
\left(\varphi_{\omega,n},\pi_x^2\varphi_{\omega,n}\right)}\right]
\le\frac{4{\cal E}_{0,>}}{\Delta E}\frac{\delta}{L_y},\ret
\label{Medge1bound}
\end{eqnarray}
where we have used the Schwarz inequality. 

\subsection{Interacting case}

Using the Schwarz inequality and Lemma~\ref{Psixbound}, we have 
\begin{eqnarray}
& &\left|\sum_{i=1}^N\sum_{j=1}^N
\left\langle\Phi_{\omega,0}^{(N)},\pi_{s,j}\frac{[1-G_\omega^{(N)}]}
{E_{\omega,0}^{(N)}-H_{\omega,0}^{(N)}}\pi_{x,i}\Phi_{\omega,0}^{(N)}
\right\rangle\right|\ret
&\le&\frac{1}{\Delta E}\sqrt{\left\langle\Phi_{\omega,0}^{(N)},\sum_{i=1}^N
\pi_{s,i}[1-G_\omega^{(N)}]\sum_{j=1}^N\pi_{s,j}
\Phi_{\omega,0}^{(N)}\right\rangle
\left\langle\Phi_{\omega,0}^{(N)},\sum_{m=1}^N
\pi_{x,m}[1-G_\omega^{(N)}]\sum_{n=1}^N\pi_{x,n}
\Phi_{\omega,0}^{(N)}\right\rangle}\ret
&\le&\frac{Nm_e\hbar\omega_c}{\Delta E}{\cal C}_2. 
\end{eqnarray}
Substituting this into the right-hand side of (\ref{Msedge1bound}), we obtain 
\begin{equation}
\left|{\bf E}_\omega\left[{\cal M}_{s,{\rm edge}}\right]\right|
\le\frac{2\hbar\omega_c}{\Delta E}\frac{\delta}{L_y}{\cal C}_2. 
\label{EMsedge1fbound}
\end{equation}

\Section{Decay estimate of wavefunctions}

In this appendix, we obtain a decay estimate for 
the Fourier component of a wavefunction for both 
non-interacting and interacting electrons gases. 

\subsection{Non-interacting case}
\label{DecaywaveNon}

The aim of this subsection is to give a proof of the following 
proposition in the non-interacting case $U^{(2)}=0$: 

\begin{pro}
\label{decayproNon}
Let $\varphi$ be a wavefunction such that 
$\left\Vert{\cal H}_\omega\varphi\right\Vert<\infty$. 
Then 
\begin{equation}
\left|\left(\varphi,[\pi_x/(eB)]^4\varphi\right)\right|
\le \ell_B^4\left(\frac{2}{\hbar\omega_c}\right)^2
\left\{\left(\Vert{\cal H}_\omega\varphi\Vert
+\Vert V_\omega\Vert\Vert\varphi\Vert\right)^2
+4\hbar\omega_c\left[\left(\varphi,{\cal H}_\omega\varphi\right)
+\Vert V_\omega\Vert\Vert\varphi\Vert^2\right]\right\}.
\label{4momentBound}
\end{equation}
\end{pro}
In order to see the physical meaning, we write 
\begin{equation}
\varphi(x,y)=L_x^{-1/2}\sum_k e^{ikx}{\hat \varphi}(k,y), 
\label{Fourierformvarphi}
\end{equation}
in terms of the Fourier transform. Clearly one has 
\begin{eqnarray}
& &\sum_k\int_{-L_y/2}^{L_y/2}dy(y-y_k)^4|{\hat \varphi}(k,y)|^2\ret
&\le& \ell_B^4\left(\frac{2}{\hbar\omega_c}\right)^2
\left\{\left(\Vert{\cal H}_\omega\varphi\Vert
+\Vert V_\omega\Vert\Vert\varphi\Vert\right)^2
+4\hbar\omega_c\left[\left(\varphi,{\cal H}_\omega\varphi\right)
+\Vert V_\omega\Vert\Vert\varphi\Vert^2\right]\right\} 
\label{strongdecay}
\end{eqnarray}
with $y_k=\hbar k/(eB)$. 
This implies that ${\hat \varphi}(k,y)$ decays more rapidly than 
$|y-y_k|^{-4}$ when $\Vert{\cal H}_\omega\varphi\Vert<\infty$. 
\medskip

Before giving the proof of Proposition~\ref{decayproNon}, 
we shall see a fairly trivial decay estimate for a wavefunction. 
Let $\varphi$ be a wavefunction. Then we formally have 
\begin{equation}
\frac{1}{2m_e}\left(\varphi,(p_x-eBy)^2\varphi\right)
+\frac{1}{2m_e}\left(\varphi,p_y^2\varphi\right)
+\left(\varphi,V_\omega\varphi\right)=
\left(\varphi,{\cal H}_\omega\varphi\right). 
\end{equation}
Clearly we get 
\begin{equation}
\frac{1}{2m_e}\left(\varphi,(p_x-eBy)^2\varphi\right)
\le \left(\varphi,{\cal H}_\omega\varphi\right)
+\Vert V_\omega\Vert\ \Vert\varphi\Vert^2,
\label{2pbound}
\end{equation}
and
\begin{equation}
\frac{1}{2m_e}\left(\varphi,p_y^2\varphi\right)
\le \left(\varphi,{\cal H}_\omega\varphi\right)
+\Vert V_\omega\Vert\ \Vert\varphi\Vert^2 .
\label{pybound}
\end{equation}
Combining the first inequality with the Fourier form 
(\ref{Fourierformvarphi}), we get 
\begin{equation}
\sum_k\int_{-L_y/2}^{L_y/2}dy \ (y-y_k)^2
\left|{\hat \varphi}(k,y)\right|^2\le 
\frac{2m_e}{e^2B^2}
\left[\left(\varphi,{\cal H}_\omega\varphi\right)
+\Vert V_\omega\Vert\ \Vert\varphi\Vert^2\right], 
\label{Fourierbound}
\end{equation}
where $y_k=\hbar k/(eB)$. This implies that the Fourier 
component ${\hat \varphi}(k,y)$ decays more rapidly than 
the inverse square of the distance $|y-y_k|$ when the wavefunction 
satisfies the condition 
$|\left(\varphi,{\cal H}_\omega\varphi\right)|<\infty$. 

In order to obtain the stronger decay bound (\ref{strongdecay}), 
we consider a formal identity, 
\begin{eqnarray}
\left(\varphi,{\cal H}_\omega[y-p_x/(eB)]^2\varphi\right)
&=&\frac{e^2B^2}{2m_e}
\left(\varphi,[y-p_x/(eB)]^4\varphi\right)
+\frac{1}{2m_e}\left(\varphi,p_y^2[y-p_x/(eB)]^2\varphi\right)\ret
& &+\left(\varphi,V_\omega[y-p_x/(eB)]^2\varphi\right).
\end{eqnarray}
Since the second term in the right-hand side can be written as 
\begin{eqnarray}
& &\frac{1}{2m_e}\left(\varphi,p_y^2[y-p_x/(eB)]^2\varphi\right)\ret
&=&\frac{1}{2m_e}\left(\varphi,p_y[y-p_x/(eB)]^2p_y\varphi\right)
-\frac{i\hbar}{m_e}\left(\varphi,p_y[y-p_x/(eB)]\varphi\right), 
\end{eqnarray}
we have
\begin{eqnarray}
& &\frac{e^2B^2}{2m_e}\left(\varphi,[y-p_x/(eB)]^4\varphi\right)
+\frac{1}{2m_e}\left(\varphi,p_y[y-p_x/(eB)]^2p_y\varphi\right)\ret
&=&\left(\varphi,{\cal H}_\omega[y-p_x/(eB)]^2\varphi\right)
+\frac{i\hbar}{m_e}\left(\varphi,p_y[y-p_x/(eB)]\varphi\right)\ret
& &-\left(\varphi,V_\omega[y-p_x/(eB)]^2\varphi\right). 
\label{4peq}
\end{eqnarray}
In order to get a bound for the first term in the left-hand side, 
we estimate the right-hand side as follows. 
The first term in the right-hand side of (\ref{4peq}) can be evaluated as
\begin{equation}
\left|\left(\varphi,{\cal H}_\omega[y-p_x/(eB)]^2\varphi\right)\right|
\le \left\Vert{\cal H}_\omega\varphi\right\Vert 
\sqrt{\left(\varphi,[y-p_x/(eB)]^4\varphi\right)} 
\end{equation}
by using the Schwartz inequality. Similarly the second term 
in the right-hand side of (\ref{4peq}) can be evaluated as 
\begin{eqnarray}
\left|\frac{\hbar}{m_e}\left(\varphi,p_y[y-p_x/(eB)]\varphi\right)\right|
&\le&\frac{\hbar}{m_e}\Vert p_y\varphi\Vert\sqrt{
\left(\varphi,[y-p_x/(eB)]^2\varphi\right)}\ret 
&\le&\frac{2\hbar}{eB}\left[
\left(\varphi,{\cal H}_\omega\varphi\right)
+\Vert V_\omega\Vert\ \Vert\varphi\Vert^2\right], 
\end{eqnarray}
where we have used (\ref{2pbound}), and (\ref{pybound}). Finally we have 
\begin{eqnarray}
\left|\left(\varphi,V_\omega[y-p_x/(eB)]^2\varphi\right)\right|
&\le&\Vert V_\omega\Vert\ \Vert\varphi\Vert\sqrt{ 
\left(\varphi,[y-p_x/(eB)]^4\varphi\right)} 
\end{eqnarray}
for the third term in the right-hand side of (\ref{4peq}). {From} these 
three bounds, we formally obtain 
\begin{eqnarray}
\frac{e^2B^2}{2m_e}\left(\varphi,[y-p_x/(eB)]^4\varphi\right)
&\le&\left(\Vert{\cal H}_\omega\varphi\Vert
+\Vert V_\omega\Vert\ \Vert\varphi\Vert\right)
\sqrt{\left(\varphi,[y-p_x/(eB)]^4\varphi\right)}\ret
&+&\frac{2\hbar}{eB}\left[\left(\varphi,{\cal H}_\omega\varphi\right)
+\Vert V_\omega\Vert\ \Vert\varphi\Vert^2\right],
\label{sqr4moment}
\end{eqnarray}
where we have used the fact that the second term in the left-hand 
side of (\ref{4peq}) is non-negative. From this (\ref{sqr4moment}), 
one can easily obtain 
\begin{eqnarray}
\sqrt{\left(\varphi,[y-p_x/(eB)]^4\varphi\right)}
&\le&\frac{m_e}{e^2B^2}\left(\Vert{\cal H}_\omega\varphi\Vert
+\Vert V_\omega\Vert\Vert\varphi\Vert\right)\ret
&+&\frac{m_e}{e^2B^2}\sqrt{\left(\Vert{\cal H}_\omega\varphi\Vert
+\Vert V_\omega\Vert\Vert\varphi\Vert\right)^2
+4\hbar\omega_c\left[\left(\varphi,{\cal H}_\omega\varphi\right)
+\Vert V_\omega\Vert\Vert\varphi\Vert^2\right]}\ret
&\le&\frac{2m_e}{e^2B^2}\sqrt{\left(\Vert{\cal H}_\omega\varphi\Vert
+\Vert V_\omega\Vert\Vert\varphi\Vert\right)^2
+4\hbar\omega_c\left[\left(\varphi,{\cal H}_\omega\varphi\right)
+\Vert V_\omega\Vert\Vert\varphi\Vert^2\right]}. \ret
\end{eqnarray}
Thus we have obtained the desired bound (\ref{4momentBound}) 
which is justified for $\varphi$ satisfying 
$\Vert{\cal H}_\omega\varphi\Vert<\infty$. 

\subsection{Interacting case}
\label{decayestInt}

Next we consider the interacting case. 
Our goal of this subsection is to give a proof of 
Proposition~\ref{proPhipi4Phibound} below which is an extension of the decay 
bound (\ref{4momentBound}) to the interacting electrons gas. We write 
\begin{equation}
N_{\rm min}=\left[\frac{2\nu}{\hbar\omega_c}
\left({\tilde {\cal E}}+\Vert V_\omega\Vert\right)\right]^{3/2},
\label{Nmin}
\end{equation}
where ${\tilde {\cal E}}$ is an upper bound for the ground state energy 
per electron $E_{\omega,0}^{(N)}/N$ of the Hamiltonian $H_{\omega,0}^{(N)}$ of 
(\ref{unperturbedHam}). 
The constant ${\tilde {\cal E}}$ is independent of the linear 
dimensions $L_x,L_y$ of the system as we show in Lemma~\ref{EUbound} in 
Appendix~\ref{EstEU}. 

\begin{pro}
\label{proPhipi4Phibound}
Let $\Phi_{\omega,0}^{(N)}$ be the ground state eigenvector of 
the Hamiltonian $H_{\omega,0}^{(N)}$ with norm one. Then 
\begin{equation}
\left(\frac{1}{eB}\right)^4
\left\langle\pi_{x,j}^2\Phi_{\omega,0}^{(N)},\pi_{x,j}^2
\Phi_{\omega,0}^{(N)}\right\rangle
\le {\ell_B}^4\left({\cal C}_3N^{2/3}+{\cal C}_4\right)\quad 
\mbox{for} \ N\ge N_{\rm min}, 
\label{Phipi4Phibpund}
\end{equation}
where the constants ${\cal C}_3$ and ${\cal C}_4$ are independent of 
the linear dimensions $L_x,L_y$ of the system. 
\end{pro}

Consider an identity 
\begin{eqnarray}
E_{\omega,0}^{(N)}\left\langle\pi_{x,j}^2\Phi_{\omega,0}^{(N)},
\Phi_{\omega,0}^{(N)}\right\rangle
&=&\left\langle\pi_{x,j}^2\Phi_{\omega,0}^{(N)},
H_{\omega,0}^{(N)}\Phi_{\omega,0}^{(N)}\right\rangle\ret
&=&\left\langle\pi_{x,j}^2\Phi_{\omega,0}^{(N)},
H_{\omega,0}^{(N,j)}\Phi_{\omega,0}^{(N)}\right\rangle
+\left\langle\pi_{x,j}\Phi_{\omega,0}^{(N)},
\left(H_{\omega,0}^{(N)}-H_{\omega,0}^{(N,j)}\right)
\pi_{x,j}\Phi_{\omega,0}^{(N)}\right\rangle,\ret
\label{identityEpi2}
\end{eqnarray}
where 
\begin{equation}
H_{\omega,0}^{(N,j)}:=\frac{1}{2m_e}\left(\pi_{x,j}^2+p_{y,j}^2\right)
+V_\omega({\bf r}_j)+U^{(N,j)}
\end{equation}
with 
\begin{equation}
U^{(N,j)}({\bf r}_j;{\bf r}_1,{\bf r}_2,\ldots,{\bf r}_{j-1},{\bf r}_{j+1},
\ldots,{\bf r}_N)
=\sum_{\ell\ne j}U^{(2)}(x_j-x_\ell,y_j-y_\ell).
\end{equation}
Clearly the first term in the right-hand side of (\ref{identityEpi2}) 
is written as 
\begin{eqnarray}
\left\langle\pi_{x,j}^2\Phi_{\omega,0}^{(N)},
H_{\omega,0}^{(N,j)}\Phi_{\omega,0}^{(N)}\right\rangle
&=&\frac{1}{2m_e}\left\langle\pi_{x,j}^2\Phi_{\omega,0}^{(N)},\pi_{x,j}^2
\Phi_{\omega,0}^{(N)}\right\rangle+
\frac{1}{2m_e}\left\langle\Phi_{\omega,0}^{(N)},\pi_{x,j}^2
p_{y,j}^2\Phi_{\omega,0}^{(N)}\right\rangle\ret
&+&\left\langle\pi_{x,j}^2\Phi_{\omega,0}^{(N)},
V_\omega({\bf r}_j)\Phi_{\omega,0}^{(N)}\right\rangle
+\left\langle\Phi_{\omega,0}^{(N)},\pi_{x,j}^2
U^{(N,j)}\Phi_{\omega,0}^{(N)}\right\rangle. 
\label{pi2HNj}
\end{eqnarray}
Note that 
\begin{equation}
\pi_{x,j}^2p_{y,j}^2=\pi_{x,j}p_{y,j}^2\pi_{x,j}-\hbar^2e^2B^2
-i\hbar eB \left(\pi_{x,j}p_{y,j}+p_{y,j}\pi_{x,j}\right),
\end{equation}
and
\begin{equation}
\pi_{x,j}^2U^{(N,j)}=\pi_{x,j}U^{(N,j)}\pi_{x,j}
-\frac{i\hbar}{2}\left(\pi_{x,j}\frac{\partial}{\partial x_j}U^{(N,j)}
+\frac{\partial}{\partial x_j}U^{(N,j)}\pi_{x,j}\right)
-\frac{\hbar^2}{2}\frac{\partial^2}{\partial x_j^2}U^{(N,j)}.
\end{equation}
Here we have used the commutation relation $\left[p_{y,j},\pi_{x,j}\right]=
i\hbar eB$ for getting the first relation. 
Since the left-hand side of (\ref{pi2HNj}) is real from (\ref{identityEpi2}), 
we have 
\begin{eqnarray}
\left\langle\pi_{x,j}^2\Phi_{\omega,0}^{(N)},
H_{\omega,0}^{(N,j)}\Phi_{\omega,0}^{(N)}\right\rangle
&\ge&\frac{1}{2m_e}\left\langle\pi_{x,j}^2\Phi_{\omega,0}^{(N)},\pi_{x,j}^2
\Phi_{\omega,0}^{(N)}\right\rangle-\frac{\hbar^2 e^2B^2}{2m_e}\ret
&-&\frac{\hbar^2}{2}\left\langle\Phi_{\omega,0}^{(N)},
\frac{\partial^2}{\partial x_j^2}U^{(N,j)}
\Phi_{\omega,0}^{(N)}\right\rangle
-\left\Vert V_\omega\right\Vert\sqrt{\left\langle\pi_{x,j}^2
\Phi_{\omega,0}^{(N)},\pi_{x,j}^2\Phi_{\omega,0}^{(N)}\right\rangle}.\ret
\label{pi2HNjbound}
\end{eqnarray}
Here we have used the Schwarz inequality for evaluating the third term 
in the right-hand side of (\ref{pi2HNj}). 
Substituting this bound (\ref{pi2HNjbound}) into (\ref{identityEpi2}), 
we obtain 
\begin{eqnarray}
& &\left(E_{\omega,0}^{(N)}-E_{\omega,0}^{(N-1)}\right)
\left\langle\pi_{x,j}^2\Phi_{\omega,0}^{(N)},
\Phi_{\omega,0}^{(N)}\right\rangle\ret
&\ge& \frac{1}{2m_e}\left\langle\pi_{x,j}^2\Phi_{\omega,0}^{(N)},\pi_{x,j}^2
\Phi_{\omega,0}^{(N)}\right\rangle-\frac{\hbar^2 e^2B^2}{2m_e}
-\frac{\hbar^2}{2}\left\langle\Phi_{\omega,0}^{(N)},
\frac{\partial^2}{\partial x_j^2}U^{(N,j)}
\Phi_{\omega,0}^{(N)}\right\rangle\ret
&-&\left\Vert V_\omega\right\Vert\sqrt{\left\langle\pi_{x,j}^2
\Phi_{\omega,0}^{(N)},\pi_{x,j}^2\Phi_{\omega,0}^{(N)}\right\rangle}\ret
&\ge& \frac{1}{2m_e}\left\langle\pi_{x,j}^2\Phi_{\omega,0}^{(N)},\pi_{x,j}^2
\Phi_{\omega,0}^{(N)}\right\rangle-\frac{\hbar^2 e^2B^2}{2m_e}
-\frac{\hbar^2\alpha}{N}\left\langle\Phi_{\omega,0}^{(N)},
U^{(N)}\Phi_{\omega,0}^{(N)}\right\rangle\ret
&-&\left\Vert V_\omega\right\Vert\sqrt{\left\langle\pi_{x,j}^2
\Phi_{\omega,0}^{(N)},\pi_{x,j}^2\Phi_{\omega,0}^{(N)}\right\rangle},
\label{pi4boundint}
\end{eqnarray}
where we have used the assumption (\ref{U2assumption}) about $U^{(2)}$ and 
\begin{equation}
\left\langle\pi_{x,j}\Phi_{\omega,0}^{(N)},
\left(H_{\omega,0}^{(N)}-H_{\omega,0}^{(N,j)}\right)
\pi_{x,j}\Phi_{\omega,0}^{(N)}\right\rangle\ge 
E_{\omega,0}^{(N-1)}\left\langle\pi_{x,j}\Phi_{\omega,0}^{(N)},
\pi_{x,j}\Phi_{\omega,0}^{(N)}\right\rangle.
\end{equation}
Further the inequality thus obtained is rewritten as 
\begin{eqnarray}
& &
\left\langle\pi_{x,j}^2\Phi_{\omega,0}^{(N)},\pi_{x,j}^2
\Phi_{\omega,0}^{(N)}\right\rangle
-2m_e\left\Vert V_\omega\right\Vert\sqrt{\left\langle\pi_{x,j}^2
\Phi_{\omega,0}^{(N)},\pi_{x,j}^2\Phi_{\omega,0}^{(N)}\right\rangle}\ret
&\le&\hbar^2 e^2B^2+2m_e\hbar^2\alpha{\cal U}
+\max\left\{0,E_{\omega,0}^{(N)}-E_{\omega,0}^{(N-1)}\right\}\times
\frac{4m_e^2 E_{\omega,0}^{(N)}}{N}\ret
&\le&\hbar^2 e^2B^2+2m_e\hbar^2\alpha{\cal U}
+\max\left\{0,E_{\omega,0}^{(N)}-E_{\omega,0}^{(N-1)}\right\}\times
4m_e^2{\tilde {\cal E}}
\label{Phipi4Phibound}
\end{eqnarray}
by using Lemma~\ref{EUbound} in Appendix~\ref{EstEU}. 
The energy difference $E_{\omega,0}^{(N)}-E_{\omega,0}^{(N-1)}$ is evaluated 
as follows:

\begin{lemma}
\label{EN-EN-1bound}
Let $n$ be an integer such that 
\begin{equation}
n+\frac{1}{2}\ge N_{\rm min}^{2/3}. 
\end{equation}
Then 
\begin{equation}
E_{\omega,0}^{(N)}-E_{\omega,0}^{(N-1)}\le\hbar\omega_c
\left(n+\frac{1}{2}\right)+2{\cal C}_5'\frac{N}{\sqrt{2n+1}}+{\cal C}_6', 
\label{E-Ebound}
\end{equation}
where ${\cal C}_5'$ and ${\cal C}_6'$ are positive constant which are 
independent of the linear dimensions $L_x,L_y$ of the system and 
of the number $N$ of the electrons. 
\end{lemma}
The proof is given in Appendix~\ref{estE-E}. 
\medskip

\begin{proof}{Proof of Proposition~\ref{proPhipi4Phibound}}
{From} (\ref{Phipi4Phibound}), we have 
\begin{eqnarray}
& &
\left\langle\pi_{x,j}^2\Phi_{\omega,0}^{(N)},\pi_{x,j}^2
\Phi_{\omega,0}^{(N)}\right\rangle
\ret&\le&4m_e^2\left\Vert V_\omega\right\Vert^2+4\hbar^2 e^2B^2
+8m_e\hbar^2\alpha{\cal U}
+\max\left\{0,E_{\omega,0}^{(N)}-E_{\omega,0}^{(N-1)}\right\}
\times 16m_e^2{\tilde {\cal E}}. 
\label{Phipi4Phibound2}
\end{eqnarray}
{From} the bound (\ref{E-Ebound}), we have 
\begin{equation}
E_{\omega,0}^{(N)}-E_{\omega,0}^{(N-1)}
\le\hbar\omega_c\left({\cal C}_5N^{2/3}+{\cal C}_6\right)
\label{E-Ebound2}
\end{equation}
by choosing $n$ as 
\begin{equation}
N^{2/3}+1\ge n+\frac{1}{2}>N^{2/3}\ge N_{\rm min}^{2/3}. 
\end{equation}
Substituting (\ref{E-Ebound2}) into the above (\ref{Phipi4Phibound2}), 
we obtain the desired bound (\ref{Phipi4Phibpund}) in 
Proposition~\ref{proPhipi4Phibound}.
\end{proof}

\Section{Estimates of the ground state energy $E_{\omega,0}^{(N)}$ and 
the ground state expectation of $U^{(N)}$}
\label{EstEU}

The aim of this appendix is to estimate 
the ground state energy $E_{\omega,0}^{(N)}$ of the Hamiltonian 
$H_{\omega,0}^{(N)}$ of (\ref{unperturbedHam}) and 
the expectation value of $U^{(N)}$ with respect to the ground state 
$\Phi_{\omega,0}^{(N)}$. The results are summarized as follows: 

\begin{lemma}
\label{EUbound}
Let $\Phi_{\omega,0}^{(N)}$ be the ground state eigenvector of the 
Hamiltonian $H_{\omega,0}^{(N)}$ with the energy eigenvalue 
$E_{\omega,0}^{(N)}$. Then the following two bounds are valid:
\begin{equation}
\frac{E_{\omega,0}^{(N)}}{N}\le\frac{(\nu+1)^2}{2\nu}\hbar\omega_c
+\left\Vert V_\omega\right\Vert+\frac{eB(\nu+1)}{h}
\left[\left\Vert U^{(2)}\right\Vert_1+\varepsilon_1(\nu)
\right]\le{\tilde {\cal E}},
\label{E/Nbound}
\end{equation}
and
\begin{equation}
\frac{1}{N}\left\langle\Phi_{\omega,0}^{(N)},U^{(N)}\Phi_{\omega,0}^{(N)}
\right\rangle\le 2\left\Vert V_\omega\right\Vert
+\frac{eB(\nu+1)}{h}\left[\left\Vert U^{(2)}\right\Vert_1+
\varepsilon_1(\nu)\right]
\le{\cal U}.
\label{avU/Nbound}
\end{equation}
Here ${\tilde {\cal E}}$ and ${\cal U}$ are positive constants 
which are independent of 
the linear dimensions $L_x,L_y$ of the system, and $\varepsilon_1(\nu)$ 
is a small real number which tends to zero as $L_x,L_y\rightarrow +\infty$. 
The norm $\Vert\cdots\Vert_1$ is defined as 
\begin{equation}
\Vert f\Vert_1:=\int_S|f(x,y)|dxdy
\end{equation}
for a function $f$ on $S$. 
\end{lemma}

We begin with the following lemma: 

\begin{lemma}
\label{totalEbound1}
The following two bounds are valid:
\begin{equation}
E_{\omega,0}^{(N)}\le N\frac{(\nu+1)^2}{2\nu}\hbar\omega_c
+N\left\Vert V_\omega\right\Vert+
\left\langle\Phi_0^{(N)},U^{(N)}\Phi_0^{(N)}\right\rangle,
\label{totalEbound}
\end{equation}
and
\begin{equation}
\left\langle\Phi_{\omega,0}^{(N)},U^{(N)}\Phi_{\omega,0}^{(N)}\right\rangle
\le 2N\left\Vert V_\omega\right\Vert+\left\langle\Phi_0^{(N)},U^{(N)}
\Phi_0^{(N)}\right\rangle,
\label{Ubound}
\end{equation}
where the vector $\Phi_0^{(N)}$ is the $N$ electrons ground state eigenvector 
of the non-interacting Hamiltonian 
\begin{equation}
\sum_{j=1}^N {\cal H}_j=\sum_{j=1}^N\frac{1}{2m_e}
\left[(p_{x,j}-eBy_j)^2+p_{y,j}^2\right]
\end{equation}
with the periodic boundary conditions. 
\end{lemma}

\begin{proof}{Proof}
By definition, we have 
\begin{eqnarray}
E_{\omega,0}^{(N)}=
\left\langle\Phi_{\omega,0}^{(N)},H_{\omega,0}^{(N)}
\Phi_{\omega,0}^{(N)}\right\rangle
&\le&\left\langle\Phi_0^{(N)},H_{\omega,0}^{(N)}
\Phi_0^{(N)}\right\rangle\ret
&\le&\sum_{j=1}^N\left\langle\Phi_0^{(N)},{\cal H}_j\Phi_0^{(N)}\right\rangle
+N\left\Vert V_\omega\right\Vert+
\left\langle\Phi_0^{(N)},U^{(N)}\Phi_0^{(N)}\right\rangle.\ret
\label{trivialEbound}
\end{eqnarray}
Therefore the first bound (\ref{totalEbound}) follows from 
\begin{equation}
\sum_{j=1}^N\left\langle\Phi_0^{(N)},{\cal H}_j\Phi_0^{(N)}\right\rangle
=\sum_\ell \left(n_\ell+\frac{1}{2}\right)\hbar\omega_c
\le M\frac{(\nu+1)^2}{2}\hbar\omega_c,
\end{equation}
where the second sum runs over all the states $\ell$ in the Fermi sea.
Further, by combining (\ref{trivialEbound}) with
\begin{equation}
\sum_{j=1}^N\left\langle\Phi_0^{(N)},{\cal H}_j\Phi_0^{(N)}\right\rangle
\le\sum_{j=1}^N\left\langle\Phi_{\omega,0}^{(N)},{\cal H}_j
\Phi_{\omega,0}^{(N)}\right\rangle,
\end{equation}
we get the second bound (\ref{Ubound}). 
\end{proof}

\noindent
Owing to this lemma, it is sufficient to estimate the expactation 
$\left\langle\Phi_0^{(N)},U^{(N)}\Phi_0^{(N)}\right\rangle$. 
For this purpose, we use the following lemma: 

\begin{lemma}
\label{U2bound}
Let $\phi_{n,k}^{\rm P}$ be the eigenvectors (\ref{phiP}) of the single 
electron Hamiltonian ${\cal H}$ of (\ref{singleham}) with the peridic 
boundary conditions (\ref{PBC}). Then 
\begin{equation}
\sum_k \int_S dx_idy_i \ U^{(2)}(x_i-x_j,y_i-y_j)
\left|\phi_{n,k}^{\rm P}(x_i,y_i)\right|^2
=\frac{eB}{h}\left[\left\Vert U^{(2)}\right\Vert_1+\varepsilon_1^{(n)}\right] 
\label{U2boundexp}
\end{equation}
for any $(x_j,y_j)\in{\bf R}^2$. 
Here the sum is over all the wavenumbers $k$ for a fixed Landau index $n$, 
and the small real number $\varepsilon_1^{(n)}$ tends to zero uniformly 
in the Landau index $n$ as $L_x,L_y\rightarrow +\infty$. 
\end{lemma}

\begin{proof}{Proof}
Consider the function 
\begin{equation}
\rho_n(x,y):=\sum_k\left|\phi_{n,k}^{\rm P}(x,y)\right|^2.
\end{equation}
{From} the definition (\ref{phiP}) of the vector $\phi_{n,k}^{\rm P}$, 
the function $\rho_n$ is periodic in both $x$ and $y$ directions as 
\begin{equation}
\rho_n(x,y)=\rho_n(x+\Delta x,y)=\rho_n(x,y+\Delta y),
\end{equation}
where 
\begin{equation}
\Delta x=\frac{h}{eB}\frac{1}{L_y} \quad \mbox{and} \quad 
\Delta y=\frac{h}{eB}\frac{1}{L_x}. 
\end{equation}
{From} this periodicity and the periodicity (\ref{PBCU2}) of 
the two-body interaction $U^{(2)}$, we can assume $|x_j|\le\Delta x/2, 
|y_j|\le\Delta y/2$. 
The integral of $\rho_n$ on the unit cell $\Delta_{\ell,m}$ becomes 
\begin{equation}
\int_{\Delta_{\ell,m}}dxdy\ \rho_n(x,y)=\frac{1}{M},
\end{equation}
where 
\begin{equation}
\Delta_{\ell,m}:=[x_\ell,x_{\ell+1}]\times[y_m,y_{m+1}]
\end{equation}
with
\begin{equation}
x_\ell=-\frac{L_x}{2}+(\ell-1)\Delta x \quad \mbox{for} \ 
\ell=1,2,\ldots,M
\end{equation}
and
\begin{equation}
y_m=-\frac{L_y}{2}+(m-1)\Delta y \quad \mbox{for} \ 
m=1,2,\ldots,M.
\end{equation}
Since the function $U^{(2)}$ is continuous by the assumption, there exists 
a point $(\xi^{\ell,m},\eta^{\ell,m})\in \Delta_{\ell,m}$ such that 
\begin{eqnarray}
& &\int_{\Delta_{\ell,m}}dx_idy_i \ U^{(2)}(x_i-x_j,y_i-y_j)\rho_n(x_i,y_i)\ret
&=&U^{(2)}(\xi^{\ell,m}-x_j,\eta^{\ell,m}-y_j)\int_{\Delta_{\ell,m}}dx_idy_i \ 
\rho_n(x_i,y_i)\ret
&=&\frac{U^{(2)}(\xi^{\ell,m}-x_j,\eta^{\ell,m}-y_j)}{M}.
\label{localintUrho}
\end{eqnarray}
Using (\ref{localintUrho}) and the definitions of $\Delta x,\Delta y$, we get 
\begin{eqnarray}
& &\sum_k\int_S dx_idy_i\ U^{(2)}(x_i-x_j,y_i-y_j)
\left|\phi_{n,k}^{\rm P}(x_i,y_i)\right|^2\ret
&=&\int_S dx_idy_i \ U^{(2)}(x_i-x_j,y_i-y_j)\rho_n(x_i,y_i)\ret
&=&\frac{eB}{h}\sum_{\ell,m}U^{(2)}(\xi^{\ell,m}-x_j,\eta^{\ell,m}-y_j)
\Delta x\Delta y \ret
&=&\frac{eB}{h}\sum_{\sqrt{(\xi^{\ell,m})^2+(\eta^{\ell,m})^2}\le R'}
U^{(2)}(\xi^{\ell,m}-x_j,\eta^{\ell,m}-y_j)\Delta x\Delta y\ret
&+&\frac{eB}{h}\sum_{\sqrt{(\xi^{\ell,m})^2+(\eta^{\ell,m})^2}>R'}
U^{(2)}(\xi^{\ell,m}-x_j,\eta^{\ell,m}-y_j)
\Delta x\Delta y 
\end{eqnarray}
with a large positive number $R'$. 
Since $U^{(2)}$ is continuous, the first term in the last line 
converges to 
\begin{equation}
\frac{eB}{h}\int_{\sqrt{(x_i)^2+(y_i)^2}\le R'} 
dx_idy_i\ U^{(2)}(x_i-x_j,y_i-y_j)
\end{equation}
as $L_x,L_y\rightarrow +\infty$. The second term is vanishing uniformly 
in $n,L_x,L_y$ as $R'\rightarrow +\infty$ from the assumption 
(\ref{powerdecayassumptionU2}) about $U^{(2)}$. 
Thus the statement of the lemma is proved. 
\end{proof}

\begin{proof}{Proof of Lemma~\ref{EUbound}}
Note that 
\begin{eqnarray}
& &\left\langle\Phi_0^{(N)},U^{(N)}\Phi_0^{(N)}\right\rangle\ret
&=&\frac{1}{2}\sum_{m,k,n,k'}\int dx_idy_i \int dx_jdy_j \
\left[\phi_{m,k}^{\rm P}({\bf r}_i)\right]^\ast
\left[\phi_{n,k'}^{\rm P}({\bf r}_j)\right]^\ast
U^{(2)}({\bf r}_{ij})
\phi_{m,k}^{\rm P}({\bf r}_i)\phi_{n,k'}^{\rm P}({\bf r}_j)\ret
&-&\frac{1}{2}\sum_{m,k,n,k'}\int dx_idy_i \int dx_jdy_j \
\left[\phi_{m,k}^{\rm P}({\bf r}_i)\right]^\ast
\left[\phi_{n,k'}^{\rm P}({\bf r}_j)\right]^\ast 
U^{(2)}({\bf r}_{ij})
\phi_{n,k'}^{\rm P}({\bf r}_i)\phi_{m,k}^{\rm P}({\bf r}_j)\ret
&\le&\sum_{m,k,n,k'}\int dx_idy_i \int dx_jdy_j \ 
U^{(2)}({\bf r}_{ij})
\left|\phi_{m,k}^{\rm P}({\bf r}_i)\right|^2
\left|\phi_{n,k'}^{\rm P}({\bf r}_j)\right|^2,
\label{UNexpectation}
\end{eqnarray}
where we have written ${\bf r}_{ij}=(x_i-x_j,y_i-y_j)$ for simplicity. 

On the other hand we have 
\begin{equation}
\sum_{m,k}\int dx_i dy_i\ U^{(2)}({\bf r}_{ij})
\left|\phi_{m,k}^{\rm P}({\bf r}_i)\right|^2
\le\frac{eB(\nu+1)}{h}\left[\left\Vert U^{(2)}\right\Vert_1+
\varepsilon_1(\nu)\right]
\end{equation}
{from} Lemma~\ref{U2bound}. Here $\varepsilon_1(\nu)$ is a small real 
number which tends to zero as $L_x,L_y\rightarrow +\infty$. 
Substituting this inequality into 
the right-hand side of (\ref{UNexpectation}), we gat 
\begin{equation}
\left\langle\Phi_0^{(N)},U^{(N)}\Phi_0^{(N)}\right\rangle
\le N\left\{\frac{eB(\nu+1)}{h}\left[\left\Vert U^{(2)}\right\Vert_1+
\varepsilon_1(\nu)\right]\right\}. 
\end{equation}
Combining this with Lemma~\ref{totalEbound1}, we obtain 
the bounds in Lemma~\ref{EUbound}. 
\end{proof}

\Section{Estimate of $E_{\omega,0}^{(N)}-E_{\omega,0}^{(N-1)}$}
\label{estE-E}

In this appendix, we prove Lemma~\ref{EN-EN-1bound}. For this purpose 
we consider 
\begin{equation}
E_{\omega,0}^{(N)}\le\frac{\eta\left(H_{\omega,0}^{(N)}\right)}
{\eta(1)},
\label{ENvariation}
\end{equation}
where 
\begin{equation}
\eta(\cdots)=\frac{1}{M}\sum_k \left\langle\Phi_{\omega,(n,k)}^{(N)}, 
(\cdots)\Phi_{\omega,(n,k)}^{(N)}\right\rangle
\end{equation}
with 
\begin{equation}
\Phi_{\omega,(n,k)}^{(N)}={\rm Asym}
\left[\Phi_{\omega,0}^{(N-1)}\otimes\phi_{n,k}^{\rm P}\right]. 
\end{equation}
Here ${\rm Asym}[\cdots]$ is the antisymmetrization of a wavefunction, 
whose definition is given in (\ref{defAsym}), $\Phi_{\omega,0}^{(N-1)}$ is 
the $N-1$ electrons ground state eigenvector of the Hamiltonian 
$H_{\omega,0}^{(N-1)}$ with norm one, 
and $\phi_{n,k}^{\rm P}$ are the normalized eigenvectors (\ref{phiP}) of 
the single electron Hamiltonian ${\cal H}$ of (\ref{singleham}) 
with the periodic boundary conditions (\ref{PBC}). 
We introduce an orthogonal decomposition of the vector 
$\Phi_{\omega,0}^{(N-1)}$ as 
\begin{equation}
\Phi_{\omega,0}^{(N-1)}=\Psi_{1,(n,k)}^{(N-1)}+\Psi_{2,(n,k)}^{(N-1)}
\label{decomPhiomega12}
\end{equation}
with 
\begin{equation}
\Psi_{1,(n,k)}^{(N-1)}=\prod_{j=1}^{N-1}\left[1-P_j^{(n,k)}\right]
\Phi_{\omega,0}^{(N-1)}, 
\end{equation}
where $P^{(n,k)}$ is the orthogonal projection onto the vector 
$\phi_{n,k}^{\rm P}$. 
Then ${\rm Asym}\left[\Psi_{2,(n,k)}^{(N-1)}\otimes\phi_{n,k}^{\rm P}\right]$ 
is identically zero because the vector $\Phi_{\omega,0}^{(N-1)}$ 
is expanded as 
\begin{equation}
\Phi_{\omega,0}^{(N-1)}=
\sum_{\{\xi_j\}} a_{\{\xi_j\}}{\rm Asym}\left[\phi_{\xi_1}^{\rm P}
\otimes\phi_{\xi_2}^{\rm P}\otimes\cdots\otimes\phi_{\xi_{N-1}}^{\rm P}
\right]
\end{equation}
in terms of the vectors $\{\phi_{n,k}^{\rm P}\}$. Here we denote by $\xi$ 
the pair of a Landau index $n$ and a wavenumber $k$, i.e., $\xi_j=(n_j,k_j)$. 
{From} this observation, we have 
\begin{eqnarray}
\eta(1)&=&\frac{1}{M}\sum_k\left\langle\Phi_{\omega,(n,k)}^{(N)},
\Phi_{\omega,(n,k)}^{(N)}\right\rangle\ret
&=&\frac{1}{M}\sum_k
\left\langle{\rm Asym}\left[\Phi_{\omega,0}^{(N-1)}\otimes\phi_{n,k}^{\rm P}
\right],
{\rm Asym}\left[\Phi_{\omega,0}^{(N-1)}\otimes\phi_{n,k}^{\rm P}\right]
\right\rangle\ret
&=&\frac{1}{M}\sum_k
\left\langle{\rm Asym}\left[\Psi_{1,(n,k)}^{(N-1)}\otimes\phi_{n,k}^{\rm P}
\right],
{\rm Asym}\left[\Psi_{1,(n,k)}^{(N-1)}\otimes\phi_{n,k}^{\rm P}\right]
\right\rangle,\ret
&=&\frac{1}{M}\sum_k\left\Vert\Psi_{1,(n,k)}^{(N-1)}\right\Vert^2
=1-\frac{1}{M}\sum_k\left\Vert\Psi_{2,(n,k)}^{(N-1)}\right\Vert^2.
\label{etaidentity}
\end{eqnarray}

\begin{lemma}
The following bound is valid:
\begin{equation}
\frac{1}{M}\sum_k\left\Vert\Psi_{2,(n,k)}^{(N-1)}\right\Vert^2
\le\frac{N_{\rm min}^{2/3}}{2n+1}
\label{avPsinormbound}
\end{equation}
with the positive constant $N_{\rm min}$ is given by (\ref{Nmin}). 
\end{lemma}

\begin{proof}{Proof}
By definition, we have 
\begin{equation}
\left\Vert\Psi_{2,(n,k)}^{(N-1)}\right\Vert^2
=\sum_{\{\xi_j\}}\left|a_{\{\xi_j\}}\right|^2
\sum_{\xi'\in\{\xi_1,\xi_2,\ldots,\xi_{N-1}\}}
\left(\phi_{\xi'},P^{(n,k)}\phi_{\xi'}\right). 
\end{equation}
Clearly, 
\begin{equation}
\frac{1}{M}\sum_k\left\Vert\Psi_{2,(n,k)}^{(N-1)}\right\Vert^2
=\frac{1}{M}\sum_{\{\xi_j\}}\left|a_{\{\xi_j\}}\right|^2
\sum_{\xi'\in\{\xi_1,\xi_2,\ldots,\xi_{N-1}\}}
\left(\phi_{\xi'},P^{(n)}\phi_{\xi'}\right), 
\label{repavPsinorm}
\end{equation}
where $P^{(n)}=\sum_k P^{(n,k)}$, i.e., the orthogonal projection 
onto the Landau level with the index $n$. 

On the other hand we have, for the ground state energy $E_{\omega,0}^{(N-1)}$, 
\begin{eqnarray}
\frac{E_{\omega,0}^{(N-1)}}{N-1}&=&\frac{1}{N-1}
\left\langle\Phi_{\omega,0}^{(N-1)},H_{\omega,0}^{(N-1)}\Phi_{\omega,0}^{(N-1)}
\right\rangle\ret
&=&\frac{1}{N-1}\sum_{j=1}^{N-1}\left\langle\Phi_{\omega,0}^{(N-1)},{\cal H}_j
\Phi_{\omega,0}^{(N-1)}\right\rangle
+\frac{1}{N-1}\sum_{j=1}^{N-1}\left\langle\Phi_{\omega,0}^{(N-1)},
V_\omega({\bf r}_j)\Phi_{\omega,0}^{(N-1)}\right\rangle\ret
&+&\frac{1}{N-1}\left\langle\Phi_{\omega,0}^{(N-1)},U^{(N-1)}
\Phi_{\omega,0}^{(N-1)}\right\rangle\ret
&\ge& \frac{1}{N-1}\sum_{j=1}^{N-1}\left\langle\Phi_{\omega,0}^{(N-1)},
P_j^{(n)}{\cal H}_j\Phi_{\omega,0}^{(N-1)}\right\rangle
-\left\Vert V_\omega\right\Vert\ret
&=&\frac{1}{N-1}\hbar\omega_c\left(n+\frac{1}{2}\right)\sum_{j=1}^{N-1}
\left\langle\Phi_{\omega,0}^{(N-1)},P_j^{(n)}\Phi_{\omega,0}^{(N-1)}
\right\rangle-\left\Vert V_\omega\right\Vert\ret
&=&\frac{1}{N-1}\hbar\omega_c\left(n+\frac{1}{2}\right)
\sum_{\{\xi_j\}}\left|a_{\{\xi_j\}}\right|^2
\sum_{\xi'\in\{\xi_1,\xi_2,\ldots,\xi_{N-1}\}}
\left(\phi_{\xi'},P^{(n)}\phi_{\xi'}\right)
-\left\Vert V_\omega\right\Vert.\ret
\end{eqnarray}
Combining this with the above (\ref{repavPsinorm}), we get 
\begin{equation}
\frac{1}{M}\sum_k\left\Vert\Psi_{2,(n,k)}^{(N-1)}\right\Vert^2
\le\frac{\nu}{\hbar\omega_c}\left(\left\Vert V_\omega\right\Vert
+\frac{E_{\omega,0}^{(N-1)}}{N-1}\right)\frac{1}{n+1/2}.
\end{equation}
Using the bound (\ref{E/Nbound}) in Lemma~\ref{EUbound} in the preceding 
appendix and $N_{\rm min}$ of (\ref{Nmin}), we obtain 
the desired bound (\ref{avPsinormbound}). 
\end{proof}

{From} (\ref{etaidentity}) and (\ref{avPsinormbound}), we have 

\begin{coro}
\begin{equation}
\eta(1)\ge \frac{1}{2}\quad \mbox{for}\ \ n+\frac{1}{2}\ge N_{\rm min}^{2/3}.
\label{etalbound}
\end{equation}
\end{coro}

Next consider the numerator of the right-hand side of (\ref{ENvariation}), 
\begin{equation}
\eta\left(H_{\omega,0}^{(N)}\right)
=\frac{1}{M}\sum_k\left\langle\Phi_{\omega,(n,k)}^{(N)},H_{\omega,0}^{(N)}
\Phi_{\omega,(n,k)}^{(N)}\right\rangle. 
\label{expreEtaH}
\end{equation}
By definition, we have 
\begin{eqnarray}
\left\langle\Phi_{\omega,(n,k)}^{(N)},H_{\omega,0}^{(N)}
\Phi_{\omega,(n,k)}^{(N)}\right\rangle
&=&\left\langle{\rm Asym}\left[\Phi_{\omega,0}^{(N-1)}\otimes
\phi_{n,k}^{\rm P}\right],H_{\omega,0}^{(N)}{\rm Asym}
\left[\Phi_{\omega,0}^{(N-1)}\otimes\phi_{n,k}^{\rm P}\right]\right\rangle\ret
&=&\left\langle{\rm Asym}\left[\Phi_{\omega,0}^{(N-1)}\otimes
\phi_{n,k}^{\rm P}\right],{\rm Asym}\left[H_{\omega,0}^{(N)}
\Phi_{\omega,0}^{(N-1)}\otimes\phi_{n,k}^{\rm P}\right]\right\rangle\ret
&=&E_{\omega,0}^{(N-1)}\left\Vert\Phi_{\omega,(n,k)}^{(N)}\right\Vert^2
+\hbar\omega_c\left(n+\frac{1}{2}\right)
\left\Vert\Phi_{\omega,(n,k)}^{(N)}\right\Vert^2\ret
&+&\left\langle{\rm Asym}\left[\Phi_{\omega,0}^{(N-1)}\otimes
\phi_{n,k}^{\rm P}\right],{\rm Asym}\left[\Phi_{\omega,0}^{(N-1)}\otimes
V_\omega\phi_{n,k}^{\rm P}\right]\right\rangle\ret
&+&\left\langle{\rm Asym}\left[\Phi_{\omega,0}^{(N-1)}\otimes
\phi_{n,k}^{\rm P}\right],{\rm Asym}\left[U^{(N,\cdots)}
\Phi_{\omega,0}^{(N-1)}\otimes\phi_{n,k}^{\rm P}\right]\right\rangle,\ret
\end{eqnarray}
where the operator $U^{(N,\cdots)}$ is defined as 
\begin{eqnarray}
& &\left(U^{(N,j)}\Phi_{\omega,0}^{(N-1)}\otimes\phi_{n,k}^{\rm P}\right)
({\bf r}_1,{\bf r}_2,\ldots,{\bf r}_{j-1},{\bf r}_{j+1},\ldots,{\bf r}_N,
{\bf r}_j)\ret
&=&\sum_{i\ne j}U^{(2)}(x_i-x_j,y_i-y_j)\Phi_{\omega,0}^{(N-1)}
({\bf r}_1,{\bf r}_2,\ldots,{\bf r}_{j-1},{\bf r}_{j+1},\ldots,{\bf r}_N)
\phi_{n,k}^{\rm P}({\bf r}_j). 
\end{eqnarray}
Substituting this into the right-hand side of (\ref{expreEtaH}), we obtain 
\begin{eqnarray}
\eta\left(H_{\omega,0}^{(N)}\right)
&=&\left[E_{\omega,0}^{(N-1)}+\hbar\omega_c\left(n+\frac{1}{2}\right)\right]
\eta(1)\ret
&+&\frac{1}{M}\sum_k\left\langle{\rm Asym}\left[\Phi_{\omega,0}^{(N-1)}
\otimes\phi_{n,k}^{\rm P}\right],{\rm Asym}\left[\Phi_{\omega,0}^{(N-1)}\otimes
V_\omega\phi_{n,k}^{\rm P}\right]\right\rangle\ret
&+&\frac{1}{M}\sum_k\left\langle{\rm Asym}\left[\Phi_{\omega,0}^{(N-1)}\otimes
\phi_{n,k}^{\rm P}\right],{\rm Asym}\left[U^{(N,\cdots)}\Phi_{\omega,0}^{(N-1)}
\otimes\phi_{n,k}^{\rm P}\right]\right\rangle.\ret
\label{expPhiHPhink}
\end{eqnarray}
The first sum in the right-hand side is written as 
\begin{eqnarray}
& &\frac{1}{M}\sum_k\left\langle{\rm Asym}\left[\Phi_{\omega,0}^{(N-1)}\otimes
\phi_{n,k}^{\rm P}\right],{\rm Asym}\left[\Phi_{\omega,0}^{(N-1)}\otimes
V_\omega\phi_{n,k}^{\rm P}\right]\right\rangle\ret
&=&\frac{1}{M}\sum_k\left\langle{\rm Asym}\left[\Psi_{1,(n,k)}^{(N-1)}\otimes
\phi_{n,k}^{\rm P}\right],{\rm Asym}\left[\Psi_{3,(n,k)}^{(N-1)}\otimes
V_\omega\phi_{n,k}^{\rm P}\right]\right\rangle,
\end{eqnarray}
where
\begin{equation}
\Psi_{3,(n,k)}^{(N-1)}=\prod_{j=1}^{N-1}
\left[1-P_j^{(n,k)}(V_\omega)\right]\Phi_{\omega,0}^{(N-1)}.
\end{equation}
Here $P^{(n,k)}(V_\omega)$ is the orthogonal projection onto the vector 
$V_\omega\phi_{n,k}^{\rm P}$. Using the Schwarz inequality, we have 
\begin{eqnarray}
& &\left|\frac{1}{M}\sum_k\left\langle{\rm Asym}\left[\Phi_{\omega,0}^{(N-1)}
\otimes\phi_{n,k}^{\rm P}\right],{\rm Asym}\left[\Phi_{\omega,0}^{(N-1)}\otimes
V_\omega\phi_{n,k}^{\rm P}\right]\right\rangle\right|\ret
&\le&\frac{1}{M}\sum_k\left\Vert\Psi_{1,(n,k)}^{(N-1)}\right\Vert\ 
\left\Vert\Psi_{3,(n,k)}^{(N-1)}\right\Vert\ 
\left\Vert V_\omega\phi_{n,k}^{\rm P}\right\Vert\ret
&\le&\frac{1}{M}\sum_k\left\Vert\Phi_{\omega,0}^{(N-1)}\right\Vert^2
\left\Vert V_\omega\right\Vert=\left\Vert V_\omega\right\Vert.
\label{APhiVAPhi}
\end{eqnarray}

The second sum in the right-hand side of (\ref{expPhiHPhink}) is written as 
\begin{eqnarray}
& &\frac{1}{M}\sum_k\left\langle{\rm Asym}\left[\Phi_{\omega,0}^{(N-1)}
\otimes\phi_{n,k}^{\rm P}\right],{\rm Asym}\left[U^{(N,\cdots)}
\Phi_{\omega,0}^{(N-1)}\otimes\phi_{n,k}^{\rm P}\right]\right\rangle\ret
&=&\frac{1}{M}\sum_k\left\langle{\rm Asym}\left[\Psi_{1,(n,k)}^{(N-1)}\otimes
\phi_{n,k}^{\rm P}\right],{\rm Asym}\left[U^{(N,\cdots)}\Psi_{1,(n,k)}^{(N-1)}
\otimes\phi_{n,k}^{\rm P}\right]\right\rangle\ret
&+&\frac{1}{M}\sum_k\left\langle{\rm Asym}\left[\Psi_{1,(n,k)}^{(N-1)}\otimes
\phi_{n,k}^{\rm P}\right],{\rm Asym}\left[U^{(N,\cdots)}\Psi_{2,(n,k)}^{(N-1)}
\otimes\phi_{n,k}^{\rm P}\right]\right\rangle
\label{PhiUPhi}
\end{eqnarray}
by using the decomposition (\ref{decomPhiomega12}). 
This second sum in the right-hand side is evaluated as follows: 

\begin{lemma}
The following bound is valid:
\begin{eqnarray}
& &\left|
\frac{1}{M}\sum_k\left\langle{\rm Asym}\left[\Psi_{1,(n,k)}^{(N-1)}\otimes
\phi_{n,k}^{\rm P}\right],{\rm Asym}\left[U^{(N,\cdots)}\Psi_{2,(n,k)}^{(N-1)}
\otimes\phi_{n,k}^{\rm P}\right]\right\rangle\right|\ret
&\le& N_{\rm min}^{1/3}
\sqrt{\left(\varepsilon_2+\alpha/\pi\right)
\left\Vert U^{(2)}\right\Vert_1\left\Vert U^{(2)}\right\Vert_\infty}
\times\frac{N-1}{\sqrt{2n+1}},
\label{avAPsi1APsi1bound}
\end{eqnarray}
where $\alpha$ is the positive constant given in 
the assumption (\ref{U2assumption}) on the interaction $U^{(2)}$, and 
$\varepsilon_2$ is a positive number which tends to zero as 
$L_x,L_y\rightarrow +\infty$. The norm $\Vert\cdots\Vert_\infty$ is given by 
\begin{equation}
\Vert f\Vert_\infty:=\sup_{(x,y)\in S}|f(x,y)|
\end{equation}
for a continuous function $f$ on $S$. 
\end{lemma}

\begin{proof}{Proof}
The interaction potential $U^{(2)}$ is written as 
\begin{equation}
U^{(2)}(x_j-x_\ell,y_j-y_\ell)
=\frac{1}{\sqrt{L_xL_y}}\sum_{k_x,k_y}{\hat U}^{(2)}(k_x,k_y)
e^{ik_xx_j+ik_yy_j}e^{-ik_xx_\ell-ik_yy_\ell}
\end{equation}
in terms of the Fourier transform of ${\hat U}^{(2)}$. Clearly, 
\begin{eqnarray}
& &U^{(N,j)}({\bf r}_j;{\bf r}_1,\ldots,{\bf r}_{j-1},{\bf r}_{j+1},
\ldots,{\bf r}_N)\ret
&=&\sum_{\ell\ne j}U^{(2)}(x_j-x_\ell,y_j-y_\ell)\ret
&=&\frac{1}{\sqrt{L_xL_y}}\sum_{k_x,k_y}{\hat U}^{(2)}(k_x,k_y)
e^{ik_xx_j+ik_yy_j}\sum_{\ell\ne j}e^{-ik_xx_\ell-ik_yy_\ell}. 
\end{eqnarray}
Using this expression, we have 
$$
\frac{1}{M}\sum_k\left\langle{\rm Asym}\left[\Psi_{1,(n,k)}^{(N-1)}
\otimes\phi_{n,k}^{\rm P}\right],{\rm Asym}\left[U^{(N,\cdots)}
\Psi_{2,(n,k)}^{(N-1)}
\otimes\phi_{n,k}^{\rm P}\right]\right\rangle\hskip 3.8cm
$$
\begin{equation}
=\frac{1}{M}
\sum_{k_x,k_y,k}\frac{{\hat U}^{(2)}(k_x,k_y)}{\sqrt{L_xL_y}}
\left\langle{\rm Asym}\left[\Psi_{1,(n,k)}^{(N-1)}
\otimes\phi_{n,k}^{\rm P}\right],{\rm Asym}
\left[{\tilde \Psi}_{2,(n,k)}^{(N-1)}
(k_x,k_y)\otimes{\tilde \phi}_{n,k}^{\rm P}(k_x,k_y)\right]\right\rangle,
\label{FourierPsi1Psi2}
\end{equation}
where
\begin{equation}
{\tilde \Psi}_{2,(n,k)}^{(N-1)}(k_x,k_y)=\left\{
\prod_{\ell\ne j}\left[1-{\tilde P}_\ell^{(n,k)}(k_x,k_y)\right]\right\}
\sum_{\ell\ne j}e^{-ik_xx_\ell-ik_yy_\ell}\Psi_{2,(n,k)}^{(N-1)}, 
\end{equation}
and
\begin{equation}
{\tilde \phi}_{n,k}^{\rm P}(k_x,k_y)=e^{ik_xx+ik_yy}\phi_{n,k}^{\rm P}.
\end{equation}
Here ${\tilde P}^{(n,k)}(k_x,k_y)$ is the projection onto 
the vector ${\tilde \phi}_{n,k}^{\rm P}(k_x,k_y)$. 
Applying the Schwarz inequality to 
the right-hand side of (\ref{FourierPsi1Psi2}), we have 
\begin{eqnarray}
& &\left|\frac{1}{M}\sum_k\left\langle{\rm Asym}\left[\Psi_{1,(n,k)}^{(N-1)}
\otimes\phi_{n,k}^{\rm P}\right],{\rm Asym}\left[U^{(N,\cdots)}
\Psi_{2,(n,k)}^{(N-1)}
\otimes\phi_{n,k}^{\rm P}\right]\right\rangle\right|^2\ret
&\le&\left[\sum_{k_x,k_y}(k_x^2+k_y^2+\alpha)^2
\left|{\hat U}^{(2)}(k_x,k_y)\right|^2\frac{1}{M}\sum_k
\left\Vert\Psi_{1,(n,k)}^{(N-1)}\right\Vert^2\right]\ret
&\times&\left[\frac{1}{L_xL_y}
\sum_{k_x,k_y}\frac{1}{(k_x^2+k_y^2+\alpha)^2}\frac{1}{M}\sum_k
\left\Vert{\tilde \Psi}_{2,(n,k)}^{(N-1)}(k_x,k_y)\right\Vert^2\right]\ret
&\le&4\alpha^2\left\Vert U^{(2)}\right\Vert_1
\left\Vert U^{(2)}\right\Vert_\infty\times
\left(\frac{1}{4\pi\alpha}+\varepsilon_2'\right)\times(N-1)^2\frac{1}{M}
\sum_k\left\Vert\Psi_{2,(n,k)}^{(N-1)}\right\Vert^2, 
\label{APsi1APsi1bound}
\end{eqnarray}
where we have used the following three bounds:
\begin{equation}
\left\Vert{\tilde \Psi}_{2,(n,k)}^{(N-1)}(k_x,k_y)\right\Vert^2
\le(N-1)^2\left\Vert\Psi_{2,(n,k)}^{(N-1)}\right\Vert^2, 
\end{equation}
\begin{eqnarray}
\sum_{k_x,k_y}(k_x^2+k_y^2+\alpha)^2
\left|{\hat U}^{(2)}(k_x,k_y)\right|^2&=&\int dx_jdy_j 
\left|\left(\frac{\partial^2}{\partial x_j^2}+\frac{\partial^2}{\partial y_j^2}
+\alpha\right)U^{(2)}({\bf r}_{j\ell})\right|^2\ret
&\le&4\alpha^2\int dx_jdy_j\left|U^{(2)}({\bf r}_{j\ell})\right|^2
\ret
&\le&4\alpha^2\left\Vert U^{(2)}\right\Vert_1
\left\Vert U^{(2)}\right\Vert_\infty,
\label{boundkU2}
\end{eqnarray}
and
\begin{equation}
\frac{1}{L_xL_y}
\sum_{k_x,k_y}\frac{1}{(k_x^2+k_y^2+\alpha)^2}=\frac{1}{4\pi\alpha}
+\varepsilon_2'. 
\end{equation}
Clearly $\varepsilon_2'$ defined by the above equation 
is a real number which tends to zero as $L_x,L_y\rightarrow +\infty$. 
The bound (\ref{boundkU2}) is easily derived 
from the assumption (\ref{U2assumption}) about $U^{(2)}$. 
Combining (\ref{APsi1APsi1bound}) with 
(\ref{avPsinormbound}), we get the desired bound (\ref{avAPsi1APsi1bound}). 
\end{proof}

The first sum in the right-hand side of (\ref{PhiUPhi}) is evaluated as 
follows:

\begin{lemma}
The following bound is valid:
\begin{eqnarray}
& &\left|\frac{1}{M}\sum_k\left\langle{\rm Asym}\left[\Psi_{1,(n,k)}^{(N-1)}
\otimes\phi_{n,k}^{\rm P}\right],{\rm Asym}\left[U^{(N,\cdots)}
\Psi_{1,(n,k)}^{(N-1)}\otimes\phi_{n,k}^{\rm P}\right]\right\rangle\right|\ret
&\le&\frac{4\nu eB}{h}
\left[\left\Vert U^{(2)}\right\Vert_1+\varepsilon_1^{(n)}\right]
+4 N_{\rm min}^{2/3}\left\Vert U^{(2)}\right\Vert_\infty
\frac{N-1}{2n+1},
\label{Psi1Psi1bound}
\end{eqnarray}
where $\varepsilon_1^{(n)}$ is a positive number which tends to zero as 
$L_x,L_y\rightarrow \infty$. 
\end{lemma}

\begin{proof}{Proof}
Note that 
\begin{eqnarray}
& &\frac{1}{M}\sum_k\left\langle{\rm Asym}\left[\Psi_{1,(n,k)}^{(N-1)}
\otimes\phi_{n,k}^{\rm P}\right],{\rm Asym}\left[U^{(N,\cdots)}
\Psi_{1,(n,k)}^{(N-1)}\otimes\phi_{n,k}^{\rm P}\right]\right\rangle\ret
&=&\frac{N-1}{M}\sum_k\int dv^{(N)}
\left|\Psi_{1,(n,k)}^{(N-1)}({\bf r}_1,\ldots,{\bf r}_{N-1})\right|^2
U^{(2)}(x_{N-1}-x_N,y_{N-1}-y_N)\left|\phi_{n,k}^{\rm P}({\bf r}_N)\right|^2
\ret
&-&\frac{N-1}{M}\sum_k\int dv^{(N)}
\left[\Psi_{1,(n,k)}^{(N-1)}({\bf r}_1,\ldots,{\bf r}_{N-2},{\bf r}_N)\right]
^\ast\left[\phi_{n,k}^{\rm P}({\bf r}_{N-1})\right]^\ast\ret
& &\hskip 1cm\times U^{(2)}(x_{N-1}-x_N,y_{N-1}-y_N)\Psi_{1,(n,k)}^{(N-1)}
({\bf r}_1,\ldots,
{\bf r}_{N-2},{\bf r}_{N-1})\phi_{n,k}^{\rm P}({\bf r}_N).
\label{expPsi1Psi1}
\end{eqnarray}
Here $dv^{(N)}=dx_1dy_1dx_2dy_2\cdots dx_Ndy_N$.
Since the absolute value of the second term in the right-hand side of 
(\ref{expPsi1Psi1}) is bounded by the first term by using the Schwarz 
inequality, we have an inequality 
$$
\left|\frac{1}{M}\sum_k\left\langle{\rm Asym}\left[\Psi_{1,(n,k)}^{(N-1)}
\otimes\phi_{n,k}^{\rm P}\right],{\rm Asym}\left[U^{(N,\cdots)}
\Psi_{1,(n,k)}^{(N-1)}\otimes\phi_{n,k}^{\rm P}\right]\right\rangle\right|
\hspace{3cm}
$$
\begin{equation}
\le\frac{2(N-1)}{M}\sum_k\int dv^{(N)}
\left|\Psi_{1,(n,k)}^{(N-1)}({\bf r}_1,\ldots,{\bf r}_{N-1})\right|^2
U^{(2)}(x_{N-1}-x_N,y_{N-1}-y_N)\left|\phi_{n,k}^{\rm P}({\bf r}_N)\right|^2.
\label{Phi1UPhi1bound}
\end{equation}
Note that 
\begin{equation}
\left|\Psi_{1,(n,k)}^{(N-1)}({\bf r}_1,\ldots,{\bf r}_{N-1})\right|^2
\le 2\left[
\left|\Phi_{\omega,0}^{(N-1)}({\bf r}_1,\ldots,{\bf r}_{N-1})\right|^2
+\left|\Psi_{2,(n,k)}^{(N-1)}({\bf r}_1,\ldots,{\bf r}_{N-1})\right|^2
\right]
\end{equation}
which is easily obtained by using the decomposition 
$\Phi_{\omega,0}^{(N-1)}=\Psi_{1,(n,k)}^{(N-1)}
+\Psi_{2,(n,k)}^{(N-1)}$.  Substituting this inequality into 
the right-hand side of (\ref{Phi1UPhi1bound}), we get 
\begin{eqnarray}
& &\left|\frac{1}{M}\sum_k\left\langle{\rm Asym}\left[\Psi_{1,(n,k)}^{(N-1)}
\otimes\phi_{n,k}^{\rm P}\right],{\rm Asym}\left[U^{(N,\cdots)}
\Psi_{1,(n,k)}^{(N-1)}
\otimes\phi_{n,k}^{\rm P}\right]\right\rangle\right|\ret
&\le&\frac{4(N-1)}{M}\sum_k\int dv^{(N)}
\left|\Phi_{\omega,0}^{(N-1)}({\bf r}_1,\ldots,{\bf r}_{N-1})\right|^2
U^{(2)}(x_{N-1}-x_N,y_{N-1}-y_N)\left|\phi_{n,k}^{\rm P}({\bf r}_N)\right|^2
\ret
&+&\frac{4(N-1)}{M}
\sum_k\int dv^{(N)}
\left|\Psi_{2,(n,k)}^{(N-1)}({\bf r}_1,\ldots,{\bf r}_{N-1})\right|^2
U^{(2)}(x_{N-1}-x_N,y_{N-1}-y_N)\left|\phi_{n,k}^{\rm P}({\bf r}_N)\right|^2
\ret
&\le&\frac{4(N-1)}{M}\sum_k\int dv^{(N)}
\left|\Phi_{\omega,0}^{(N-1)}({\bf r}_1,\ldots,{\bf r}_{N-1})\right|^2
U^{(2)}(x_{N-1}-x_N,y_{N-1}-y_N)\left|\phi_{n,k}^{\rm P}({\bf r}_N)\right|^2
\ret&+&4(N-1)\left\Vert U^{(2)}\right\Vert_\infty
\frac{1}{M}\sum_k\left\Vert\Psi_{2,(n,k)}^{(N-1)}\right\Vert^2.
\end{eqnarray}
Combining this with (\ref{avPsinormbound}) and Lemma~\ref{U2bound}, 
we obtain the desired result (\ref{Psi1Psi1bound}). 
\end{proof}

\begin{proof}{Proof of Lemma~\ref{EN-EN-1bound}}
Combining (\ref{PhiUPhi}), (\ref{avAPsi1APsi1bound}), (\ref{Psi1Psi1bound}), 
we have 
\begin{eqnarray}
& &\left|\frac{1}{M}\sum_k\left\langle{\rm Asym}\left[\Phi_{\omega,0}^{(N-1)}
\otimes\phi_{n,k}^{\rm P}\right],{\rm Asym}\left[U^{(N,\cdots)}
\Phi_{\omega,0}^{(N-1)}\otimes\phi_{n,k}^{\rm P}\right]\right\rangle\right|\ret
&\le&N_{\rm min}^{1/3}\left[
\sqrt{\left(\varepsilon_2+\alpha/\pi\right)
\left\Vert U^{(2)}\right\Vert_1\left\Vert U^{(2)}\right\Vert_\infty}
+\frac{4N_{\rm min}^{1/3}\left\Vert U^{(2)}\right\Vert_\infty}{\sqrt{2n+1}}
\right]\frac{N-1}{\sqrt{2n+1}}\ret
&+&\frac{4\nu eB}{h}
\left[\left\Vert U^{(2)}\right\Vert_1+\varepsilon_1^{(n)}\right].
\label{APhiUAPhibound}
\end{eqnarray}
Combining this, (\ref{expPhiHPhink}) and (\ref{APhiVAPhi}), we obtain 
\begin{eqnarray}
\eta\left(H_{\omega,0}^{(N)}\right)
&\le&\left[E_{\omega,0}^{(N-1)}+\hbar\omega_c\left(n+\frac{1}{2}\right)\right]
\eta(1)+\left\Vert V_\omega\right\Vert\ret
&+&\frac{4\nu eB}{h}
\left[\left\Vert U^{(2)}\right\Vert_1+\varepsilon_1^{(n)}\right]+
{\cal C}_5'\frac{N-1}{\sqrt{2n+1}},
\end{eqnarray}
where ${\cal C}_5'$ is a positive constant.
Substituting this into the right-hand of (\ref{ENvariation}), we get 
\begin{eqnarray}
& &E_{\omega,0}^{(N)}-E_{\omega,0}^{(N-1)}\ret
&\le&\hbar\omega_c\left(n+\frac{1}{2}\right)
+2\left\Vert V_\omega\right\Vert+ \frac{8\nu eB}{h}
\left[\left\Vert U^{(2)}\right\Vert_1+\varepsilon_1^{(n)}\right]
+2{\cal C}_5'\frac{N-1}{\sqrt{2n+1}}\ret
\end{eqnarray}
for $n+1/2\ge N_{\rm min}^{2/3}$, where we have used (\ref{etalbound}). 
\end{proof}


\bigskip
\bigskip

\noindent
{\Large\bf Acknowledgements} 
\medskip

\noindent
It is a pleasure to thank the following people for discussions 
and correspondence: 
Mahito Kohmoto, Bruno Nachtergaele, Hal Tasaki, and Masanori Yamanaka. 


\end{document}